\documentclass{nature}
\usepackage[T1]{fontenc}
\usepackage{float}
\usepackage{deluxetable}
\usepackage{graphicx}
\usepackage{pdflscape}
\usepackage{hyperref}
\usepackage{pgf,pgffor}
\usepackage{textcomp}
\usepackage{threeparttable} 
\usepackage{xcolor}
\usepackage{ulem}
\usepackage{lineno}
\usepackage{caption}
\usepackage[caption = false]{subfig}
\usepackage{academicons}
\usepackage{pifont}
\usepackage{longtable}
\usepackage{amssymb}
\usepackage{cleveref}
\definecolor{orcidlogocol}{HTML}{A6CE39}
\usepackage[symbol]{footmisc}

\makeatletter

\def\@to{to}

\let\saved@includegraphics\includegraphics
\AtBeginDocument{\let\includegraphics\saved@includegraphics}
\renewenvironment*{figure}{\@float{figure}}{\end@float}
\makeatother

\newcommand{\mnras}{Mon. Not. R. Astron. Soc.}
\newcommand{\apj}{Astrophys. J.}
\newcommand{\apjs}{Astrophysical Journal, Supplement}
\newcommand{\aj}{Astronomical Journal} 
\newcommand{\apjl}{Astrophys. J. Lett.}
\newcommand{\apss}{{Astrophys. and Space Sc.}}
\newcommand{\aap}{Astron. Astrophys.}
\newcommand{\nat}{Nature}
\newcommand{\araa}{Annual Review of Astron and Astrophysis} 
\newcommand{\pasp}{Publications of the ASP}
\newcommand{\baas}{Bulletin of the AAS}

\newcommand{\msun}{M_\odot}

\newcommand{\farcs}{$.\!\!^{\prime\prime}$}

\newcommand{\ionf}[2]{[#1\,\textsc{#2}]}
\newcommand{\ionp}[2]{#1\,\textsc{#2}}

\newcommand{\fermilat}{{Fermi}-LAT }

\newcommand{\FIG}[1] {Figure~\ref{#1}}

\newcommand{\AFF}[1]{$^{\foreach\d[count=\ni]in{#1}{\ifnum\ni=1\ref{\d}\else,\ref{\d}\fi}}$}

\title{A 12.4 day periodicity in a close binary system after a supernova}
\author{Ping Chen\AFF{aff:WIS}\thanks{E-mail: chen.ping@weizmann.ac.il,\href{https://orcid.org/0000-0003-0853-6427}{\textcolor{orcidlogocol}{} \hspace{2mm} orcid.org/0000-0003-0853-6427}},
Avishay Gal-Yam\AFF{aff:WIS},
Jesper Sollerman\AFF{aff:OKC_astro},   
Steve Schulze\AFF{aff:OKC_phy},
Richard S. Post\AFF{aff:PO},
Chang Liu\AFF{aff:NU_DPA, aff:NU_CIERA}, 
Eran O. Ofek\AFF{aff:WIS},
Kaustav K. Das\AFF{aff:CaltechCCA},     
Christoffer Fremling\AFF{aff:CaltechOO, aff:Caltech_astro}, 
Assaf Horesh\AFF{aff:HebrewU}, 
Boaz Katz\AFF{aff:WIS},
Doron Kushnir\AFF{aff:WIS},
Mansi M. Kasliwal\AFF{aff:CaltechCCA},  
Shri R. Kulkarni\AFF{aff:CaltechCCA}, 
Dezi Liu\AFF{aff:yunnan},
Xiangkun Liu\AFF{aff:yunnan},
Adam A. Miller \AFF{aff:NU_DPA, aff:NU_CIERA}, 
Kovi Rose\AFF{aff:USydney},
Eli Waxman\AFF{aff:WIS},
Sheng Yang\AFF{aff:OKC_astro, aff:HAS}, 
Yuhan Yao\AFF{aff:CaltechCCA}, 
Barak Zackay\AFF{aff:WIS},
Eric C. Bellm\AFF{aff:UW},                   
Richard Dekany\AFF{aff:CaltechOO},
Andrew J. Drake\AFF{aff:Caltech_astro},
Yuan Fang\AFF{aff:yunnan},
Johan P. U. Fynbo\AFF{aff:DAWN, aff:NBI}, 
Steven L. Groom\AFF{aff:IPAC},          
George Helou\AFF{aff:IPAC},
Ido Irani\AFF{aff:WIS},
Theophile Jegou du Laz\AFF{aff:Caltech_astro},
Xiaowei Liu\AFF{aff:yunnan},
Paolo A. Mazzali\AFF{aff:Liverpool, aff:MPIA},
James D. Neill\AFF{aff:Caltech_astro}, 
Yu-Jing Qin\AFF{aff:Caltech_astro},
Reed L. Riddle\AFF{aff:CaltechOO},
Amir Sharon\AFF{aff:WIS},
Nora L. Strotjohann\AFF{aff:WIS},
Avery Wold\AFF{aff:IPAC},                   
Lin Yan\AFF{aff:CaltechOO}
}

\graphicspath{{./}{figures/}}

\begin{document}
\maketitle


\begin{affiliations}

\item Department of Particle Physics and Astrophysics, Weizmann Institute of Science, Rehovot 7610001, Israel\label{aff:WIS}
\item The Oskar Klein Centre, Department of Astronomy, Stockholm University, Albanova University Center, 106 91 Stockholm, Sweden\label{aff:OKC_astro}
\item The Oskar Klein Centre, Department of Physics, Stockholm University, Albanova University Center, 106 91 Stockholm, Sweden\label{aff:OKC_phy}
\item Post Observatory, Lexington, MA 02421, USA\label{aff:PO}
\item Department of Physics and Astronomy, Northwestern University, 2145 Sheridan Rd, Evanston, IL 60208, USA \label{aff:NU_DPA}
\item Center for Interdisciplinary Exploration and Research in Astrophysics (CIERA), Northwestern University, 1800 Sherman Ave, Evanston, IL 60201, USA \label{aff:NU_CIERA}
\item Cahill Center for Astrophysics, California Institute of Technology, MC 249-17, 1200 E California Boulevard, Pasadena, CA, 91125, USA\label{aff:CaltechCCA}
\item Caltech Optical Observatories, California Institute of Technology, Pasadena, CA, USA \label{aff:CaltechOO} 
\item Division of Physics, Mathematics and Astronomy, California Institute of Technology, Pasadena, CA, 91125, USA\label{aff:Caltech_astro}
\item Racah Institute of Physics, The Hebrew University of Jerusalem, Jerusalem 91904, Israel\label{aff:HebrewU}
\item South-Western Institute for Astronomy Research, Yunnan University, Chenggong District, Kunming 650500, Yunnan Province, People’s Republic of China\label{aff:yunnan}
\item Sydney Institute for Astronomy, School of Physics, The University of Sydney, New South Wales 2006, Australia\label{aff:USydney}
\item Henan Academy of Sciences, Zhengzhou 450046, Henan, China\label{aff:HAS}
\item DIRAC Institute, Department of Astronomy, University of Washington, 3910 15th Avenue NE, Seattle, WA 98195, USA\label{aff:UW}
\item The Cosmic DAWN Center, Denmark\label{aff:DAWN}
\item Niels Bohr Institute, University of Copenhagen, Jagtvej 155, Dk-2200 Copenhagen N, Denmark\label{aff:NBI}
\item IPAC, California Institute of Technology, 1200 E. California Blvd, Pasadena, CA 91125, USA\label{aff:IPAC}
\item Astrophysics Research Institute, Liverpool John Moores University, IC2 Building, Liverpool Science Park, 146 Brownlow Hill, Liverpool L3 5RF, UK\label{aff:Liverpool}
\item Max-Planck Institute for Astrophysics, Garching, Germany\label{aff:MPIA}
\end{affiliations}

\begin{abstract}

Neutron stars and stellar-mass black holes are the remnants of massive star explosions\cite{Heger2003}. Most massive stars reside in close binary systems\cite{Sana2012}, and the interplay between the companion star and the newly formed compact object has been theoretically explored\cite{Hirai2022}, but signatures for binarity or evidence for the formation of a compact object during a supernova explosion are still lacking.  Here we report a stripped-envelope supernova, SN\,2022jli, which shows 12.4-day periodic undulations during the declining light curve. Narrow H$\alpha$ emission is detected in late-time spectra with concordant periodic velocity shifts, likely arising from hydrogen gas stripped from a companion and accreted onto the compact remnant. A new \fermilat  $\gamma$-ray source is temporally and positionally consistent with SN\,2022jli. The observed properties of SN\,2022jli, including periodic undulations in the optical light curve, coherent H$\alpha$ emission shifting, and evidence for association with a $\gamma$-ray source, point to the explosion of a massive star in a binary system leaving behind a bound compact remnant. Mass accretion from the companion star onto the compact object powers the light curve of the supernova and generates the $\gamma$-ray emission. 





\end{abstract}

\vspace{0.7cm}

SN\,2022jli was discovered by Libert Monard on 2022 May 5 (JD=2459704.67) and was later confirmed by several surveys (Method Section $\S~1$). It was classified as a Type Ic supernova (SN Ic). The supernova explosion occurred in the spiral arm of a nearby galaxy, NGC\,157 (Extended Data Fig.~\ref{fig:sn2022jli_ngc157}), at a redshift of $z=0.0055$ with a peculiar-velocity-corrected Hubble flow distance of $D = 22.5\,\mathrm{Mpc}$ (Method Section $\S~2$). The photospheric spectra of SN\,2022jli match well to those of spectroscopically regular SNe\,Ic\cite{Gal-Yam2017}. We measured an ejecta expansion velocity of around $8,200\, \mathrm{km}\,\mathrm{s}^{-1}$ from the absorption minima of prominent \ionp{Fe}{ii} lines (Extended Data Fig.~\ref{fig:spec_evolution_compare}). Remarkably, the supernova brightens again around one month after discovery, which is unusual for SNe Ic. Since then, we obtained extensive follow-up photometry and spectroscopy data (Methods Section $\S~3$ and $\S~4$). 

\begin{figure}
\centering
\includegraphics[width=0.8\textwidth]{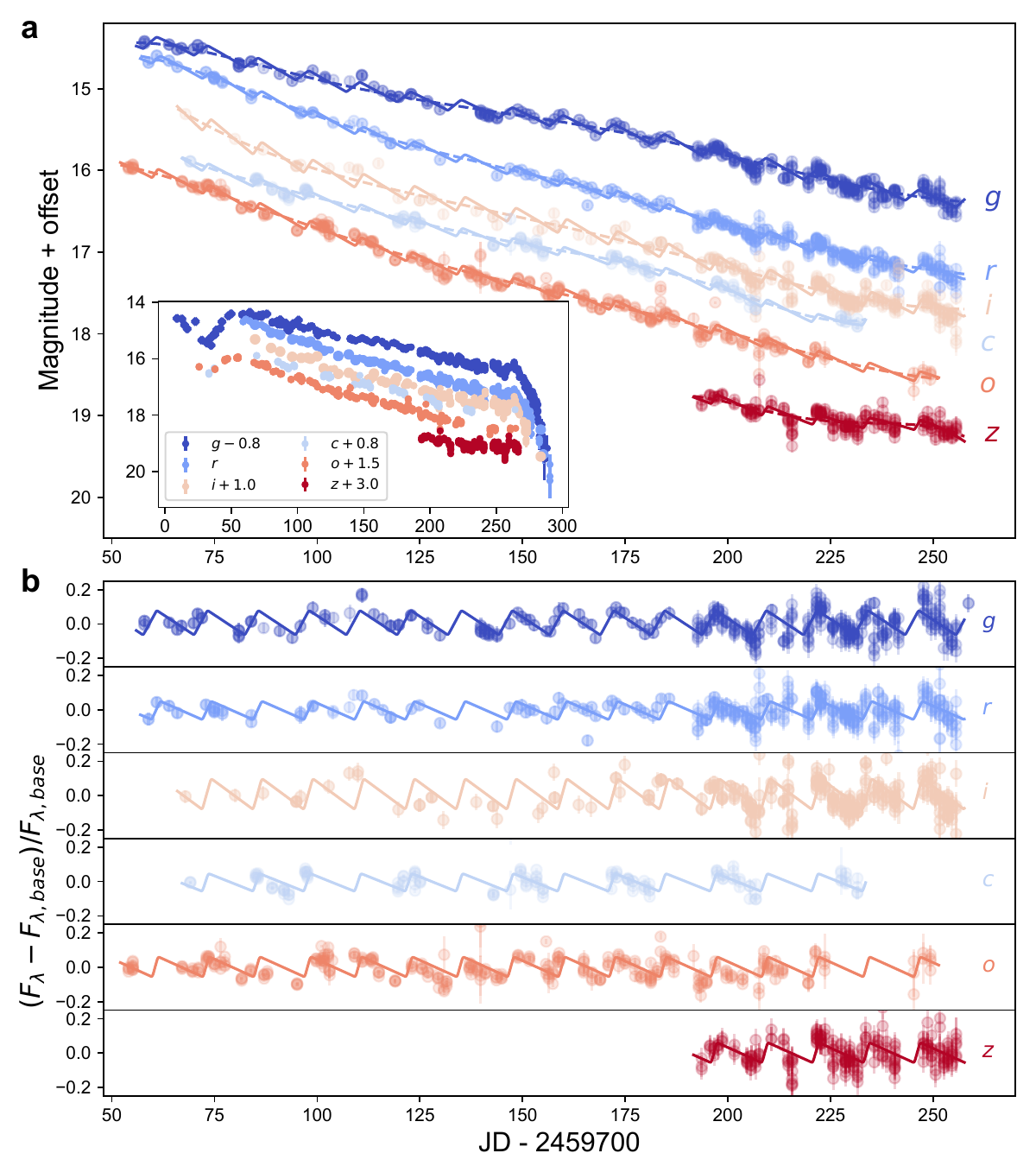}
\caption{\textbf{Multiband light curves of SN\,2022jli showing periodic undulations.} {\bf (a)} $g, r, i, z, c$, and $o$-band light curves of SN\,2022jli during phases between +50 and +260 days after discovery. The dashed lines show the polynomial fit to the light curve, which serves as the ``baseline''. An empirical model with a 12.4-day period is shown as a solid line for each band. The inset panel shows the whole range of the multiband light curves. {\bf (b)} The relative undulations in $g, r, i, z, c$, and $o$-band light curves. All the error bars of the data points are $1\sigma$ confidence intervals.
\label{fig:fig1} 
}
\end{figure}

\begin{figure}
\centering
\includegraphics[width=0.8\textwidth]{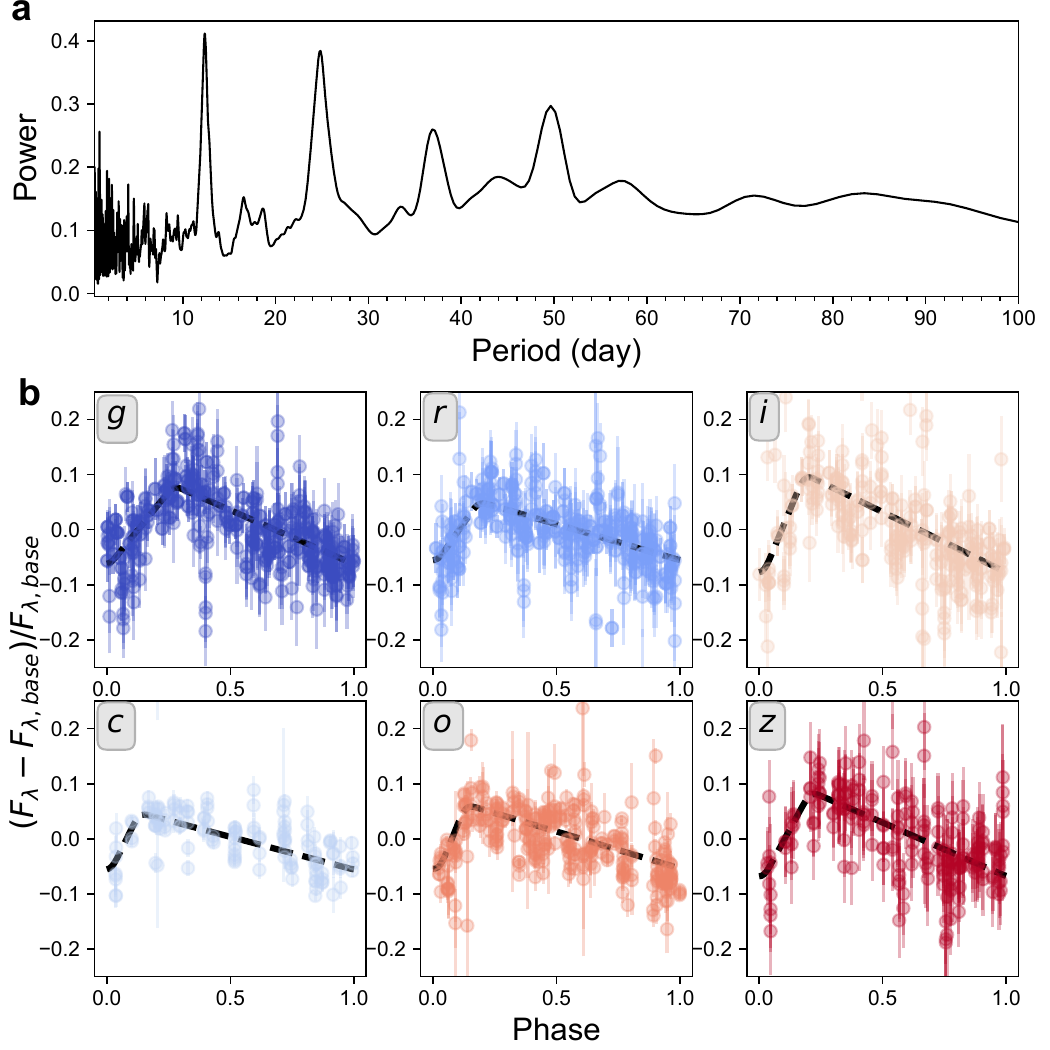}
\caption{\textbf{Multiband periodogram and the undulation profile of SN\,2022jli.} {\bf (a)} Power spectrum of the multiband light curve by jointly fitting of $g, r, i, z, c$, and $o$-band light curves shown in Fig.~\ref{fig:fig1}. The power spectrum shows clear peaks at 12.4 days and its lower-frequency harmonic aliases at around 24.8, 37.2, and 49.6 days. {\bf (b)} The folded undulation profile using the period of 12.4 days. The black dashed lines show the best-fit empirical model (Section $\S~5$) describing the undulation profile by a fast rise and gradual decline. Error bars of the data points are $1\sigma$ confidence intervals.
\label{fig:fig2}
}
\end{figure}

The light curve of SN\,2022jli (inset panel of Fig.~\ref{fig:fig1}) shows three distinct evolutionary phases: \ionp{Phase}{i}, the first decline phase; \ionp{Phase}{ii}, the rebrightening and gradual decline phase; and \ionp{Phase}{iii}, the late-time fast decline phase. During the gradual decline in \ionp{Phase}{ii} starting around two months after discovery, SN\,2022jli shows periodic undulations in the light curves (Fig.~\ref{fig:fig1}). These bumps appear in all the observed optical bands and last around 200 days until the onset of the late-time rapid decline. We performed a periodogram analysis of the multiband light curves and found a prominent peak at 12.4 days in the power spectrum (Fig.~\ref{fig:fig2}). The false alarm probability for the detected periodicity is less than $10^{-9}$ (Method Section $\S~5$). We also performed a periodicity analysis on the individual band light curves and found the same prominent period in each band around 12.4 days (Extended Data Fig.~\ref{fig:period_v1}). The phase-folded light curve adopting $P=12.4$ days (Fig.~\ref{fig:fig2}) reveals a constant profile composed of a fast rise lasting $\sim$ 3 days ($\sim$ 0.25 phase) and a relatively slow decline. We divided the nearly 200-day undulating light curve into two halves with equal time spans and repeated the periodicity analysis performed for the whole dataset. We did not find significant differences in either the bump period or profile between the two parts. Individual bumps have been observed in hydrogen-poor superluminous supernovae\cite{Nicholl2016, Yan2017, Hosseinzadeh2022, West2023, ChenZH2023}, and evidence for rapid variability in the optical light curve has also been found in Type Ia SN 2014J\cite{Bonanos2016}, but this is the first time multiple bumps with such a strong periodic signal have been detected.   

\begin{figure}
\centering
\includegraphics[width=0.9\textwidth]{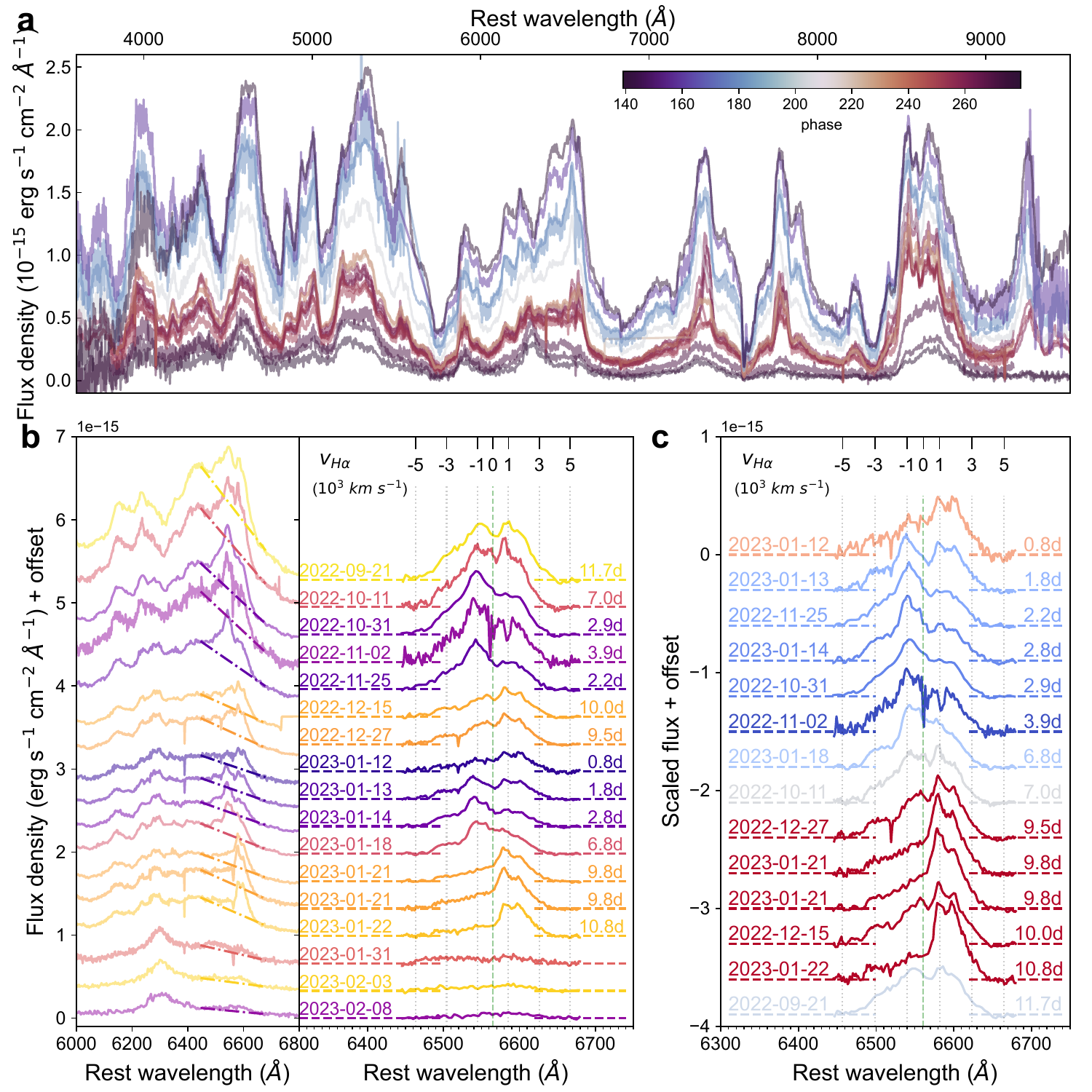}
\caption{\textbf{The spectral evolution of SN\,2022jli between +139 and +280 days after discovery.} {\bf (a)} Flux calibrated optical spectra of SN\,2022jli. The spectral phase in days relative to the discovery time is shown with the color bar. {\bf (b)} The evolution of H$\alpha$ emission. The left panel shows the zoom-in view around 6000--6800\,\AA. The right panel shows the narrow H$\alpha$ after subtracting the pseudo-continuum (indicated by the dash-dotted line in the left panel). For each spectrum, the observation date is shown on the left, and the phase relative to the minimum of the bump profile is shown on the right. The H$\alpha$ velocity is shown at the top. {\bf (c)} The same as the right panel of (b), but the flux of each spectrum has been scaled to have the same integrated flux within 6430\,--\,6680\,\AA, and the spectra are sorted by the phase relative to the minimum of the bump profile.
\label{fig:fig4}
}
\end{figure}

Figure~\ref{fig:fig4} shows a selected sample of spectra between +139 and +280 days after discovery. Prior to the rapid decline phase ($\lesssim 270$\,d), these spectra show multiple emission features as well as noticeable continuum emission. Compared with the nebular spectra of other stripped-envelope supernovae (SESNe), our spectra show stronger iron-group-element emission and many oxygen emission lines (Extended Data Figs.~\ref{fig:spec_evolution_compare}, \ref{fig:spec_lineident}). The comparison reveals that SN\,2022jli resembles some hydrogen-poor superluminous supernovae\cite{Gal-Yam2009, Nicholl2016}, transients that might have an association with Gamma-ray Burst (GRB)\cite{Matheson2001, Mazzali2004}, and other peculiar long-lasting supernovae, such as SN\,2012au\cite{Milisavljevic2013} and iPTF15dtg\cite{Taddia2019}, which are suggested to be powered by central engines. During the gradual decline phase, SN\,2022jli shows unique emissions around $6500$\,\AA\ with a relatively 
narrow feature superposed on a broad component. We attribute the narrow feature to hydrogen H$\alpha$ emission (Method Section $\S~9$). The H$\alpha$ lines show remarkable shifting around zero velocity in a repetitive pattern consistent with having a 12.4-day period (bottom panels of Fig.~\ref{fig:fig4}).
SN\,2022jli experiences significant spectroscopic evolution as it enters the fast decline phase (Extended Data Fig.~\ref{fig:spec_fast_decline}). One conspicuous change is that the relatively narrow emissions, for example, the permitted \ionp{O}{i} lines with a full width at half maximum (FWHM) around 2,600 km\,s$^{-1}$,  disappear, and broader \ionf{O}{i} $\lambda\lambda$6300, 6363 with an FWHM around 5500 km\,s$^{-1}$ emerge (Extended Data Fig.~\ref{fig:specfit}).

\begin{figure}
\centering
\includegraphics[width=0.8\textwidth]{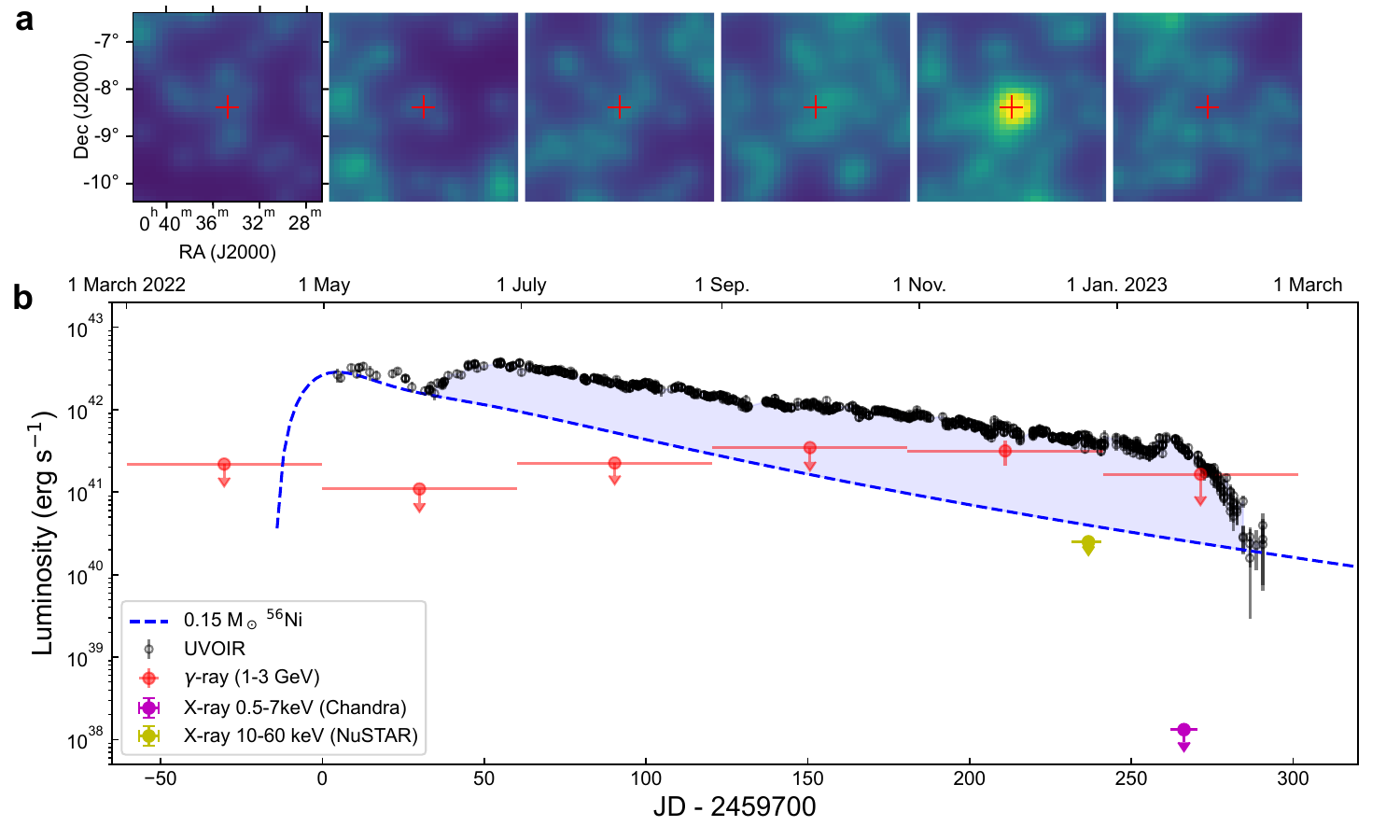}
\caption{\textbf{The pseudo-bolometric light curve and multi-frequency data of SN\,2022jli.} {\bf (a)} The $\gamma$-ray source detection map generated with bimonthly \fermilat observation in the energy band of 1\,--\,3\,GeV using the Poisson noise matched filter method (Section $\S~15$). The red plus symbol indicates the position of SN\,2022jli. A clear $\gamma$-ray source is detected in the 2022 November 1 to 2023 January 1 bin. {\bf (b)} The multi-frequency light curve of SN\,2022jli. The black points show the pseudo-bolometric light curve from 3750\,\AA\, to 25000\,\AA. The blue line shows the radioactive decay model with $0.15\, M_\odot$ $^{56}$Ni. The light curve of high-energy (1\,--\,3\,GeV) $\gamma$-ray emission associated with SN\,2022jli is shown with red points. Two epochs of X-ray observations with NuSTAR (yellow) and Chandra (magenta) around 250 days resulted in non-detections. All the error bars are $1\sigma$ uncertainties. All the non-detections are shown as 3$\sigma$ upper limits.
\label{fig:fig3}
}
\end{figure}

The pseudo-bolometric light curve of SN\,2022jli (Method Section $\S~6$) displays an evolution similar to those of the individual bands. It displays two peaks (Fig.~\ref{fig:fig3}), making it a double-peaked SESN, though the rising part of the first peak is missing. The \ionp{Phase}{i} light curve could be explained by $^{56}\mathrm{Ni}$ radioactive decay as in normal SNe Ic (Method Section $\S~7$; see discussion below). The \ionp{Phase}{ii} light curve requires another energy source of around $2\times10^{49}$ ergs, and the dramatic drop in \ionp{Phase}{iii} implies a sudden shutoff of the extra energy input. To probe the properties of such a late-time energy source, we obtained two epochs of radio observation with the Australian Telescope Compact Array (ATCA) at $+213$ and $+228$ days after the discovery, respectively. Neither epoch had a clear detection of SN\,2022jli, and the second epoch with the longer exposure time gave $5\sigma$ limits of $<0.074$\,mJy\,beam$^{-1}$ and $<0.055$\,mJy\,beam$^{-1}$ at central wavelengths of 5.5\,GHz and 9.0\,GHz, respectively. We also performed X-ray observations of SN\,2022jli with NuSTAR ($+227$ to $+237$ days after discovery) and the Chandra X-ray Observatory ($+257$ to $+266$ days after discovery), but did not detect any emission in soft or hard X-rays (Methods Section $\S~11$) with upper limits of $\mathrm{L_{30-60\,KeV} < 2.5\times10^{40}\,erg\,s^{-1}}$ and $\mathrm{L_{0.5-7\,KeV} < 1.3\times10^{38}\,erg\,s^{-1}}$.

We searched for high-energy $\gamma$-ray emission using data from Large Area Telescope on board the Fermi Gamma-ray Space Telescope (\fermilat) and found a $\gamma$-ray source in the direction of SN\,2022jli (Method Section $\S~15$). The new $\gamma$-ray source is detected after the supernova explosion, and there was no detection in the past 13.5 years of archival data before the supernova explosion (Extended Data Fig.~\ref{fig:fermi_TS_lc}). SN\,2022jli is within the 68\% confidence localization area of the new source, and the detection time of the $\gamma$-ray photons from the new source shows evidence for a correlation with the 12.4-day optical flux undulation (Extended Data Fig.~\ref{fig:fermi_localization_periodicity}). The temporal and spatial coincidence, together with potential periodicity, suggests an association of the new $\gamma$-ray emission with SN\,2022jli. The new source is most significantly detected in the 1\,--\,3\,GeV energy band with $L_{1-3\,GeV} = 3.1\times10^{41}$ erg\,s$^{-1}$ in November and December 2022, and the whole $\gamma$-ray light-curve is shown in Fig.~\ref{fig:fig3}. We suggest the $\gamma$-ray emission was only detected several months after the supernova explosion due to high pair-production opacities to $\gamma$-ray photons at early time\cite{Zdziarski1989, Acharyya2023}. The observed $\gamma$-ray emission falls below the \fermilat sensitivity at the end of December 2022, one month prior to the fast drop in optical flux.

Below we discuss the energy source that powers the unique light curves of SN\,2022jli. Due to the photometric and spectroscopic resemblance to normal SESNe in \ionp{Phase}{i} (Extended Data Figs.~\ref{fig:spec_evolution_compare} and \ref{fig:lc_compare}), it is natural to attribute the first peak to the same origin as in normal SESNe, a $^{56}$Ni-decay powered peak.  It is more intriguing to think about the energy source of the second peak and the powering mechanism of the undulations. Supernova ejecta interaction with hydrogen-poor circumstellar medium (CSM) has been commonly adopted to explain the observed bumps in the light curves of SESNe\cite{Chatzopoulos2012, Chevalier2017, Chen2017, West2023, ChenZH2023}. It is appealing to connect the periodic bumps to ejecta-CSM interaction (ECI) with a CSM having density fluctuation, such as the nested dust shells surrounding the Wolf-Rayet binary WR 140\cite{Lau2022}, and a late-time drop in the light curve is also expected after the ejecta sweeps through the confined dense CSM\cite{Ofek2014, Zhu2023}. However, the persistence and short period of the bumps, especially the short time scale of around three days for the rising part, put tight constraints on the feasibility of the ECI scenario to explain the observed properties. For example, consider a CSM composed of nested spherical shells. In that case, the light travel time difference between the near and far sides of the ECI is $\Delta t \sim \frac{R}{c} = 10 \times \frac{v_{\rm ej}}{10^4\,{\rm km\,s^{-1}}} \frac{t}{300\,{\rm days}}$ days, where $v_{\rm ej}$ is the effective ejecta expansion velocity and $t$ is the time from the explosion. For the observed ejecta velocity of 8,200 km\,s$^{-1}$, at 260 days, the light travel time difference is around 7 days, which would smear out any periodic signal from ECI due to CSM density fluctuation. Moreover, the light travel time difference is phase-dependent as the ejecta expand and interact with CSM at different radii. The observed periodic bumps span around 200 days, showing no significant evolution, arguing against the ECI explanation for the periodic undulations.  

The observed bumps have a short time scale, a constant profile, and a constant ratio to the continuum flux, i.e., the fluctuating light is not being diluted even when the ejecta expands. This requires an emitting region that covers a constant fraction of the expanding ejecta and that behaves coherently. The corresponding power source must reside in a spatially confined region to remain coherent and be located close to or at the center of the ejecta. In particular, the quick rise of the bump profile requires that the diffusion time of optical photons through the ejecta is short ($<3$\,d).

The periodic undulation first appears around the second peak, requiring that the diffusion time of the ejecta is already short at this time. This means that the \ionp{Phase}{ii} light curve could not be powered by $^{56}$Ni decay because the diffusion time would be similar to the rise time to peak (tens of days), which is inconsistent with powering the much shorter bumps. This argument applies to any other energy source that was produced during the explosion (such as a magnetar) with a decreasing or constant energy output after the explosion. A possible mechanism that could provide extra energy to power the later evolution of this supernova is accretion onto a newly-formed compact object\cite{Michel1988, Chevalier1989, Zhang2008, Dexter2013, Moriya2018, Moriya2019}, which we propose to explain the observed light curves of SN\,2022jli (Method Section $\S~7$). In this model, the deferred onset of the energy input could be due to the time it takes to form an accretion disk, while the abrupt luminosity drop-off at the late time is accounted for by the central object running out of infalling gas to fuel the accretion.   

Given the periodic signals observed in three different aspects of the supernova, the undulation in the optical light curve, the shifting H$\alpha$ emission, and the high-energy $\gamma$-ray photons, the most viable explanations seem to inevitably involve a binary system to provide the ``clocking'' mechanism. The luminosity of the narrow H$\alpha$ emission closely follows the evolution of the bolometric luminosity (Extended Data Fig.~\ref{fig:Ha_luminosity_velocity}), even during the transition phase when the supernova luminosity shows an enhancement before diving down. This means the H$\alpha$ emission shares the same origin as the excess luminosity of the supernova, implying the H$\alpha$ also comes from the central region of the system. Therefore, a companion star with a hydrogen-rich envelope is necessary to provide the hydrogen, which corroborates the existence of a binary system. Such a binary system provides direct evidence supporting the binary origins of some SESNe. 

We suggest that the compact supernova remnant and the companion star remain bound in a binary system after the supernova explosion. One plausible idea to power the supernova is through the accretion from the bloated companion star onto the compact object, forming an accretion disk. An eccentric orbit is expected (Method Section $\S~12$), which modulates the accretion rate generating the observed undulation. With $L=\epsilon \dot{M} c^2$, and adopting the optical-NIR pseudo-bolometric luminosity for the accretion luminosity $L$, we calculated a peak accretion rate $\dot{M}$ around $4\times10^{-3} \left(\frac{\epsilon}{0.01}\right)^{-1}\msun\,{\rm yr^{-1}}$, and a total mass to account for the accretion-powered energy around $10^{-3} \left(\frac{\epsilon}{0.01}\right)^{-1} \msun$, where $\epsilon$ is the radiative efficiency of accretion. Such an extreme accretion rate could give rise to an extreme outflow of wind and/or a jet. The generation of a jet and/or wind favors the idea that the detected $\gamma$-ray emission of SN\,2022jli shares a similar origin to that seen in black hole X-ray binary systems\cite{Zanin2016}. Jets have been predicted in several supernova explosion scenarios\cite{Akashi2020, Hober2022}.

The unprecedented properties of SN\,2022jli tell that whatever happens in the system should be a rare phenomenon, which might be explained by the rarity of a bound binary system surviving a supernova explosion \cite{Renzo2019, Chrimes2022}.  SN\,2022jli provides direct observational evidence for the survival of such a binary system after a supernova explosion. SN\,2022jli builds a direct link between the supernova explosion and the formation of a compact object. It highlights the importance of the interplay between the companion star and the newly formed compact object in shaping the appearance of a core-collapse supernova.  




\clearpage

\bibliographystyle{naturemag}

\clearpage

\noindent \FIG{fig:fig1}: {\textbf{Multiband light curves of SN\,2022jli showing periodic undulations.} {\bf (a)} $g, r, i, z, c$, and $o$-band light curves of SN\,2022jli during phases between +50 and +260 days after discovery. The dashed lines show the polynomial fit to the light curve, which serves as the ``baseline''. An empirical model with a 12.4-day period is shown as a solid line for each band. The inset panel shows the whole range of the multiband light curves. {\bf (b)} The relative undulations in $g, r, i, z, c$, and $o$-band light curves. All the error bars of the data points are $1\sigma$ confidence intervals.} 

\noindent \FIG{fig:fig2}: {\textbf{Multiband periodogram and the undulation profile of SN\,2022jli.} {\bf (a)} Power spectrum of the multiband light curve by jointly fitting of $g, r, i, z, c$, and $o$-band light curves shown in Fig.~\ref{fig:fig1}. The power spectrum shows clear peaks at 12.4 days and lower-frequency harmonic aliases at around 24.8, 37.2, and 49.6 days. {\bf (b)} The folded undulation profile using the period of 12.4 days. The black dashed lines show the best-fit empirical model (Section $\S~5$) describing the undulation profile by a fast rise and gradual decline. Error bars of the data points are $1\sigma$ confidence intervals.}

\noindent \FIG{fig:fig4}: {\textbf{The spectral evolution of SN\,2022jli between +139 and +280 days after discovery} {\bf (a)} Flux calibrated optical spectra of SN\,2022jli. The spectral phase relative to the supernova discovery time is shown with the color bar. {\bf (b)} The evolution of H$\alpha$ emission. The left panel shows the zoom-in view around 6000--6800\,\AA. The right panel shows the narrow H$\alpha$ after subtracting the pseudo-continuum (indicated by the dash-dotted line in the left panel). For each spectrum, the observation date is shown on the left, and the phase relative to the peak of the bump profile is shown on the right. The H$\alpha$ velocity is shown at the top. {\bf (c)} The same as the right panel of (b), but the flux of each spectrum has been scaled to have the same integrated flux within 6430\,--\,6680\,\AA, and the spectra are sorted by the phase relative to the minimum of the bump profile.}

\noindent \FIG{fig:fig3}: {\textbf{The pseudo bolometric light curve and multi-frequency data of SN\,2022jli.} {\bf (a)} The $\gamma$-ray source detection map generated with bimonthly \fermilat observation in the energy band of 1\,--\,3\,GeV using the Poisson noise matched filter method (Section $\S~15$). The red plus symbol in the center of the field indicates the position of SN\,2022jli. A clear $\gamma$-ray source is detected in the 2022 November 1 to 2023 January 1 bin. {\bf (b)} The multi-frequency light curve of SN\,2022jli. The black points show the pseudo-bolometric light curve from 3750\,\AA\, to 25000\,\AA. The blue line shows the radioactive decay model with $0.15\, M_\odot$ $^{56}$Ni. The light curve of high-energy (1\,--\,3\,GeV) $\gamma$-ray emission associated with SN\,2022jli is shown with blue points. Two epochs of X-ray observations with NuSTAR (yellow) and Chandra (magenta) around 250 days resulted in non-detections. All the error bars are $1\sigma$ uncertainties. All the non-detections are shown as 3$\sigma$ upper limits. }

\clearpage

\clearpage

\clearpage

\clearpage

\clearpage

\begin{methods}



\subsection{1. Discovery of SN\,2022jli:}

SN\,2022jli was first discovered and reported to the Transient Name Server (TNS; \url{https://www.wis-tns.org/object/2022jli}) by Libert Monard\cite{Monard2022} on 2022 May 5 . It was later recovered by wide-field transient surveys including the Asteroid Terrestrial-impact Last Alert System (ATLAS\cite{Tonry2018}; ATLAS22oat), the Gaia transient survey\cite{Hodgkin2021} (Gaia22cbu), the Panoramic Survey Telescope and Rapid Response System (PanSTARRS\cite{Chambers2016}; PS22gwo), and the Zwicky Transient Facility (ZTF\cite{Bellm2019a, Graham2019}; ZTF22aapubuy). The transient was discovered as it was rising in the morning sky. Solar conjunction before the discovery of SN\,2022jli hindered any direct constraint on the explosion time so we set the discovery epoch as the reference time for phase definition whenever the reference epoch is not explicitly specified throughout this paper. The Gaia detection gives a transient coordinate of right ascension $\alpha = 00^{\rm h}34^{\rm m}45.690^{\rm s}$ and declination $\delta = -08^{\circ}23^\prime12.16^{\prime\prime}$ (J2000.0). The Galactic extinction in the direction of SN\,2022jli is E($B-V$)$_{\rm MW}$ = 0.039 mag. The latest non-detection was on 2022 February 6, from Gaia, 87.5 days before the discovery. Therefore, due to the large seasonal gap, we barely have any constraint on the explosion time of SN\,2022jli.

\subsection{2. Host galaxy and extinction:}
SN\,2022jli is located on a spiral arm of NGC\,157 (Extended Data Fig.~\ref{fig:sn2022jli_ngc157}). NGC\,157 has a large number of \ionp{H}{ii} regions, resulting in a complex background with strong nebular emission from the host galaxy, which commonly causes over-subtraction of host-galaxy lines. The distance to NGC 157 is uncertain. The peculiar-velocity-corrected Hubble flow distance gives D (Virgo + GA + Shapley) = $22.5\pm1.6$ Mpc ($\mu=31.76$~mag)\cite{Mould2000} using a standard $\Lambda$CDM cosmology\cite{Komatsu2011} with $\Omega_{\rm M} = 0.27$, $\Omega_{\Lambda} = 0.73$ and $H_0 = 73$~km~s$^{-1}$~Mpc$^{-1}$ \cite{Riess2022}. The Tully-Fisher method gives smaller values of 12--13 Mpc\cite{Nasonova2011, Tully2013, Erwin2017}. In this work, we adopt the distance of 22.5 Mpc for our analysis but caution that distance-related quantities are subject to large uncertainties. 

We detected prominent sodium absorption from the host galaxy in the supernova spectra, distinguished as a doublet in our medium resolution spectra taken with IMACS and X-Shooter (Extended Data Fig.~\ref{fig:sn2022jli_ngc157}). We measured a sodium absorption equivalent width EW(\ionp{Na}{i} D1+D2) = $1.07\pm 0.11$ \AA, which converts to E($B-V$)$_{\rm host} = 0.25_{-0.07}^{+0.11}$ mag \cite{Poznanski2012}. We did not detect significant sodium absorption from the Milky Way, consistent with the low Galactic reddening. We also detected narrow diffusion interstellar band (DIB) absorption at 6283 \AA\, (Extended Data Fig.~\ref{fig:sn2022jli_ngc157}), from which we measured an equivalent width of EW(DIB6283) = $0.27\pm 0.03$ \AA, which converts to E($B-V$)$_{\rm host} = 0.25\pm0.03$ mag\cite{Lan2015} or E($B-V$)$_{\rm host} = 0.28\pm0.04$ mag\cite{Fan2019}. The extinction uncertainty derived from DIB absorption only includes the uncertainty of the equivalent width measurement but not the uncertainty from the conversion. In this work, we adopted  E($B-V$)$_{\rm host} = 0.25$ mag and the conservative uncertainty obtained from sodium absorption for the analysis.

\subsection{3. Photometry:} 

SN\,2022jli was detected by ZTF starting from 2022 June 27. ZTF $gri$ photometry obtained with the ZTF survey \cite{Dekany2020} camera was processed with the ZTF image reduction pipeline\cite{Masci2019} employing the ZOGY image-subtraction method\cite{Zackay2016}. We obtained additional $gri$ images with the robotic 60-inch telescope at Palomar (P60\cite{Cenko2006}), using the Spectral Energy Distribution Machine (SEDM\cite{Blagorodnova2018}). {\tt FPipe}\cite{Fremling2016} was used to extract PSF photometry from image subtraction against Sloan Digital Sky Survey (SDSS) templates. ZTF and SEDM photometry were obtained and reduced in real-time mode and streamed to the Fritz SkyPortal \cite{vanderWalt2019, Coughlin2023} to aid further follow-up observations. The SEDM photometry obtained above with {\tt FPipe} was considered preliminary and was refined as described below. 

We included photometry data reported to TNS by Libert Monard at the Kleinkaroo Observatory (KKO), which were taken on 2022 May 5, 6, 11, and 22. We obtained All-Sky Automated Survey for Supernovae (ASAS-SN) $g$-band photometry \cite{Shappee2014} from the ASAS-SN Sky Patrol \cite{Kochanek2017}. We adopted the option of ``Image Subtraction (No reference flux added)'' for the photometry method, which performs aperture photometry on the coadded image subtracted data for each epoch but does not add the flux of the source on the reference image to the light curve. We noted a constant offset between ASAS-SN $g$-band and ZTF $g$-band photometry in flux. We added a constant flux of 360 mJy to the measured flux in the ASAS-SN data to minimize the difference between ASAS-SN and ZTF in the overlapping time range. We extracted ATLAS-$c$ and ATLAS-$o$ band light curve data from ATLAS forced photometry server \cite{atlasfp, Tonry2018, Smith2020}. We also used three epochs of Gaia-$G$ band photometry obtained by Gaia transient survey \cite{Hodgkin2021} through the Gaia Alert service \cite{gaiaalert}.

After discovering the periodic bumps, we started more follow-up observations with the 0.8m RC32 telescope operated by the Post Observatory (PO). We performed aperture photometry on the reference-subtracted images and calibrated them with the SDSS standard catalog. For image subtraction, we built template images from SDSS images \cite{sdssdr12}. The {\tt FPipe} pipeline did not successfully reduce all P60 SEDM images. We performed the same photometry for SEDM images with some manual assistance. The detailed photometry procedures are described in ref\cite{Chen2022}.

We also observed SN\,2022jli with the Multi-channel Photometric Survey Telescope (Mephisto\cite{LiuXW2019, Yuan2020}) during the telescope commissioning phase from 2022 December 10 to 2023 February 6. The observations were done in $uv$ and $iz$ bands. We conducted image subtraction and then performed PSF photometry on the subtracted images. The images taken on the night of 2023 February 6 were used as template images. We obtained the flux of SN\,2022jli with PSF photometry from the template images and added to the flux obtained from the difference images. 

Supplementary Information Fig.~\ref{fig:lc_sn2022jli} shows the observed light curves, and all photometry is listed in Supplementary Information Table~\ref{tab:phot}.

\subsection{4. Spectroscopy:}
We obtained 46 low-to-medium resolution (R $\sim$ 100 to 6000) spectra of SN\,2022jli, taken with the instruments listed in Supplementary Information Table~\ref{tab:spec}. 
The spectra are shown in Supplementary Information Figs.~\ref{fig:spec_opt_sedm}, \ref{fig:spec_opt_nonsedm}, and \ref{fig:spec_nir}. All spectra will be made publicly available through the Weizmann Interactive Supernova Data Repository (WISeREP\cite{wiserep, Yaron2012}). Detailed information on the spectroscopic observations and data reduction is listed below. 
\paragraph{P60/SEDM:} The Spectral Energy Distribution Machine (SEDM \cite{BenAmi2012, Blagorodnova2018}) is an integral field unit (IFU) spectrograph mounted on the 60-inch robotic telescope (P60 \cite{Cenko2006}) at Palomar Observatory. We conducted 25 epochs of spectroscopy observation with the SEDM between 2022 June 29 and 2023 February 2. The SED Machine has a very low resolution ($R\sim100$) covering the wavelength range from 3650 to 10000~\AA. All SEDM IFU data were reduced using the pipeline described in \cite{Rigault2019}, and new modules for the SEDMachine described in \cite{Kim2022} were used to remove contamination from cosmic rays and non-target light.

\paragraph{NOT/ALFOSC:} The Alhambra Faint Object Spectrograph and Camera (ALFOSC) is mounted on the 2.56m Nordic Optical Telescope (NOT). We acquired six epochs of low-resolution spectra with NOT/ALFOSC between 2022 August 6 and 2023 February 8. The spectra were obtained with a slit width of either 1\farcs0 or 1\farcs3 depending on the seeing, and using grism \#4. The data were reduced using the pipeline {\tt foscgui}\cite{foscgui}. The reduction includes cosmic-ray rejection, bias corrections, flat fielding, and wavelength calibration using HeNe arc lamps imaged immediately after the target. The relative flux calibration was done with spectrophotometric standard stars observed on the same night or nights before the observation. 

\paragraph{P200/DBSP:} The Double Beam SPectrograph (DBSP\cite{Oke1982a}) is mounted on the 200-inch Hale telescope at Palomar Observatory (P200). The DBSP uses a dichroic (at 5500 \AA\ for the used D55 dichroic) to split light into separate red and blue channels (``sides''), observed simultaneously. We obtained three epochs of spectroscopy of SN\,2022jli  with DBSP. The observations were taken using a blue grating with 600 lines per mm blazed at 4000~\AA, a red grating with 316 lines per mm blazed at 7500~\AA, and a 1\farcs5 wide slit on 2022 August 20 and 2022 November 2, and a 1\farcs0 wide slit on 11 2022 October 11. The data are reduced using the python package {\tt DBSP\_DRP}\cite{dbspdrp} that is primarily based on {\tt PypeIt} \cite{Prochaska2020b, Prochaska2020a}.

\paragraph{MMT/BINOSPEC:} We obtained 5 spectra of SN\,2022jli between 2022 December 27 and 2023 January 31 with Binospec\cite{Fabricant2019} on the MMT Observatory 6.5m telescope. All data were acquired with a grating of 270 lines/mm and a 1\farcs0 slit mask. The Binospec spectra have a wavelength coverage of 3900 -- 9240 \AA. The basic data processing (bias subtraction, flat fielding) is done using the Binospec pipeline\cite{Kansky2019}. The processed images are downloaded from the MMTO queue observation data archive. All the spectra are reduced with {\tt IRAF}, including cosmic-ray removal, wavelength calibration (using arc lamp frames taken immediately after the target observation), and relative flux calibration with archived spectroscopic standards observation. 

\paragraph{Magellan/IMACS \& FIRE}

We observed SN\,2022jli with the Folded-port InfraRed Echellette (FIRE \cite{Simcoe2013}) spectrograph on 2022 August 25, and with both the FIRE and the Inamori-Magellan Areal Camera and Spectrograph (IMACS\cite{Dressler2011}) on 2022 December 14. Both FIRE and IMACS are mounted on the 6.5m Magellan Baade telescope. The first epoch of the FIRE spectrum was taken with the long-slit mode and the second with the echelle mode. The long-slit mode FIRE spectrum was reduced with the IDL pipeline
{\tt firehose}\cite{Simcoe2013}. The echelle FIRE spectrum was reduced with {\tt PypeIt} \cite{Prochaska2020b, Prochaska2020a}. The IMACS spectrum was taken with the 1200 lines/mm grating at two different tilt angles, covering 5130 -- 6780\,\AA\, and 7290 -- 8920\,\AA. The IMACS spectrum was reduced with {\tt IRAF} in the same way as for the BINOSPEC spectra.

\paragraph{VLT/XSHOOTER:} We obtained three intermediate resolution spectra with the X-shooter echelle spectrograph \cite{Vernet2011} on 2023 January 14, 18, 21 through a DDT program (Program ID: 110.25A6, P.I.: P. Chen). 
These data were executed in ToO mode in order to spectroscopically monitor the supernova during the different phases of the light curve undulations.
All observations were performed in nodding mode and with 1\farcs3/1\farcs2/1\farcs2 wide slits (UVB/VIS/NIR). The observations covered the entire spectral range of the X-shooter spectrograph from 3000 to 24800 \AA. We first removed cosmic rays with the tool {\tt astroscrappy}\cite{astroscrappy}, which is based on the cosmic-ray removal algorithm by \cite{vanDokkum2001}. Then the data were processed with the X-shooter pipeline v3.3.5, and the ESO workflow engine ESOReflex \cite{Goldoni2006, Modigliani2010, Freudling2013}. All data from three arms were reduced in nodding mode. The nodding mode reduction is critical for NIR data to ensure a good sky-line subtraction. The spectra of the individual arms were stitched by averaging the overlap regions. The atmospheric absorption in the VIS and NIR arms was corrected with the software tool {\tt molefit}\cite{Smette2015} (v4.2.3).

\subsection{5. Periodic undulation in the light curve}
\paragraph{Multiband periodicity analysis}
We adopt the multiband periodogram method \cite{VanderPlas2015}, a general extension of the well-known Lomb–Scargle approach\cite{Lomb1976, Scargle1982, VanderPlas2018}, to quantitatively detect the periodic signal in the multiband light curves of SN\,2022jli. The light curves in each band are modeled as arbitrary truncated Fourier series with the period and/or phase shared across all bands. The model contains two parts: 1) an N$_\mathrm{base}$-term truncated Fourier series that models the shared variability among all six bands, i.e., the ``base model''; 2) a set of N$_\mathrm{band}$-term truncated Fourier series for an individual band, which models the residual from the base model. The model can be described as follows $y_k(t|\omega, \theta) = \theta_0 + \sum_{n=1}^{N_{\mathrm{base}}} [\theta_{2n-1} \mathrm{sin}(n\omega t) + \theta_{2n} \mathrm{cos}(n\omega t)] + \theta_0^{(k)} + \sum_{n=1}^{N_{\mathrm{band}}} [\theta_{2n-1}^{(k)} \mathrm{sin}(n\omega t) + \theta_{2n}^{(k)} \mathrm{cos}(n\omega t)]$ where $\theta_0$ is the constant offset of the base model, $\theta_0^{(k)}$ is the residual component of each filter $k$, and $[\theta_{2n-1} (\mathrm{or}\, \theta_{2n-1}^{(k)}), \theta_{2n} (\mathrm{or}\, \theta_{2n}^{(k)})] = [A\,\mathrm{cos}\phi, A\,\mathrm{sin}\phi]$ describes the amplitude and phase of the single-component sinusoidal model of $d(t)=A\,\mathrm{sin}(\omega t+\phi)$. 
We used the {\tt gatspy}\cite{gatspy} tool to fit the $g$, $r$, $i$, $z$, $c$, and $o$ band light curve observed between +50 and +250 days after discovery. In practice, the light curves have been detrended and normalized in flux before periodicity analysis. We first subtracted fluxes in each band, $F_{\lambda}$, with the mean flux light curve and then normalized the residual flux by the mean flux. The mean flux in each band was modeled with a polynomial ``baseline'', $F_{\lambda, \rm{base}}$. The detrending and normalization procedure can be described as $\psi = (F_\lambda-F_{\lambda, \rm{base}})/F_{\lambda, \rm{base}}$. The adopted polynomial orders for the baseline model in each band depend on the baseline's length and smoothness. As a result, polynomial orders of 6, 6, 4, 4, 6, and 5 have been used for $g$, $r$, $i$, $z$, $c$, and $o$ bands, respectively. We adopted N$_{\rm{base}}=4$ and $N_{\rm{band}}=2$ in our fitting and obtained a significant peak in the power spectrum corresponding to 12.4 days which is accompanied by three low-frequency harmonic aliases. We applied the bootstrap method to study the significance of the detected periodicity by shuffling the data points and calculating the power spectrum as done for the original data. We performed $10^9$ experiments, and none of them revealed any similar peaks in the power spectrum, which means the false alarm probability (FAP) of the 12.4-day period is smaller than $10^{-9}$. To test whether there is a significant phase-dependent periodicity evolution, we divided the above data into two parts to perform the same analysis on each part as done for the whole data. We obtained a best-fit period of $12.23\pm0.10$ days for the first-half data and a best-fit period of $12.40\pm0.08$ days for the second-half data. $3\,\sigma$ uncertainties are reported. No significant period change was detected in our data. 


\paragraph{Empirical model of the undulation profile}

The relative undulation of SN\,2022jli (Fig.~\ref{fig:fig2}) shows characteristic profiles composing a fast rise and then a gradual decline. We constructed an empirical model for the undulation profile to more quantitatively characterize its features. The basis of the empirical model is a piecewise function with two linear components, as given below: 

\begin{equation}
\psi(t) = 
\left\{
    \begin{array}{lr}
        \frac{A}{t_{\mathrm{rise}}}t + C, &  0 \leq t < t_{\mathrm{rise}} \\
        \frac{-A}{P-t_{\mathrm{rise}}} (t-P) + C, &  t_{\mathrm{rise}} \leq t < P
    \end{array}
\right.
\end{equation}
where $t_{\mathrm{rise}}$ is the time for the rising phase, A is the whole amplitude from the lowest to the highest point, P is the period, and C is a constant.  The empirical model is defined within the range of one period $t\in [0, P]$, which starts from the point with the lowest value of $\psi$. To smooth the model, we convolved the above function with a narrow Gaussian kernel ($\sigma=0.2$ days). We applied the above empirical model to fit the phase-folded undulation profile in each band and obtained a rising phase duration of $3.5\pm 0.2$, $2.5\pm 0.2$, $2.5\pm 0.2$, $2.7\pm 0.3$, $1.9\pm 0.2$,  and $1.9\pm 0.2$ days for $g$, $r$, $i$, $z$, $c$, and $o$ band, respectively. The best-fit empirical models are shown with black dashed lines in Fig.~\ref{fig:fig2} and are also used in Fig.~\ref{fig:fig1}.






\paragraph{Individual band periodicity analysis}
We also performed periodicity analysis with the Lomb-Scargle method\cite{Lomb1976, Scargle1982, VanderPlas2018} in individual bands.  The left panels of Extended Data Fig.~\ref{fig:period_v1} show the Lomb-Scargle power spectrum for each $g$, $r$, $i$, $c$, $o$, and $z$ band with blue lines.  We also compute the power spectrum of the window function shown with the yellow lines. The diurnal peaks of the window power are pronounced, which is typical for ground-based data with nightly observations. The strong diurnal component from the window function causes each frequency signature $f_0$ to be partially aliased at $f_0 + n\delta f$, for integers $n$ and $\delta f = 1$ cycle day$^{-1}$. A significant peak of around 12.4 days is detected in each band, as shown in the inset panel, where the period corresponding to the peak power is given $P_g=12.5$ days, $P_r=12.4$ days, $P_i=12.4$ days, $P_c=12.8$ days, $P_o=12.2$ days, and $P_z=12.2$ days. All the detected peaks have FAP smaller than $10^{-9}$. The power spectra of the observational window functions do not show any significant peaks around the detected periods, confirming the periodicity's authenticity.  The right panels show the phase-folded light curves after detrending and normalization. 


\subsection{6. Bolometric light curve:}
\label{sec:bol}

We built the pseudo-bolometric light curve with the photometry data (Method $\S~4$) and spectral sequence (Methods $\S~4$) of SN\,2022jli. We first get the integrated flux in the wavelength range of 3750 -- 9150\,\AA, which most of our optical spectra cover. This wavelength range corresponds roughly to the $BVRI$ bands, and we denote the luminosity obtained from such a wavelength range as $L_{BVRI}$. Then we derive the fraction of NIR contribution (9150--25000\,\AA) to the bolometric luminosity which is defined as $f_{\rm NIR}=\frac{L_{NIR}}{L_{BVRI}+L_{NIR}}$, and apply the correction to get $L_{bol}= L_{BVRI}+L_{NIR}$. The $U$-band contribution to the bolometric luminosity is not considered, but is known to be small for SESNe, around 10\% around peak light and then quickly declining to less than 5\% \cite{Lyman2014}. To get the integrated optical flux, for each epoch of photometry, the closest spectrum was used as the spectral energy distribution template and scaled by a constant value to match the observed flux, then the scaled spectrum was corrected for the Milky Way foreground extinction with $E(B-V)_{\rm MW} = 0.039$\,mag\cite{Schlafly2011} and host extinction with $E(B-V)_{\rm host} = 0.25$\,mag (Methods $\S~2$) with the CCM extinction law adopting $R_V=3.1$\cite{Cardelli1989}. For the NIR fractional contribution, we got $f_{+107d} = 0.28$ and $f_{+224d}=0.29$ for SN\,2022jli. We derived the same ratios for two epochs of spectra of the SESN SN\,2013ge and found $f_{\mathrm{SN2013ge,+9d}} = 0.24$ and $f_{\mathrm{N2013ge,+20d}}=0.36$. In the end, we used a uniform faction of 0.3 for the NIR contribution and added a 10\% uncertainty to account for the variation. The distance uncertainty was not included in the uncertainty budget. The resulting bolometric light curve is shown in Fig.~\ref{fig:fig3} and Extended Data Fig.~\ref{fig:Lbol_undulation}. 

The pseudo-bolometric light curve of SN\,2022jli shows clear undulations as seen in the individual optical light curves. We constructed the empirical undulation profile the same way as done in Method Section $\S~5$ by adopting a 6th-order polynomial for the baseline. We notice an enhancement in the pseudo-bolometric luminosity during the last bump prior to the rapid decline (panels b and c of Extended Data Fig.~\ref{fig:Lbol_undulation}). If we assume the last bump followed the evolution as the previous undulations, the integrated luminosity of the extrapolation in the last bump would be $\sim 3.0\times10^{47}$ ergs while the integrated luminosity of observed data in the last bump is $\sim 3.9\times10^{47}$ ergs. In the accretion-powered scenario (Method Section $\S~7$), the excess radiation in the last bump converts to an accretion mass of $M\mathrm{_{excess}}=5\times10^{-6} \left(\frac{\epsilon}{0.01}\right)^{-1}$ M$_\odot$, where $\epsilon$ is the radiative efficiency of accretion. The undulation profile of the last bump is significantly different from the previous undulations, with a peak time delayed by around 4 days.

\subsection{7. The energy source of the SN luminosity}

In the standard model of SNe Ic, the SN luminosity is dominantly powered by radioactive decay of unstable isotopes, in particular $^{56}$Ni, and its daughter element $^{56}$Co, with some contribution from previously stored kinetic energy in the early time through shock cooling emission. During the first month after discovery, SN\,2022jli looks like a normal Type Ic supernova both in terms of the spectral evolution (Extended Data Fig.~\ref{fig:spec_evolution_compare}) and from the light curve evolution (Extended Data Fig.~\ref{fig:lc_compare}). Due to the spectral and photometric similarities, it is natural to attribute the first decline phase to $^{56}$Ni decay as in normal SNe Ic. We found that around $0.15\,\mathrm{\msun\,^{56}Ni}$ produced in the explosion combined with a characteristic diffusion time scale of around 17.5 days and a $\gamma$-ray escape time scale of around 100 days can explain the first peak of SN\,2022jli (Fig.~\ref{fig:fig3}). The $\gamma$-ray escape time scale is adopted arbitrarily with a typical value for SNe Ib/c \cite{Sharon2020}. The characteristic diffusion time is a function of ejecta mass, velocity, and opacity. Due to the limited amount of data around the first peak, we do not attempt to constrain the ejecta properties by modeling the light curve.

After the first decline phase, SN\,2022jli brightens again around +20 days after discovery and reaches the second peak around +50 days after discovery. This makes SN\,2022jli a double-peaked SESN. After the second peak, the luminosity of SN\,2022jli declines around ten times from +70 to +270 days. After subtracting the underlying radioactive decay emission from $0.15\,\mathrm{\msun\,^{56}Ni}$, the integrated energy is $2\times 10^{49}$ erg, which requires an additional energy source. 
Double-peaked light curves have been observed in other SESNe, for example, SN\,2005bf\cite{Folatelli2006}, PTF11mnb \cite{Taddia2018}, SN\,2019cad\cite{Gutierrez2021}, SN\,2019stc\cite{Gomez2021, Chugai2022}, SN\,2021uvy\cite{Gomez2022}, SN\,2022xxf\cite{Kuncarayakti2023}, and a sample of such objects in ref\cite{Das2023}. The morphology of those double-peaked light curves shows a large diversity. Except SN\,2021uvy, all of them do not have a long-lasting gradual decline phase after the second peak as seen in SN\,2022jli (Extended Data Fig.~\ref{fig:lc_compare}). The popular explanation for the double-peaked light curves of SESNe includes interaction between the ejecta and either extended material at the outskirts of the progenitor \cite{Piro2015} or detached circumstellar material surrounding the progenitor \cite{Jin2021}; double-peaked distribution of radioactive $^{56}$Ni\cite{Folatelli2006, Orellana2022}; and enhanced magnetar power\cite{Chugai2022}.  
Below we discuss the possible origins of the second peak of SN\,2022jli, and test whether they can explain the undulations observed after the second peak. 

The light curve of SN\,2022jli after the second peak declines at a rate similar to that of light curves powered by fully trapped $\gamma$-ray from the decay of $^{56}$Co, i.e., around 0.01 mag/day. It is appealing to attribute the double-peaked light curves to the result of the double-peaked distribution of $^{56}$Ni. In this scenario, the first peak is powered by $^{56}$Ni carried out by a jet-like phenomenon and deposited in the outer layers of the ejecta\cite{Bersten2013}, and the second peak is powered by $^{56}$Ni residing in the deep layers of the ejecta. One problem with the $^{56}$Ni explanation is the plummet of luminosity at very late times. The luminosity declined by one order of magnitude during around 20 days, which can not be explained by, for example, $\gamma$-ray photon leakage or the formation of cold dust in the ejecta. Another issue with $^{56}$Ni comes when considering the diffusion time scale. The timescale of the $^{56}$Ni-powered peak is determined by the time it takes the emission to diffuse through the ejecta. The long time before the second peak means there must be a large ejecta mass, causing a long diffusion time. Such a long diffusion time will smear out short-time scale signals generated from the center of the ejecta (Method Section $\S~12$). The other possibility for the energy source is the spin-down energy of the newborn magnetar. Similarly, the magnetar model also has problems with the late-time drop of the supernova flux and with the diffusion time. 

The interaction between ejecta and CSM could explain the rise and fall of the second peak if the CSM is distributed in a confined distance range from the progenitor. The interaction could generate periodic energy input if the CSM has evenly distributed density fluctuation. However, such periodic energy fails to produce the observed bumps in the light curves of SN\,2022jli due to the light travel time difference (Method Section $\S~12$). 


Accretion onto the supernova remnant has been proposed as another energy source to power supernova light curves, and most of the models consider the fallback ejecta as the fuel\cite{Chevalier1989, Zhang2008, Dexter2013, Moriya2019}. Accretion power could provide the extra energy of SN\,2022jli. In light of the existence of a bloated companion star in a close orbit (Method Section $\S~13$) and the requirement of the hydrogen-rich material to explain the observed H$\alpha$ emission (Method Section $\S~9$), we propose that the companion star with hydrogen-rich envelope is the donor to fuel the accretion. The accreted mass required to account for the extra $2\times 10^{49}$ erg is $M_{{\rm acc}} = 10^{-3} \left(\frac{\epsilon}{0.01}\right)^{-1}\msun$ where $\epsilon$ is the radiative efficiency of accretion. In the accretion scenario, an extremely high Eddington luminosity ratio is inevitable to explain the observed light curve. For a 1.4 $\msun$ neutron star, the Eddington luminosity ratio is around $10^4$. For a $10~\msun$ black hole, the Eddington luminosity ratio decreases to around $10^3$, alleviating the super-Eddington tension. Many ultraluminous X-ray sources\cite {Walton2011, Israel2017} have been observed to have Eddington luminosity ratios above 100, but still, the mechanism of super-Eddington accretion remains an open question\cite{Brightman2019}. If the accretion origin is valid for SN\,2022jli, it provides a new environment and opportunity to study super-Eddington accretion.


Under the scenario that accretion powers the excess emission of SN\,2022jli, it is intriguing to consider the possibility of one pre-existing accreting compact object in the system. Regardless of the origin of the compact object, one strong constraint on the accretion is that we need hydrogen-rich material to fuel the accretion to explain the hydrogen emission and its evolution in the late-time spectra (see below Method Section $\S~9$ and $\S~10$). However, the non-detection of hydrogen lines in the photospheric spectra (see below Method Section $\S~8$) indicates hydrogen-poor ejecta, which excludes the supernova ejecta as the main fuel for accretion. This means that accretion of the supernova ejecta onto a pre-existing compact companion to the progenitor star can not explain the spectroscopic properties of SN\,2022jli. Therefore, we suggest a newly formed compact object in the supernova explosion and a companion star with a hydrogen-rich envelope is the most likely binary system.

\subsection{8. Spectral analysis}
\paragraph{Photospheric spectra:}
Two spectra taken on 2022 May 11 (+6.0 days after discovery) and 2022 May 24 (+19.2 days after discovery) were used to classify SN\,2022jli\cite{Grzegorzek2022, Cosentino2022}; these spectra are available on TNS. We estimated an ejecta velocity of around 8,200 km s$^{-1}$ from the absorption minimum of the identified absorption lines. We compared the +19.2d spectrum of SN\,2022jli to other supernovae (Extended Data Fig.~\ref{fig:spec_evolution_compare}) and found that SN\,2022jli resembles normal SNe Ib/c well, and it is a genuinely good match to the SE SN\,2013ge. Helium absorption was found in the early-time spectra of SN\,2013ge through careful analysis of both the optical and NIR spectral sequence, which likely results from a thin layer of helium remaining at the time of core collapse\cite{Drout2016}. Actually, a complete stripping of the He layer from the progenitor stars of SNe Ic is not expected in many models, and the contribution of the remaining helium layer to the spectra of SNe Ic has been long debated\cite{Hachinger2012, Dessart2020b, Williamson2021}. Helium absorption might also exist in early-time spectra of SN\,2022jli, but we do not have an extensive enough early-time spectral sequence to explore this fully. We point out that the absorption signature of Helium was indeed reported in the NIR spectra of SN\,2022jli presented in ref\cite{Tinyanont2023}.


\paragraph{Nebular spectra:} Normal SNe Ib/c start to enter the nebular phase some months after the explosion, during which the supernova ejecta become optically thin, and emission lines with little continuum emission dominate the spectra. The spectra of SN\,2022jli are well sampled at late times during the long-lasting gradual decline phase and the fast-decline phase. The late-time spectra of SN\,2022jli show significant differences with those of normal SNe Ib/c, for example, SN\,2013ge in Supplementary Information Fig.~\ref{fig:compare_2013ge}. Before the fast-decline phase, the spectra of SN\,2022jli are good matches to the other two long-lasting SESNe (SN\,2012au and iPTF15dtg), showing prominent permitted Oxygen emission and iron-plateau, and they look more similar to some SLSNe, GRB-SNe than normal SNe Ib/c (Extended data Fig.~\ref{fig:spec_evolution_compare}). We identify the emission lines in Extended data Fig.~\ref{fig:spec_lineident}. Besides the aforementioned iron plateau and Oxygen emission, the other elements contributing to the plethora of emission lines in SN\,2023jli spectra can be identified as Ca, Mg, C, and Na. The \ionp{O}{i} $\lambda$7774 line seems to have two different components manifesting as a relatively narrow feature on a broader base. The \ionp{O}{i} $\lambda 9263$ line appears to be an isolated line without significant blending from other emissions. We measured the width of the \ionp{O}{i} $\lambda 9263$ line by fitting a simple Gaussian profile and obtained the velocities as follows: $v_{\rm FWHM}({\rm \ionp{O}{i}}\lambda 9263, +203.8\,{\rm d}) = 2660\pm 90$ km\,s$^{-1}$, and $v_{\rm FWHM}({\rm \ionp{O}{i}}\lambda 9263, +260.9\,{\rm d}) = 2720\pm 50$ km\,s$^{-1}$. We also measured the width of the narrow component of the \ionp{O}{i}$ \lambda$7774 line, and obtained $v_{\rm FWHM}({\rm \ionp{O}{i}} \lambda 7774, +203.8\,{\rm d}) = 2580\pm 200$ km\,s$^{-1}$, and $v_{\rm FWHM}({\rm \ionp{O}{i}} \lambda 7774, +260.9\,{\rm d}) = 2480\pm 30$ km\,s$^{-1}$. Before transitioning to the fast-decline phase ($\lesssim +270$ days), SN\,2022jli shows prominent emission around 6500\,\AA, which differs from all the other comparison objects (see Methods $\S~9$ for discussion on this feature).  


Accompanying the fast photometric evolution of SN\,2022jli from the gradual decline to the rapid drop in luminosity, significant spectral evolution was also noticed (Extended Data Fig.~\ref{fig:spec_fast_decline}). One conspicuous change is the disappearance of those narrow Oxygen lines, including \ionp{O}{i} $\lambda$6158, \ionp{O}{i} $\lambda$7774, \ionp{O}{i} $\lambda$8446, and \ionp{O}{i} $\lambda$9263, and probably also \ionf{O}{ii} $\lambda\lambda$7320,7330 after the supernova enters the fast-declining phase ($>270$ days after discovery). The narrow features of the \ionp{Ca}{ii} NIR triplets also disappeared. In the meantime, the \ionf{O}{i} $\lambda\lambda$6300,6363 lines emerge. We measured an \ionf{O}{i} $\lambda$6300 line width of $v_{\rm FWHM}(+280\,{\rm d}) = 5520\pm 190$ km\,s$^{-1}$. Theoretically, the fast change of \ionp{O}{i} emission from permitted emission to forbidden emission indicates a dramatic change of density or temperature in the ejecta. In the case of SN\,2022jli, the transition happened rather quickly, during which the ejecta density is not expected to change much. Therefore, the most plausible explanation for the spectral change is a decrease in temperature due to the quenching of the central energy source, which is consistent with the contemporary fast decline of the light curve. The narrower widths of the disappearing lines corroborate the idea that they were emitted from the inner parts of the ejecta with lower velocities. 

Theoretical calculations of nebular phase spectra of SESNe powered by both Nickel decay and a central engine are rare\cite{Dessart2019, Omand2023}. Detailed modeling of SN\,2022jli spectra considering both energy sources and their temporal evolution might shed more light on the explosion mechanism, but is beyond the scope of this paper.

\subsection{9. Hydrogen emission in the late-time spectra}
\label{sec:hydrogen_emission}

The spectra of SN\,2022jli show unique strong emission around 6500\,\AA\,(Extended Data Fig.~\ref{fig:spec_evolution_compare}). The emission seems to comprise two components, with one narrow feature sitting on top of a broad component. The wavelength range of interest in the supernova spectrum is contaminated by host galaxy emission from nearby \ionp{H}{ii} regions, but we confirm that the narrow feature does not come from any artifact due to host galaxy line contamination (Supplementary Information Fig.~\ref{fig:xsh_Ha}). We attribute the narrow emission feature as H$\alpha$ emission, and the co-evolution with the emission at the wavelength of H$\beta$ supports this identification (Supplementary Information Fig.~\ref{fig:HaHb_coevolution}). The narrow feature shows back-and-forth shifts in wavelength, i.e., moving to shorter and longer wavelengths around the rest-wavelength of H$\alpha$ (bottom left panel of Fig.~\ref{fig:fig4}). The shifting behavior shows a cyclical pattern, consistent with the 12.4-day period as derived for the undulation period in the optical light curves.  The luminosity of the narrow H$\alpha$ line closely follows the total luminosity of the supernova as ${\rm L_{H\alpha} = 0.004 \times L_{bol} }$ (Extended Data Fig.~\ref{fig:Ha_luminosity_velocity}). There is a transition phase in the bolometric luminosity, an overshooting before it drops, during which the H$\alpha$ luminosity also follows the bolometric luminosity closely. The narrow feature disappears after the SN enters the fast-decline phase (Extended Data Fig.~\ref{fig:spec_fast_decline}). The tight connection between the narrow H$\alpha$ feature and the light curve of the supernova, both the periodicity and the luminosity, implies that the energy that powers the extra emission of the supernova is also responsible for the H$\alpha$ emission. 

The region producing the periodic undulation in the light curve has been limited to a relatively small size in the center of the ejecta, which implies the emission of H$\alpha$ also comes from the center of the ejecta. The hydrogen material that gives rise to the narrow H$\alpha$ and H$\beta$ emission most likely comes from the envelope of the companion star that is accreted onto the newly formed compact object. H$\alpha$ and H$\beta$ emission lines have been commonly observed in binary systems with accretion 
disks\cite{Budaj2005, Miller2007, Atwood-Stone2012}, which could serve as an analogy in the low accretion rate regime to understand the emission mechanism and structure of the H$\alpha$ in SN\,2022jli.

After the narrow H$\alpha$ emission vanishes, there is still a prominent emission around 6500\,\AA~(Extended Data Fig.~\ref{fig:spec_fast_decline}). Similar emission has been observed in other SESNe\cite{Patat1995, Maeda2007, Taubenberger2011}, mainly SNe IIb or Ib. The origin of such emissions has been debated. For example, the emission has been explained as either \ionf{N}{ii}$ \lambda\lambda6548,6583$ \cite{Jerkstrand2015, Fang2018} or H$\alpha$ \cite{Matheson2000a, Matheson2000b}. In SNe IIb, the hydrogen could be leftovers in the outer layer, and the ejecta-wind interaction could be a possible power source for the ionization. However, Dessart et al. (2021)\cite{Dessart2021} argue that the hydrogen envelopes of Type IIb SNe are too small and dilute to produce any noticeable H$\alpha$ emission or absorption after $\sim$150 days. We performed spectral decomposition of the spectra around 6400\,\AA. The spectra can be well decomposed into four Gaussian emission profiles, among which two are emission lines of \ionf{O}{i} $\lambda\lambda6300,6363$ (Supplementary Information Fig.~\ref{fig:specfit}), but the derived velocities for the different lines are not consistent for \ionf{O}{i} and the \ionf{N}{ii} or H$\alpha$ components. We get $v_{[{\rm OI}]} = -890\,{\rm km\,s^{-1} (+271d)}$, $v_{[{\rm OI}]} = -480\,{\rm km\,s^{-1} (+275d)}$, $v_{[{\rm OI}]} = -310\,{\rm km\,s^{-1} (+280d)}$; $v_{[{\rm NII}]} = -3770\,{\rm km\,s^{-1} (+271d)}$, $v_{[{\rm NII}]} = -2690\,{\rm km\,s^{-1} (+275d)}$, $v_{[{\rm NII}]} = -2200\,{\rm km\,s^{-1} (+280d)}$; $v_{{\rm H}\alpha} = -3260\,{\rm km\,s^{-1} (+271d)}$, $v_{{\rm H}\alpha} = -2170\,{\rm km\,s^{-1} (+275d)}$, $v_{{\rm H}\alpha} = -1680\,{\rm km\,s^{-1} (+280d)}$. It is difficult to explain the velocity difference if the emission is dominated by \ionf{N}{ii}. Another possible origin of the emission could be hydrogen stripped from the companion star by the supernova ejecta\cite{Marietta2000, LiuZW2015, Dessart2020a}. Late-time H$\alpha$ emission with the potential origin of stripped hydrogen from the companion star has been observed in several Type Ia supernovae\cite{Kollmeier2019, Prieto2020, Elias-Rosa2021}. However, the observed H$\alpha$ profiles in those SN Ia spectra have much narrower line widths around $\mathrm{1000\,km\,s^{-1}}$ while the component on the red side of \ionf{O}{i} $\lambda\lambda6300,6363$ in SN\,2022jli has a width of FWHM $\sim\mathrm{10,000\,km\,s^{-1}}$ which is twice the width of the \ionf{O}{i} $\lambda6300$ line. The hydrogen origin from the companion star has also been proposed to explain the nebular H$\alpha$ emission in the Type Ic SLSN iPTF13ehe\cite{Yan2015, Moriya2015}, for which other works argue for a hydrogen origin from hydrogen-rich CSM produced in mass loss before the supernova explosion\cite{Yan2015,Yan2017}.



\subsection{10. Evolution of the accretion-powered H$\alpha$ emission}

 In the accretion-powered supernova scenario, the compact remnant and the companion star are bound in an eccentric orbit, where the hydrogen-rich material is accreted from the envelope of the companion star to the newborn compact remnant every time the compact remnant passes through the pericenter of the orbit.  Now we consider whether the orbital motion of such an eccentric orbit can explain the observed velocity shift. The light-of-sight velocity of the compact remnant can be written as 
 \begin{equation}
    V_r = V_z + K(\cos(w+f)+e \cos(w))\, ,
 \end{equation}
 where  $V_Z$ is the proper motion velocity of the binary system barycenter, K is the characteristic velocity amplitude, $w$ is the argument of periapse,  $f$ is the true anomaly, and $e$ is the orbital eccentricity. We adopted the same nomenclature for the orbital elements as used by ref \cite{Murray2010}. 
The velocity amplitude can be written as
\begin{equation}
    K = \frac{m_2}{m_c+m_2}\frac{n a \sin i}{\sqrt{1-e^2}} = 214 (\frac{m_2}{15\msun})^2 (\frac{m_c+m_2}{1.4\msun+15\msun})^{-\frac{2}{3}}\frac{\sin i}{\sqrt{1-e^2}} \mathrm{km\,s^{-1}}\, ,
\end{equation}
where $m_c$ is the compact remnant mass, $m_2$ is the companion star mass, $n=\frac{2\pi}{P}$, $a$ is the semi-major axis of the elliptical orbit, and $i$ is the inclination of the orbit. We can see the velocity amplitude is highly degenerate with respect to the binary masses and the orbital eccentricity. 
 

 We extracted the accretion-powered H$\alpha$ by simply subtracting the emission by a pseudo continuum. The continuum model is constructed by fitting a linear model to the arbitrarily selected continuum region on both sides of the emission feature.  We estimated the H$\alpha$ velocities by measuring the flux-weighted centroid of the emission feature without considering the nontrivial velocity structure. We estimated the uncertainties for the velocity by considering both the flux uncertainty and the systematic uncertainty due to the different choices on the continuum region. The uncertainty budget did not include the systematic uncertainty that would be introduced by the unknown nature of the underlying baseline for the Halpha feature. The result is shown in Extended Data Fig.~\ref{fig:Ha_luminosity_velocity}. Only the last bump period was well sampled with decent spectral resolutions. We caution the unusual rebrightening of the last bump (Method Section $\S~6$), which might introduce other contributions to $H\alpha$ velocity besides the orbital motion. Generally speaking, the Keplerian orbit with high eccentricity can roughly account for the velocity evolution trend, as shown by four models with representative parameters. In this exercise, we simply subtracted one arbitrarily chosen linear ``continuum'' to get the H$\alpha$ emission without considering the potential contamination from other lines. The spectral data analyzed here span more than 120 days, and the spectral evolution could cause problems when we combine data from different periods with significant separation in time. This might explain the obvious outlier from the earliest spectrum.

In this heuristic experiment, we assume the orbital velocity is the major component. We need to point out that other possible kinematic processes may complicate the velocity evolution. These complications include the velocity structure of the accretion disk, the mass flow from the companion star to the compact object, and the outflow of winds. The potential eclipse of the accretion disk by the companion star and the emission from the companion envelope could also complicate the velocity structure of the H$\alpha$ emission line. We stress that the velocity measurement with the current method under simplified assumptions is helpful in understanding the property of the binary system qualitatively. We caution the readers to refrain from deriving the exact orbital parameters from the current analysis.



\subsection{11. X-ray and radio observations:}
\paragraph{Chandra}

We obtained three epochs of X-ray observation with the Advanced CCD Imaging Spectrometer (ACIS) of the Chandra X-ray Observatory from 2023 January 17 to 2023 January 25 (+257.4, +261.8, and +265.8 days after discovery) under an approved Director Discretionary Time Proposal (PI: Chen). Each epoch has an exposure time of 10.06 ks. The Chandra ACIS-S data were reduced with the CIAO\cite{Fruscione2006} software package (v4.14) and relevant calibration files (CALDB version 4.10.2), applying standard filtering criteria. SN\,2022jli was not detected in any of the three epochs. We measured the count rate within a 5'' radius aperture at the supernova position and obtained absorbed flux upper limits in 95\% confidence interval of $\mathrm{5.27\times10^{-15}\,erg\,s^{-1}\,cm^{-2}}$ (+257.4 days), $\mathrm{5.32\times10^{-15}\,erg\,s^{-1}\,cm^{-2}}$ (+261.8 days), and $\mathrm{5.27\times10^{-15}\,erg\,s^{-1}\,cm^{-2}}$ (+265.8 days).  We merged the three epochs and obtained an absorbed flux upper limit of $\mathrm{f_{ul}(0.5-7\,keV; 95\%)} = 1.77\times10^{-15}\,\mathrm{erg\,s^{-1}\,cm^{-2}}$.

The Galactic neutral hydrogen column density in the direction of the SN is $\mathrm{N_{H_{gal}}}=3.49\times {10}^{20}$ cm$^{-2}$ (ref\cite{Kalberla2005}). From our optical spectra, we estimate $E{(B-V)}_{{\rm{host}}}=0.25$ mag. Assuming a Galactic dust-to-gas ratio\cite{Guver2009}, $\mathrm{N_H(cm^{-2}) = 2.21\times 10^{21} A_V (mag)}$, the extinction value corresponds to an intrinsic neutral hydrogen column density of $\mathrm{N_{H_{host}}}\sim 1.7\times {10}^{21}\;{\mathrm{cm}}^{-2}$. For an assumed simple power-law spectral model with spectral photon index ${\rm{\Gamma }}=2$, we find an unabsorbed flux limit in 95\% confidence interval of $2.18\times {10}^{-15}$ erg s$^{-1}$ cm$^{-2}$ (0.5–7 keV). At the distance of 22.5 Mpc, this flux translates into a luminosity of $1.32\times {10}^{38}$ erg s$^{-1}$.

The supernova ejecta may be optically thick to soft X-rays for decades. The opacities of X-ray flux absorption are dominantly due to photoelectric absorption (photon-ionization and photon-excitation) below around 100 keV. The X-ray optical depth at a given epoch can be described as $\mathrm{\tau(E) \simeq \frac{3\kappa_{bf}M_{ej}}{4\pi(v_{ej}t)^2}}$ where $M_{ej}$ is the supernova ejecta mass, $v_{ej}$ is the ejecta velocity, and $\kappa_{bf}$ is the bound-free opacity which is largely determined by the ejecta abundance. Alp et al.\cite{Alp2018} calculated a typical optical depth of $\tau(t, E) \simeq 100t_4^{-2}E^{-2}$ for core-collapse supernovae, where $t_4$ is the time since the explosion in units of 10000 days, and E is the energy in units of keV. In the case of SN\,2022jli, we have $\tau(\mathrm{+250\,d, 1\,keV}) = 1.6\times10^5$,  $\tau(\mathrm{+250\,d, 10\,keV}) = 1.6\times10^3$. The supernova progenitor in their calculation has an ejecta mass of around 10 $\msun$. Even if SN\,2022jli has a smaller ejecta mass, the X-ray optical depth in the Chandra energy band is still very high.


In the context of interacting supernovae, our upper limits constrain the X-ray luminosity to be more than three orders of magnitude lower than the optical/NIR luminosity at the same epoch. Such a ratio of X-ray to optical/NIR luminosity puts a strong constraint on the interaction contribution to the bolometric luminosity. Higher X-ray to optical/NIR luminosity have been observed in supernovae with strong interaction,  for example,  Type IIn SN 2010jl\cite{Ofek2014, Chandra2015}, Type Ibn SN 2006jc\cite{Immler2008}, and Type IIn SN 2006jd\cite{Chandra2012, Stritzinger2012}, where strong X-ray emission from ejecta-CSM interaction is detected by the Chandra observatory. 

\paragraph{NuSTAR}

SN\,2022jli was also observed with the Nuclear Spectroscopic Telescope Array (NuSTAR \cite{Harrison2013}). NuSTAR has two coaligned X-ray telescopes, with corresponding focal plane modules FPMA and FPMB. The observations were conducted in three epochs spanning from +227 to +237 days after discovery. The first epoch (ID 90801535002) was on 2022 December 18, with an effective exposure time of 20,353 seconds (FPMA) and 20,142 seconds (FPMB). The second epoch (ID 90801535004) was on 2022 December 23, with an effective exposure time of 20,231 seconds (FPMA) and 20,008 seconds (FPMB). The third epoch (ID 90801535006) was on 2022 December 27, with an exposure time of 20,257 seconds (FPMA) and 20,036 seconds (FPMB).   
The data were reduced using HEASoft v.6.31 and the NuSTAR Data Analysis Software (NuSTARDAS) v.2.1.2, in particular
the {\tt nupipeline (version 0.4.9)} and {\tt nuproducts} routines. No source was detected at the supernova position.  
We focused our analysis on the 30\,--\,60\,keV energy range because the probability of detecting the supernova is higher in the hard X-ray energy range with lower optical depth. We estimated an upper limit of $\mathrm{4.5\times10^{-4}\,counts \,s^{-1}}$ by calculating $\mathrm{3\times\sqrt{B_{tot}}/t_{exp}}$ where $\mathrm{B_{tot}}$ is the total count in a circular aperture with 50'' radius at the position of SN\,2022jli from both epochs and both instruments and $\mathrm{t_{exp}=121\,ks}$ is the total effective exposure time. Assuming a power-law model with photon index $\Gamma=2$, the above count rate upper limit corresponds to an unabsorbed flux of  $\mathrm{f_{ul}(30-60\,keV)=4.1\times10^{-13}\,erg\,s^{-1}\,cm^{-2}}$ or an upper limit of $\mathrm{2.5\times10^{40}\,erg\,s^{-1}}$ at a luminosity distance of 22.5 Mpc.




\paragraph{ATCA}

We observed SN\,2022jli with the Australian Telescope Compact Array 
(ATCA\cite{Wilson2011}) in the C/X-band ($3.9$--$11.0$\,GHz) on $\textrm{MJD } 59916$ for 2 hours and on $\textrm{MJD } 59931$ for 5 hours using the extended 6C array configuration (CX517). For both observations, we used the ATCA calibrator source PKS 1934-638 as the primary flux calibrator and the calibrator source PKS 0003-066 for phase calibration scans. We used the \texttt{Miriad} software\cite{Sault1995} to reduce and image the data from these observations. We used \texttt{mfclean} with Briggs weighting and a \texttt{robust} parameter of $0.0$, with multi-frequency synthesis (\texttt{mfs}) deconvolution. 

Neither observation resulted in a clear detection of SN\,2022jli in the observed frequency range. We use \texttt{imfit} in C-band ($5.5$\,GHz) to extract $5\sigma$ non-detection limits of $<0.176$\,mJy beam$^{-1}$ and $<0.074$\,mJy beam$^{-1}$, respectively on $\textrm{MJD } 59916$ and $\textrm{MJD } 59931$. At the X-band $9.0$\,GHz central observing frequency we obtain a $5\sigma$ limits of $<0.057$\,mJy beam$^{-1}$ and $<0.055$\,mJy beam$^{-1}$, respectively on $\textrm{MJD } 59916$ and $\textrm{MJD } 59931$.

\subsection{12. Light travel time and diffusion time}
We can use the time scale of the 12.4-day period and the time scale of around three days in the rising part of the undulation profile to constrain the size and location of the varying energy source. If the process that generates the energy powering the periodically undulating light curve happens outside of the supernova ejecta, for example, the ejecta and CSM interaction (ECI), the finite size of the ejecta can cause different light travel time from different parts of the ejecta, i.e., the emission from the closer side to the observer arrive earlier. In the case of ECI, if we consider a simple model where the optically thin ejecta expand homologously and the CSM has a spherical shell-like structure, the light travel time difference between the earliest and latest arrival of emission from the same sphere is $\Delta t \sim \frac{R}{c} = 10\times \frac{v}{10^4\,\mathrm{km\,s^{-1}}} \frac{t}{300\,\mathrm{day}} $days. Such a phase-dependent light travel time difference will smear out any potential periodic signals.

If the energy source is located inside the ejecta, any sudden change in the input luminosity is subject to delay and smear effects by the ejecta that the emission goes through before reaching the observer. The short rising time of around 3 days of the bump requires the diffusion time to be less than that. The diffusion time through spherical ejecta with opacity $\kappa$ from a radius $r$ to the outer edge R is 
\begin{equation}
    t_{\mathrm{diff}}(r, R) = \frac{\kappa}{c}\int^{R}_r \rho(r) r dr\,.
\end{equation}
If adopting ejecta density profile $\rho(r) = \frac{M}{2\pi R^2r}$ (ref\cite{Chevalier&Soker1989}), we get

\begin{equation}
    t_{\mathrm{diff}}(r, R) = \frac{M\kappa}{2\pi c R}\left(1-\frac{r}{R}\right)\,.
\end{equation}

 For ejecta in homologous expansion, we have $R = v t $ where $v$ is the velocity of the outer edge of the ejecta, and $t$ is the time since the supernova explosion. We get the diffusion time from the center of the ejecta  
\begin{equation}
    t_{\mathrm{diff}}(0, v t) = 2\times \frac{M}{M_\odot}\frac{\kappa}{0.07\,\mathrm{cm^2\,g^{-1}}} \left(\frac{v}{10^4\, \mathrm{km\,s^{-1}}}\right)^{-1} \left(\frac{t}{50\,\mathrm{day}}\right)^{-1} \mathrm{day}\,.
\end{equation}
If we assume the discovery time of SN\,2022jli is around the first peak and assume a typical rising time of 15 days for SNe Ic, the second peak is around 70 days after the explosion. We note that the discovery time could be later than the first peak because we missed the rising part of the first peak. Therefore, at the second peak, adopting ejecta velocity of 8,200 km\,s$^{-1}$, the requirement of diffusion time less than 3 days is roughly consistent with the energy source residing in the center of the ejecta if the ejecta mass ${\rm M}_{\rm ej} \lesssim 1.7~M_\odot$. 

\subsection{13. Supernova explosion in a binary system}

Two mainstream scenarios, the stellar-wind origin, and the binary interaction origin, have been proposed to explain envelope stripping for the progenitor of SESNe. Observational evidence has been found in both directions\cite{Langer2012, Gal-Yam2014}. A supernova explosion in a compact binary system, i.e., a binary system with a small separation between two stars, significantly impacts the companion star and the further evolution of the binary system\cite{Postnov2014}. The existence of the companion star, in return, might have profound effects on the manifestation of the observed supernova.  

Assuming the mass loss in the supernova explosion is instantaneous, the equations relating the pre-explosion and post-explosion orbital parameters (in an instantaneous reference frame centered on the companion star right at the time of explosion) are \cite{Postnov2014}:

\begin{equation}
    \mu_f \frac{V_f^2}{2} - \frac{GM_cM_2}{a_i} = - \frac{GM_cM_2}{2a_f}
\end{equation}

and 
\begin{equation}
    \mu_f a_i \sqrt{w_z^2 + (V_i+w_y)^2} =  \mu_f\sqrt{G(M_c+M_2)a_f(1-e^2)}
\end{equation}

where $M_2$ is the companion mass, $M_c$ is compact remnant mass, $\mu_f=M_cM_2/(M_c+M_2)$ is the reduced mass of the system after explosion, $V_i$ is the relative velocity before explosion,  $V_f$ is the relative velocity after explosion, $a_i$ is the semi-major radius of the orbit before explosion, $a_f$ is the semi-major radius of the orbit after explosion, $\vec{w} = (w_x, w_y, w_z)$ is the kick velocity.

The above equations result in the following evolution of orbital parameters:
\begin{equation}
    \frac{a_f}{a_i} = \left[2-\chi \left(\frac{w_x^2+w_z^2+(V_i+w_y)^2}{V_i^2}\right)\right]^{-1}
\end{equation}

and 
\begin{equation}
    1-e^2 = \chi\frac{a_i}{a_f}\left(\frac{w_z^2+(V_i+w_y)^2}{V_i^2}\right)
\end{equation}
where $\chi=(M_1+M_2)/(M_c+M_2) > 1$. $M_1$ is the mass of the primary star that exploded. When the explosion is spherically symmetric, the compact remnant gains no kick velocity, i.e., $V_f = V_i$. If the binary system remains bound, we get the simplified results of the orbital parameters of 
\begin{equation}
    \frac{a_f}{a_i} = \frac{1}{2-\chi}
\end{equation}
and
\begin{equation}
    e=\frac{M_1 - M_c}{M_c+M_2}
\end{equation}
The above results tell that the supernova explosion will widen the binary orbit and introduce eccentricity to the system even when the explosion is symmetric.

In the case of SN\,2022jli, the post-explosion binary system has
\begin{equation}
    a_f = \left[\frac{G(M_2+M_c)P^2}{4\pi^2}\right]^{1/3}\, ,
\end{equation}
where $P = 12.4$ days. Before the supernova explosion, the separation between the primary and secondary stars is smaller, which can easily achieve $\frac{a_i}{R_2} < 10$, where $R_2$ is the radius of the companion star, for a large parameter space of $M_1$, $M_2$, and $M_c$, in which regime the ejecta impact has a significant influence on the companion\cite{LiuZW2015, Hirai2018, Ogata2021}. The momentum transfer and energy injection will result in an impact velocity to the companion, strip material from the companion, and bloat the envelope of the companion. The amount of removed stellar mass, the resulting impact velocity, and the companion's reaction to the impact in radius/temperature/luminosity strongly depend on the binary separation and explosion energy. For a main-sequence companion star, the radius can easily increase by one or two orders of magnitude\cite{LiuZW2015, Ogata2021} for a typical explosion energy of 10$^{51}$ erg and $\frac{a}{R_2}\lesssim 10$. The fluffy material of the bloated envelope makes it easier to get accreted onto the supernova remnant, which could form an accretion disk and provide extra energy to power the supernova\cite{Hober2022}. The accretion rate is not expected to be constant since the companion radius evolves. Once the deposited energy fully radiates away, the companion starts to contract by releasing their gravitational energy and eventually resumes their original state before the SN explosion. 

Due to accretion from the companion star to the compact remnant with likely wind and jet production in the meantime, the evolution of the surviving binary system is subject to the mass loss and mass transfer process. The detailed orbital evolution of the binary system is beyond the scope of this work. But we can argue qualitatively if any significant orbital evolution is expected during the $\sim$ 300-day evolution of SN\,2022jli. If we assume non-conservative, i.e., $\dot{M}, \dot{J_{\mathrm{orb}}}\neq 0$ where $\dot{M}$ is the total mass change of the system and  $\dot{J_{\mathrm{orb}}}$ is the orbital angular momentum, delta-function mass loss/transfer through Roche Lobe Overflow (RLOF), the secular relative change of the orbital semimajor axis has $\langle \frac{da}{a} \rangle \sim -\frac{d M_2}{M_1}$ with an order-of-magnitude estimate \cite{Dosopoulou2016}, where $M_2$ is the mass of the accretor and $M_1$ is the mass of the donor. With $dM_2 \sim M_{\mathrm{acc}} = 10^{-3} \left( \frac{\epsilon}{0.01}\right)^{-1} \msun$ for SN\,2022jli (Method Section $\S~7$), and assuming $M_1=5 \msun$, we can tell the relative change of semi-major axis during the first $\sim 300$ days of SN\,2022jli is very small which is consistent with what we found in Method Section $\S~5$. 

A surviving companion star has been important evidence for the binary origin of some SESNe. Such companion stars might have been observed in SN\,1993J\cite{Maund2004} (SN IIb), SN\,2001ig\cite{Ryder2018} (SN IIb), SN\,2006jc\cite{Maund2016, Sun2020} (SN Ibn), and SN\,2011dh\cite{Folatelli2014, Maund2019} (SN IIb). SN\,2022jli provides a good candidate to search for the surviving companion star of a Type Ib/c supernova. 

Detailed numerical simulations have shown that a supernova explosion can often unbind the secondary star in the binary system\cite{Tauris1998, Portegies2000}. In fact, ref\cite{Renzo2019} found that around 90\% of the pre-explosion binary systems become unbound after the core-collapse explosion, and the rarity of a bound binary system surviving a supernova explosion has been observationally confirmed by dedicated searches for the companion star of the compact remnant in young supernova remnants\cite{Kochanek2019, Kochanek2021} and search for companion stars of magnetars\cite{Chrimes2022}. This may be the reason why we did not observe similar properties of SN\,2022jli in other SESNe. 

\subsection{14. Possible explanations for the late-time rapid decline} 
One interesting phenomenon of SN\,2022jli is the rapid decline around 260 days after the discovery. One possible explanation is that the rapid decline happens when the accretion rate drops significantly after the bloated envelope of the companion star shrinks back (Method Section $\S~13$). The other possible explanations involve a dramatic change in the binary orbit, for example, the compact object plunges into the companion star \cite{Hirai2022}, which prevents further mass accretion. For the latter scenario to happen, the compact object is expected to penetrate the envelope of the companion star and lose significant angular momentum and mechanical energy to shrink the distance to the companion star at periapse. For an eccentric orbit, the periapsis distance is $r_{\mathrm{peri}} = a(1-e)$. The period of the binary system in SN\,2022jli, therefore the semimajor axis, did not show any significant evolution before the rapid decline (Method Section $\S~5$). The mechanical energy of a binary system is $E=-\frac{GM_1M_2}{2a}$ and the absolute value of the orbital momentum is $J_{\mathrm{orb}} = \mu \sqrt{G M a (1-e^2)}$ where $M=M_1+M_2$ and $\mu=\frac{M_1M_2}{M}$. If we simply assume an external force working on the compact object to slow the velocity instantaneously at the periapse, the mechanical energy of the binary system is more susceptible to the external force than the angular momentum. It is unlikely that the orbital eccentricity changes significantly while the semimajor axis remains constant. In the first explanation, the binary orbital parameter undergoes secular evolution, which barely changes in the comparatively short duration of the supernova. The accretion rate suffers a fast drop, but the periodicity of the orbital undulation shall persist if the compact object continues to accrete mass from the companion. We inspect the late-time evolution of the pseudo-bolometric light curve of SN\,2022jli to see if the 12.4-day periodicity continues. During the fast decline phase, there is a hint of a small bump in the light curve as indicated by the vertical dashed line in panel b of Extended Data Fig.~\ref{fig:Lbol_undulation}, which is consistent with 12.4-day separation to the peak of the last clear bump before the rapid decline. This might be a hint of evidence for the continuous accretion. The prominent accretion rate variation before the rapid-decline phase might change the disk properties significantly. We suspect this could trigger some disk instability\cite{Hameury2020}, which might be responsible for the enhanced luminosity before the rapid-decline phase, as reported in Method Section $\S~6$.  


\subsection{15. Fermi-LAT detection of $\gamma$-ray emission from the direction of SN\,2022jli}
\label{Fermi_LAT}

The Large Area Telescope (LAT) on the Fermi Gamma-Ray Space Telescope\cite{Atwood2009} has been surveying the entire sky since 2008. LAT has a large field of view of $\sim 60^\circ$, enabling it to scan the sky in about 3 hours.  We queried LAT data within a $10^\circ$ radius of SN\,2022jli with photon energies between 100 MeV to 300 GeV observed in the past 14.5 years between 2008 September 1 and 2023 March 1 using the Fermi Science Support Center data server\cite{fssc}.  We filtered the photons using source-type events ({\tt evclass=128}) with the most stringent cuts on the data quality ({\tt DATA\_QUAL==1 \&\& LAT\_CONFIG==1}), reconstructed both in the front and the back of the detector ({\tt evtype=3}) and with a maximum zenith angle of $90^{\circ}$.


\paragraph{New $\gamma$-ray source detection}
\label{sec:gtlike}

We used the standard binned likelihood analysis method {\tt gtlike} to analyze the data. The sources in the incremental Fermi Large Area Telescope Fourth Source Catalog (4FGL-DR3; \cite{Abdollahi2022}), together with the diffuse Galactic and isotropic backgrounds\cite{gammabkg} {\tt gll\_iem\_v07.fits} and {\tt iso\_P8R3\_SOURCE\_V3\_v1.txt}, are included in the model. We tried different binning strategies in time and energy to search for new $\gamma$-ray sources. A significant new source was detected in data observed after the supernova explosion from 2022 May 1 to 2023 March 1, as shown in the test statistic (TS) map in the top left panel of Extended Data Fig.~\ref{fig:fermi_TS_lc}. The energy light curve of the new source in the 100\,MeV\,--\,300\,GeV energy band is shown in the bottom left panel of Extended Data Fig.~\ref{fig:fermi_TS_lc}. The source was detected in two bins corresponding to September and October 2022 (TS=20.0) and November and December 2022 (TS=29.6). We got the spectral energy distribution of the new $\gamma$-ray source in three different time windows: September and October 2022 (top left panel of Supplementary Information Fig.~\ref{fig:fermi_sed}); November and December 2022 (top right panel of Supplementary Information Fig.~\ref{fig:fermi_sed}), and the above two combined (bottom panel of Supplementary Information Fig.~\ref{fig:fermi_sed}). The new source is most significantly detected in the 1\,--\,3\,GeV energy band. We also extracted the 1\,--\,3\,GeV light curve at the SN position, and the result is shown in the bottom right panel of Extended Data Fig.~\ref{fig:fermi_TS_lc}. 


We performed likelihood modeling with the 1\,--\,3\,GeV data observed from 2022 November 1 to 2023 January 1. The TS map is shown in the top right panel of Extended Data Fig.~\ref{fig:fermi_TS_lc}, and the localization of the new source from the above modeling is shown in Extended Data Fig.~\ref{fig:fermi_localization_periodicity}. The best localization of the new $\gamma$-ray source is RA$= 8.620(\pm 0.084)^\circ$, Dec$=-8.425(\pm0.077)^\circ$ with uncertainty given by the 68\% confidence interval.  The \fermilat source 4FGL J0035.8-0837 at RA=$8.958(\pm 0.060)^\circ$, Dec$=-8.632(\pm0.048)^\circ$ is $0.33^\circ$ East, $0.21^\circ$ South to the newly detected source. They are two different sources spatially separated from each other, but we note that they could contaminate each other when measuring their fluxes.

\paragraph{Poisson noise matched filter method}
\label{sec:matched_filter}
We also used the Poisson noise matched filter to corroborate the detection of the GeV photons associated with SN\,2022jli. Since we are dealing with the problem of detecting sources embedded in low-number-count Poisson noise,  the optimal matched filter presented in ref\cite{Ofek2018} was used. We built the Poisson-noise optimal filter as
\begin{equation}
    P_{\mathrm{poi}} = \mathrm{ln}\left(1+\frac{F}{B}P\right)\, ,
\end{equation}
where B is the expectancy for the background level, F is the unknown flux of the source we would like to detect, and P is the PSF of LAT (the observed photon distribution of a point source with LAT). We divided the 14.5-year data into two-month bins, and we then divided the photons into seven energy bands (0.1\,--\,0.3\,GeV, 0.3\,--\,1.0\,GeV, 1\,--\,3\,GeV, 3\,--\,10\,GeV, 10\,--\,30\,GeV, 30\,--\,100\,GeV, 100\,--\,300\,GeV) in each temporal bin. We generated count maps for each temporal and energy bin. Following the method in ref\cite{Ofek2018}, the log-likelihood difference image used to detect the source is 
\begin{equation}
    S=M\otimes \overleftarrow{P_{\mathrm{poi}}}
\end{equation}
where M denotes the measured data, i.e., the count map, $\otimes$ denotes convolution, $\leftarrow$ denotes coordinates reversal.
The PSF of \fermilat observation is a function of an incident photon's energy, the inclination angle, and the event class. In practice, we built the PSF by generating energy-dependent PSF data with {\tt gtpsf} for the two-month observation and then simply averaged the PSF of different energies. 

The result shows a significant source detection at the SN position in the 1\,--\,3\,GeV band in November and December 2022, as shown in Supplementary Information Fig.~\ref{fig:Smap2_E1000to3000}, which is consistent with the result obtained with {\tt gtlike}. The other \fermilat source near SN\,2022jli, 4FGL J0035.8-0837 has also been clearly revealed in the 1\,--\,3\,GeV band in March and April 2020 (Supplementary Information Fig.~\ref{fig:Smap2_E1000to3000}), and also in the 3\,--\,10\,GeV band in March and April 2020, and July and August 2020 (Supplementary Information Fig.~\ref{fig:Smap2_E3000to10000}). Emission of 4FGL J0035.8-0837 might also contribute to the photons seen in July and August 2013 (Supplementary Information Fig.~\ref{fig:Smap1_E3000to10000}).

\paragraph{Evidence for periodicity of the new $\gamma$-ray source}
We studied the temporal distribution of the photons associated with the new $\gamma$-ray source. There are eleven 1\,--\,3\,GeV photons within the half containment radius of the averaged PSF from 2022 September 1 to 2023 January 1. The estimated number of background photons within the above temporal and spatial range is two. As shown in Extended Data Fig.~\ref{fig:fermi_localization_periodicity}, there is a large avoidance region (50\%) in the phase space after folding the $\gamma$-ray photon light curve with a 12.4-day period, as shown by the shaded grey area. This hints at a correlation between the $\gamma$-ray photon arrival time and the optical light-curve bump phase. We performed a simple simulation by drawing N photons randomly distributed in four months, assuming there is no preference for when the photons arrive with respect to the bump phase. 98.9\% (98.0\%, 96.5\%) of the experiments result in the maximum separation of less than 0.5 for any two photons in the phase space for $N=11$ (N=10, N=9). We checked and found that the correlation of $\gamma$-ray photons with the 12.4-day period could not be caused by the telescope survey profile, as shown in Supplementary Information Fig.~\ref{fig:gamma_period}.

\paragraph{Association between SN\,2022jli and the new $\gamma$-ray source}

The new Fermi-LAT $\gamma$-ray source is positionally and temporally consistent with SN\,2022jli, which suggests that the new $\gamma$-ray source is associated with the supernova explosion. The correlation between the $\gamma$-ray photon detection time and the 12.4-day periodic undulation of the optical light curve provides further evidence that the new $\gamma$-ray source and SN\,2022jli share the same origin. 

Due to the potential periodicity detected in the new $\gamma$-ray source, the probability of it coming from some other contamination source, such as a blazar, within the localization area is low. Nevertheless, we inspected any potential sources within the 99\% localization confidence area which might emit $\gamma$-rays. First, the nucleus of NGC\,157 is a dormant supermassive black hole\cite{Dullo2020}, which does not belong to either flat spectrum radio quasars (FSRQs) or BL Lac objects (BL Lacs). The only blazar candidate within the 99\% localization confidence area is NVSS J003456-082820, as shown in Extended Data Fig.~\ref{fig:fermi_localization_periodicity}. NVSS J003456-082820 was listed as a blazar candidate in the Blazar Radio and Optical Survey (BROS \cite{Itoh2020}) catalog based on its spectral index between the radio bands of 0.15 GHz and 1.4 GHz and its compactness. We did not detect any optical flare from NVSS J003456-082820 in ZTF data but noted that around 20\% of the orphan $\gamma$-ray flares happen without an accompanying optical flare\cite{Liodakis2019}. The true nature of NVSS J003456-082820 requires further confirmation from multifrequency observations. We studied the expectancy of one BROS catalog object falling in the localization area of the new $\gamma$-ray source by calculating the chance coincidence probability of observing NVSS J003456-082820. We calculated $\mathrm{P_{ch} = 1- exp[-\pi(R_0^2+ 4\sigma_\gamma^2)\rho_A]}$ where $R_0$ is the angular distance between the new $\gamma$-ray source location and NVSS J003456-082820, and $\sigma_\gamma$ is the 68\% localization uncertainties radius of the new $\gamma$-ray source, and $\rho_A$ is the surface density of BROS sources. The result of $P_{ch}\simeq 0.5$ tells that having one BROS object in the observed field of the new $\gamma$-ray source is not surprising.

\paragraph{Origin of the SNe-associated $\gamma$-ray emission}
Several physical processes could generate cosmic $\gamma$ rays. These processes include radioactive decay, particle-particle collisions, acceleration of  charged particles (bremsstrahlung or synchrotron radiation), inverse Compton scattering, and matter-antimatter annihilation. One notable category of $\gamma$-ray sources is Gamma-ray Bursts (GRBs), lasting from ten milliseconds to several hours, which is different from the gamma-ray emission we observed in SN\,2022jli with a month time scale. Gamma-ray sources with radioactive decay origin are usually associated with nucleosynthesis. For example, $\gamma$-ray emissions from $^{56}$Co decay at 847 and 1248 keV have been detected in some very nearby supernovae such as SN\,1987A \cite{Matz1988, Teegarden1989} and SN\,2014J \cite{Churazov2015}. The GeV emission in SN\,2022jli apparently has different origins.

Significant efforts have been devoted to searching for other types of $\gamma$-ray emission associated with supernova explosions\cite{Ackermann2015, Renault-Tinacci2018, Yuan2018, Prokhorov2021, Acharyya2023}. These efforts are motivated by different mechanisms that could happen in the supernova environment producing $\gamma$-ray emissions, for example, proton-proton collisions in supernova ejecta and CSM interaction \cite{Ackermann2015} and inverse Compton scattering in the young neutron-star wind nebula\cite{Renault-Tinacci2018, Acharyya2023}. However, the detected $\gamma$-ray sources associated with supernovae are limited.  The only tentative detections of such objects are reported for the Type II supernova iPTF14hls \cite{Yuan2018} and Type IIP SN\,2004dj \cite{Xi2020}. This is the first time significant GeV $\gamma$-ray emission has been detected from a SESN. We have shown that ECI or a magnetar cannot explain the observed optical light curves of SN\,2022jli, which disfavors $\gamma$-ray generation from such processes. The accretion of hydrogen-rich mass from the companion star is expected to happen in the center of the supernova ejecta. Potential $\gamma$-ray emission in accretion-powered supernovae has barely been discussed in the literature, but $\gamma$-ray emission generated in accretion-related processes in microquasars has been commonly observed, for example, in Cygnus X-3 \cite{Abdo2009, Piano2012}. The ultrahigh accretion rate required to power SN\,2022jli is susceptible to launching extreme outflow winds or an accretion jet. We speculate that the observed $\gamma$-ray photons in SN\,2022jli could be produced by such a wind or a jet analogous to the observed $\gamma$-ray emission in some X-ray binaries\cite{Zanin2016}.


\end{methods}

\clearpage

\subsubsection*{Data availability}
Photometry and spectroscopy of SN\,2022jli will be made available via the WISeREP public database. Facilities that make all their data available in public archives promptly or after a proprietary period include Palomar 48-inch/ZTF,  VLT/X-Shooter, NuSTAR, Chandra X-ray observatory, and Fermi Gamma-ray Space Telescope. Data from ATLAS, Gaia, ASAS-SN, and KKO were obtained from public sources. 

\subsubsection*{Code availability}

\noindent
The code and data used to perform the analysis and produce the figures for this paper are available in a public GitHub repository (\url{https://github.com/AtomyChan/SN2022jli}).

\begin{addendum}

\item 
We thank Brad Cenko, Yuri Levin, Elena Pian, Subo Dong, Dong Lai, Ofer Yaron, Jonathan Morag, Yahel Sofer Rimalt, and Tomer Shenar for the valuable discussions. We thank David J. Thompson for his useful comments. P.C. and A.G.-Y. thank Yuri Beletsky for his assistance with Magellan telescope remote observations. 

This research has made use of the NuSTAR Data Analysis Software (NuSTARDAS) jointly developed by the ASI Space Science Data Center (SSDC, Italy) and the California Institute of Technology (Caltech, USA). 

SED Machine is based upon work supported by the National Science Foundation under Grant No. 1106171. 

Based on observations obtained with the Samuel Oschin Telescope 48-inch and the 60-inch Telescope at the Palomar Observatory as part of the Zwicky Transient Facility project. ZTF is supported by the National Science Foundation under Grant No. AST-2034437 and a collaboration including Caltech, IPAC, the Weizmann Institute of Science, the Oskar Klein Center at Stockholm University, the University of Maryland, Deutsches Elektronen-Synchrotron and Humboldt University, the TANGO Consortium of Taiwan, the University of Wisconsin at Milwaukee, Trinity College Dublin, Lawrence Livermore National Laboratories, IN2P3, University of Warwick, Ruhr University Bochum and Northwestern University. Operations are conducted by COO, IPAC, and UW. 

The Gordon and Betty Moore Foundation, through both the Data-Driven Investigator Program and a dedicated grant, provided critical funding for SkyPortal. 

This work has used data from the Asteroid Terrestrial-impact Last Alert System (ATLAS) project. The Asteroid Terrestrial-impact Last Alert System (ATLAS) project is primarily funded to search for near-earth asteroids through NASA grants NN12AR55G, 80NSSC18K0284, and 80NSSC18K1575; byproducts of the NEO search include images and catalogs from the survey area. This work was partially funded by Kepler/K2 grant J1944/80NSSC19K0112 and HST GO-15889, and STFC grants ST/T000198/1 and ST/S006109/1. The ATLAS science products have been made possible through the contributions of the University of Hawaii Institute for Astronomy, the Queen’s University Belfast, the Space Telescope Science Institute, the South African Astronomical Observatory, and The Millennium Institute of Astrophysics (MAS), Chile. 

Based on observations made with the Nordic Optical Telescope, owned in collaboration by the University of Turku and Aarhus University, and operated jointly by Aarhus University, the University of Turku, and the University of Oslo, representing Denmark, Finland and Norway, the University of Iceland and Stockholm University at the Observatorio del Roque de los Muchachos, La Palma, Spain, of the Instituto de Astrofisica de Canarias. One of the spectra from NOT was obtained as part of an NBI school where Meghana Killi, Natalie Allen, Kate Gould, Dazhi Zhou participated under the leadership of Johan Fynbo.
This work has been supported by the research project grant “Understanding the Dynamic Universe” funded by the Knut and Alice Wallenberg Foundation under Dnr KAW 2018.0067. 

Based on observations collected at the European Organisation for Astronomical Research in the Southern Hemisphere under ESO programme(s) 110.25A6. 

MMT Observatory access for part of the MMT/BINOSPEC data was supported by Northwestern University and the Center for Interdisciplinary Exploration and Research in Astrophysics (CIERA).

Mephisto is developed at and operated by the South-Western Institute for Astronomy Research of Yunnan University (SWIFAR-YNU), funded by the ``Yunnan University Development Plan for World-Class University” and ``Yunnan University Development Plan for World-Class Astronomy Discipline”. 

S. Schulze acknowledges support from the G.R.E.A.T. research environment, funded by {\em Vetenskapsr\aa det},  the Swedish Research Council, project number 2016-06012.

BZ is supported by a research grant from the Willner Family Leadership Institute for the Weizmann Institute of Science, a research grant from the Center for New Scientists at the Weizmann Institute of Science and a research grant from the Ruth and Herman Albert Scholarship Program for New Scientists.

A.H. is grateful for the support by the I-Core Program of the Planning and Budgeting Committee and the Israel Science Foundation, and support by ISF grant 647/18. This research was supported by Grant No. 2018154 from the United States-Israel Binational Science Foundation (BSF). A.H. is especially grateful to the Sir Zelman Cowen Academic Initiatives for their generous funding and support. 

D.Z.L., X.K.L,  Y.F., and X.W.L. acknowledge the support from special grants for Yunnan technology leading talents and Provincial Innovation Team. D.Z.L., X.K.L,  Y.F., and X.W.L. also acknowledge supports from the ``Science \& Technology Champion Project” (202005AB160002) and from two “Team Projects” - the ``Innovation Team” (202105AE160021) and the ``Top Team” (202305AT350002), all funded by the "Yunnan Revitalization Talent Support Program"

The Australia Telescope Compact Array is part of the \href{https://ror.org/05qajvd42}{ATNF} which is funded by the Australian Government for operation as a National Facility managed by CSIRO. We acknowledge the Gomeroi people as the Traditional Owners of the Observatory site.

The Cosmic Dawn Center (DAWN) is funded by the Danish National Research Foundation under grant No. 140.  JPUF is supported by the Independent Research Fund Denmark (DFF--4090-00079) and thanks the Carlsberg Foundation for support.


\item[Author Contributions] 
All authors reviewed the manuscript and contributed to the interpretation and/or data acquisition. 
P.C. led the studies. P.C. discovered the periodicity and the $\gamma$-ray detection, organized the follow-up observations, conducted X-ray and most optical data acquisition, conducted most of the data reduction, performed the data analysis, and wrote most of the manuscript, including making all the figures.
A.G.-Y. contributed significantly to the follow-up data (including observations performed with MMT/BINOSPEC, Magellan/FIRE, Magellan/FIRE, VLT/X-Shooter, NuSTAR, Chandra, and ATCA) and to the interpretation, wrote on the manuscript, and thoroughly reviewed the paper.  
J.S. initiated the study, obtained spectra with SEDM and NOT, helped develop the ideas in the paper, wrote on the manuscript, and thoroughly reviewed the paper.
S.S. contributed to follow-up observations with NOT and reduced the NOT spectra.
R.S.P. performed follow-up observations with the 0.8m RC32 telescope at Post Observatory.
C.L. contributed to one MMT/BINOSPEC spectrum, reduced Magellan/FIRE echelle spectra, and contributed to the interpretation. 
E.O.O. contributed to follow-up observations with NuSTAR.
B.K., D.K., E.O.O., E.W., and B.Z. contributed to the interpretation and especially suggested the analysis of Fermi-LAT data.
D.Z.L. performed Mephisto photometry. X.K.L. organized Mephisto observations. 
A.H. and K.R. performed ATCA radio observation and data reduction. 
K.K.D., Y.Y., M.M.K, and S.R.K contributed to P200/DBSP spectra. 
C.F. contributed to SEDM photometry. 
A.A.M. contributed to one MMT/BINOSPEC spectrum. 
S.Y. contributed to the analysis of light curves with the nickel decay model. 
JPUF obtained a spectrum from the NOT. 
S.R.K., A.J.D., J.D.N., E.C.B., S.L.G., A.W., T.J.L., R.L.R, G.H., and R.D. are ZTF builders.  
D.Z.L., X.W.L., X.K.L., and Y.F. contributed to the observation and reduction of Mephisto data. 
I.I., A.S., N.L.S., P.A.M., and L.Y. contributed to the interpretation.
Y.J.Q contributed to the observation of DBSP spectra.

%

 \item[Competing Interests] The authors declare no competing financial interests.

\item[Correspondence] Correspondence and requests for materials should be addressed to the Ping~Chen (E-mail: chen.ping@weizmann.ac.il)\\

\end{addendum}

\clearpage

\begin{extended_data}

\setcounter{figure}{0}
\setcounter{table}{0}
\captionsetup[table]{name={\bf Extended Data Table}}
\captionsetup[figure]{name={\bf Extended Data Figure}}

\begin{figure}
\centering
\includegraphics[width=\columnwidth]{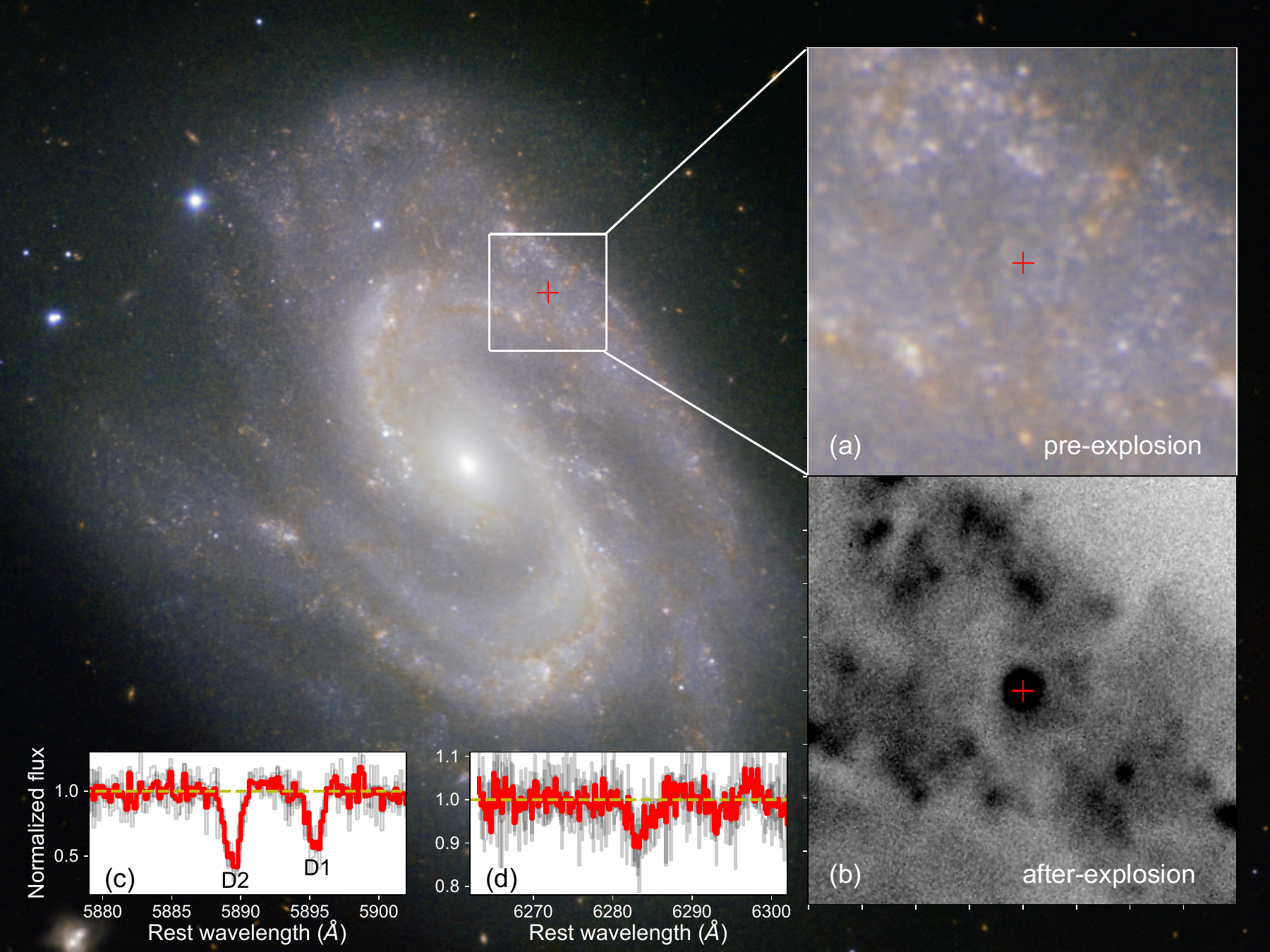}
\caption{SN\,2022jli and the host galaxy NGC 157. The background image shows the 3-color ($Y, H, K$ bands) image of NGC 157 taken before the supernova explosion with the HAWK-I instrument on ESO's Very Large Telescope (VLT) at the Paranal Observatory in Chile (Credit: ESO). The red plus symbol in the figure indicates the position of SN\,2022jli. The inset panels: (a) zoom-in view around SN\,2022jli showing the nearby environment in NIR; (b) zoom-in view around SN\,2022jli on an $r$-band image taken with Magellan/IMACS on 2022 December 15; (c) \ionp{Na}{i} absorption lines from the host galaxy; (d) the narrow diffuse interstellar band absorption (DIB6283) from the host galaxy. Panel (a) and (b) share the same field of view size. In panels (c) and (d), the red spectrum shows the averaged spectrum of three X-Shooter spectra and one IMACS spectrum, whereas the individual spectra are shown in grey in the background.}
\label{fig:sn2022jli_ngc157}
\end{figure}

\begin{figure}
\centering
\includegraphics[width=0.9\textwidth]{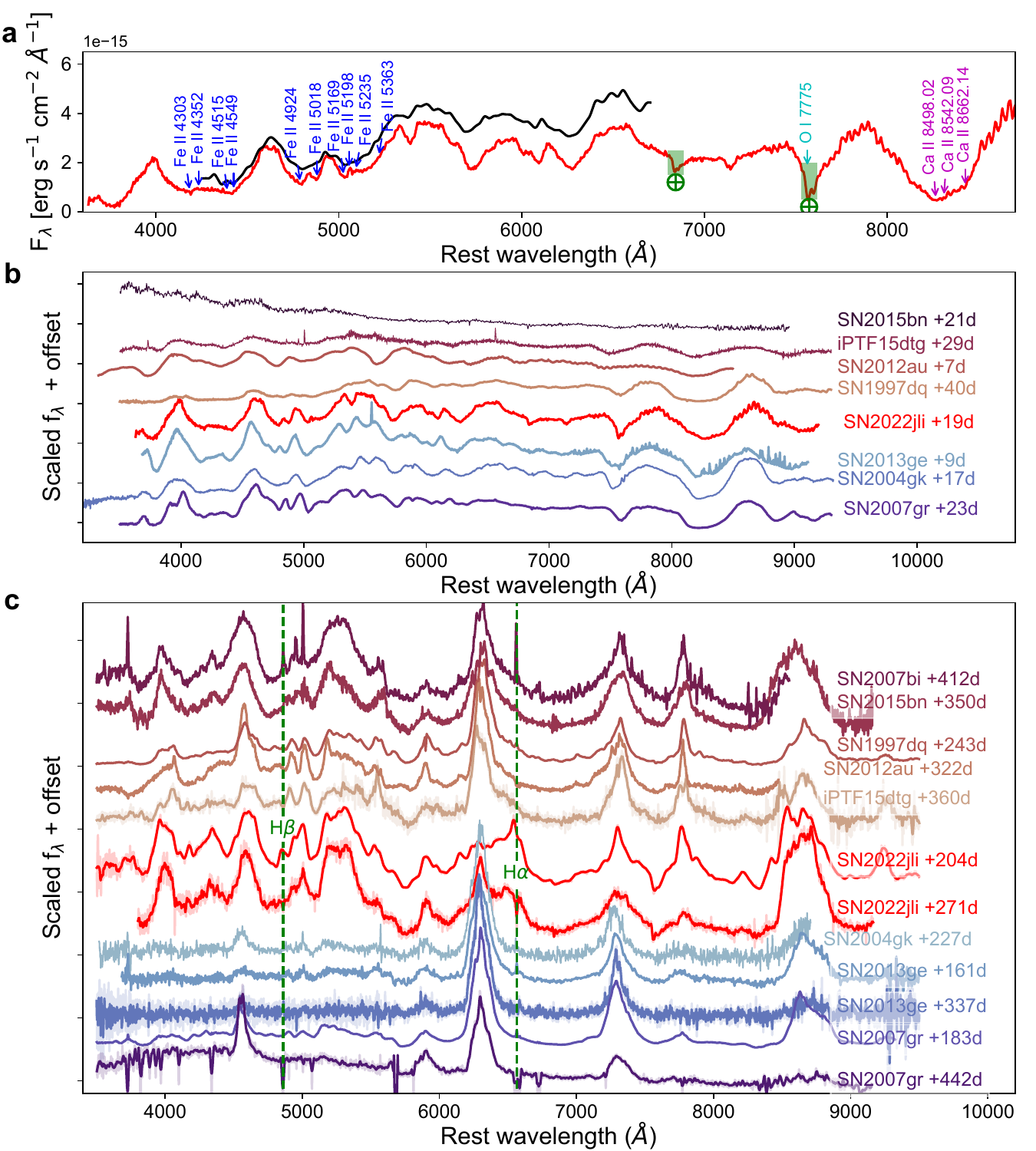}
\caption{Spectral evolution and comparison of SN\,2022jli. {\bf (a)} Photospheric spectra of SN\,2022jli. The +6d spectrum and the +19d spectrum are shown in black and red, respectively. The identified absorption lines of \ionp{Fe}{ii}, \ionp{O}{i}, and \ionp{Ca}{ii} are indicated with the arrows. A blue-shifted velocity of $8,200\, \mathrm{km\,s^{-1}}$ has been applied to all the lines. {\bf (b)} The +19d spectrum of SN 2022jli compared with photospheric spectra of other SNe. The comparison objects include: normal SNe Ic (SN\,2004gk\cite{Modjaz2014}, SN\,2007gr\cite{Hunter2009}), SN Ib/c (SN\,2013ge\cite{Drout2016}), ``Hypernovae'' (SN\,1997dq \cite{Matheson2001, Taubenberger2009, Mazzali2004}), Hydrogen-poor SLSNe (SN\,2015bn\cite{Nicholl2016, Jerkstrand2017}), and other long-lasting peculiar SESNe (SN\,2012au\cite{Milisavljevic2013}, iPTF15dtg \cite{Taddia2019}).   {\bf (c)} Late-time spectra of SN 2022jli and other comparison SNe. The comparison objects are the same as in panel (b) except that we have added another Hydrogen-poor SLSN SN\,2007bi \cite{Gal-Yam2009}. The vertical dashed lines mark the wavelength of H$\alpha$ and H$\beta$. }
\label{fig:spec_evolution_compare}
\end{figure}

\begin{figure}
\centering
\includegraphics[width=0.8\textwidth]{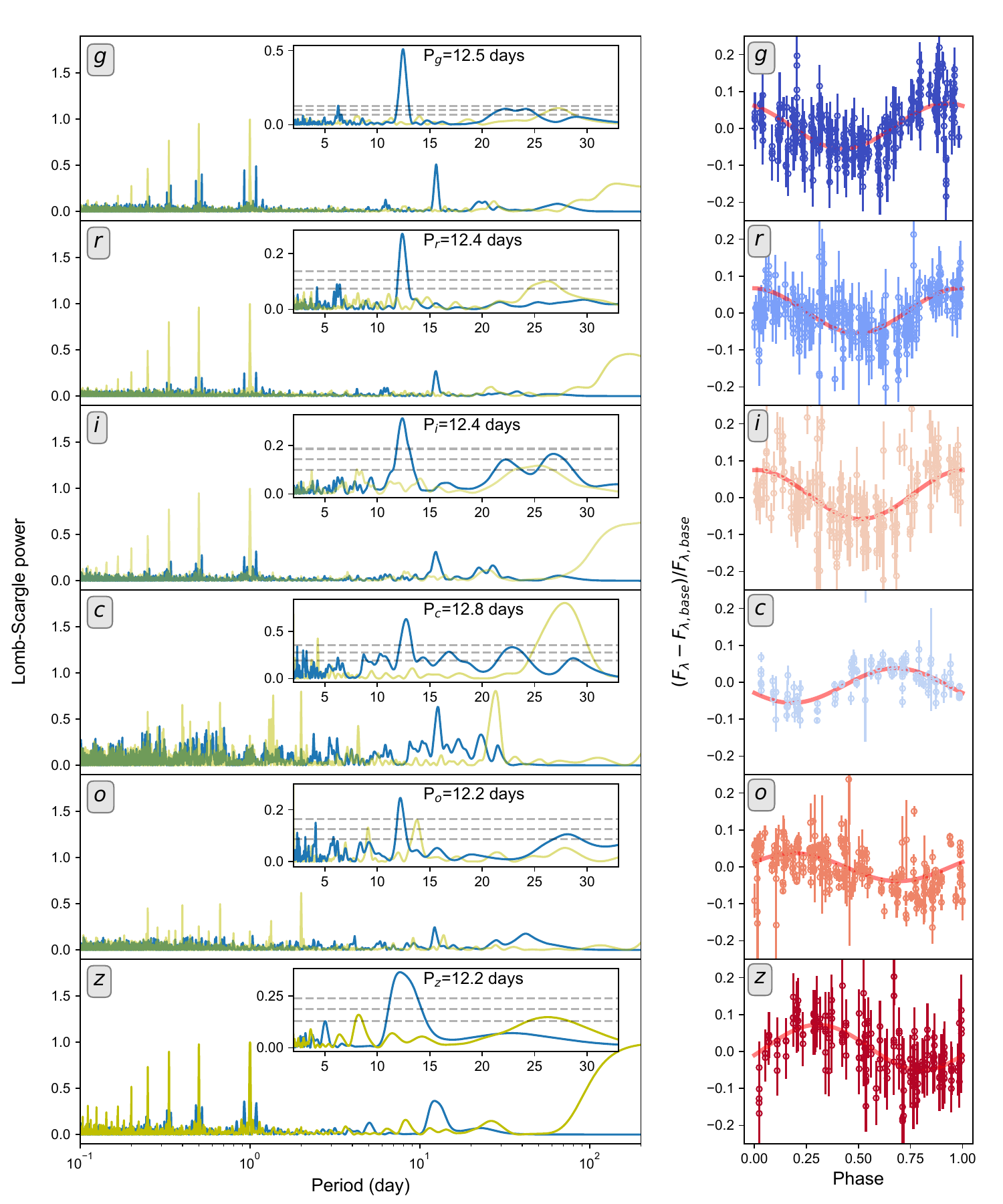} \\
\caption{Periodicity analysis of light curves in the individual bands. Left: the Lomb-Scargle power spectrum of the light curve (blue) and the observation window function (yellow). The zoom-in panel shows the region of interest around the true periodic signal. The horizontal lines indicate the false alarm probability (FAP) levels of $10^{-3}$, $10^{-6}$, and $10^{-9}$ from top to bottom. Right: the phase-folded subtracted and normalized light curve. The period adopted for the folding corresponds to the peak power in the zoom-in panel on the left.}
\label{fig:period_v1}
\end{figure}

\begin{figure}
\centering
\includegraphics[width=\textwidth]{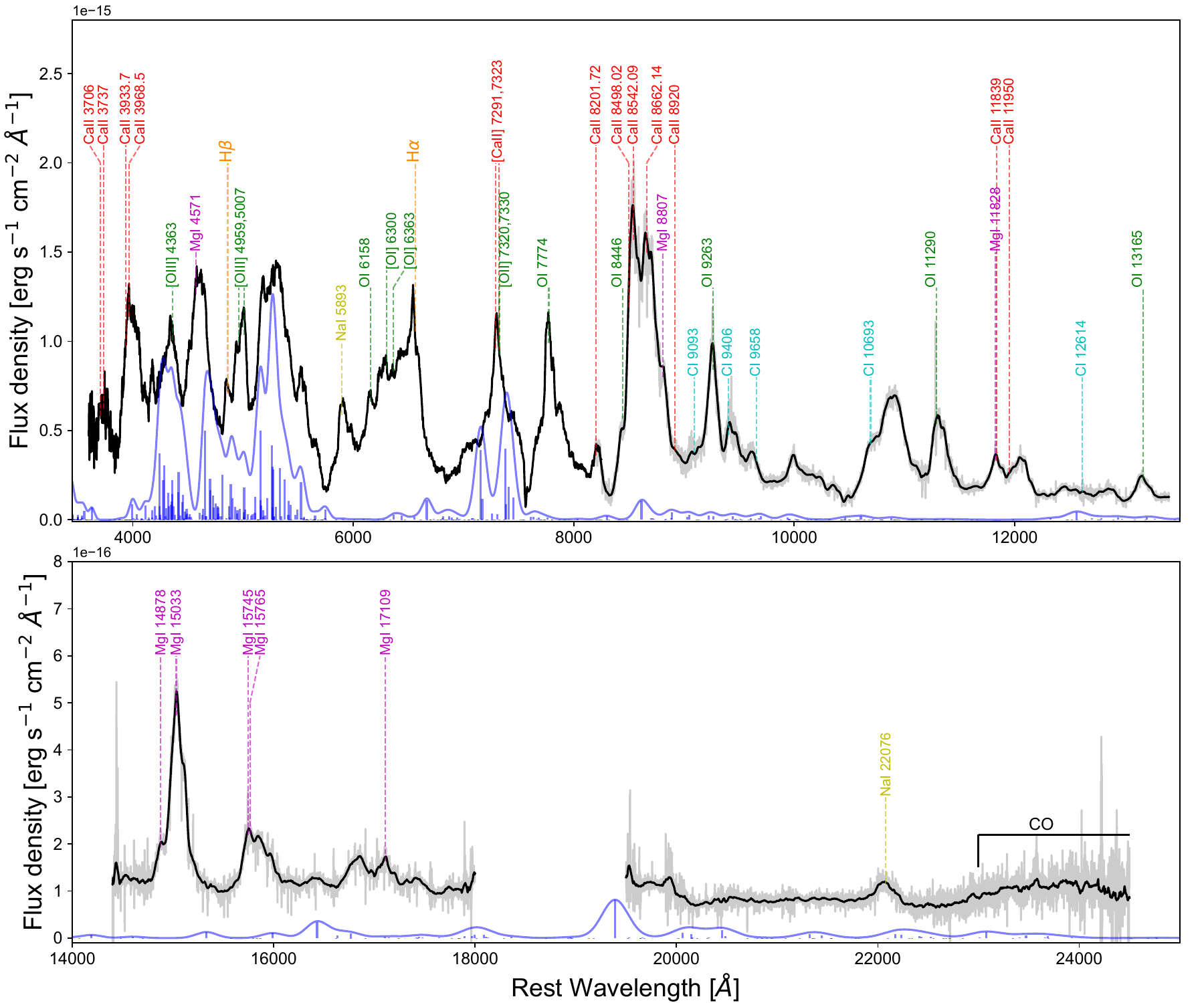}
\caption{Optical and  NIR spectra of SN\,2022jli around +210 days after discovery (optical spectrum taken with NOT/ALFOSC on 2022 November 25 and NIR spectrum taken with Magellan/FIRE on 2022 December 15). The NIR spectrum has been scaled to match the flux of the optical spectrum in the overlapping region. The prominent features and the corresponding ions (or molecules) that likely contributed to the emission lines are marked. The spectrum in blue shows the mock spectrum of \ionp{Fe}{ii}, \ionp{Fe}{iii} and \ionp{Ni}{ii} emission, and the vertical lines indicate emission from individual transitions which give relatively strong emission.  The absolute strength of the mock spectrum is arbitrary, and the relative strength between different transitions for the same ion was calculated for a temperature of $10^4$\,K, and an electron density of $10^7\mathrm{cm}^{-3}$.}
\label{fig:spec_lineident}
\end{figure}

\begin{figure}
\centering
\includegraphics[width=\textwidth]{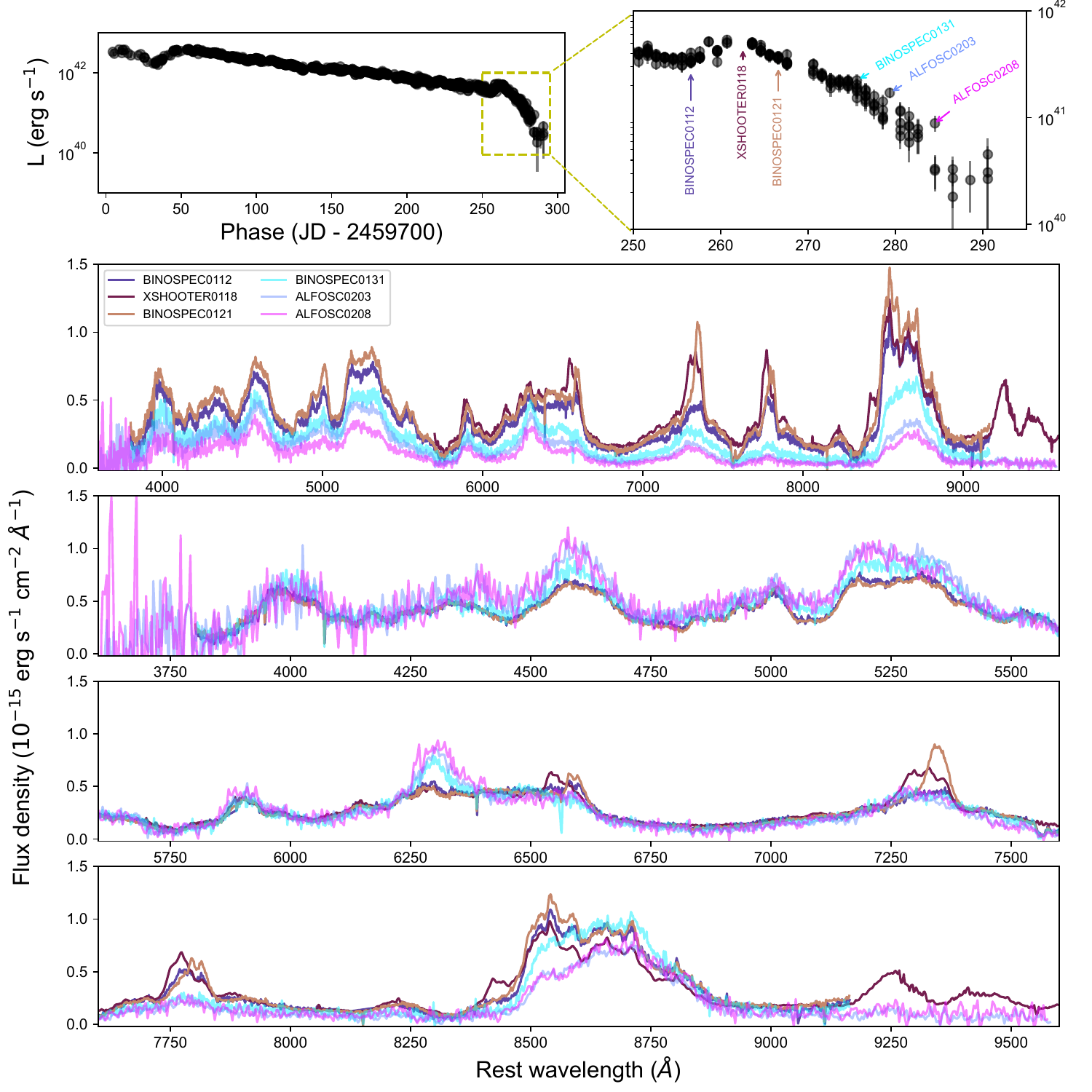}
\caption{Spectral evolution of SN\,2022jli before and during the fast decline phase of the light curve. The top left panel shows the bolometric light curve, and the top right panel shows the zoom-in view of the fast decline phase of the light curve. The second panel shows the six epochs of spectra as indicated in the top right panel. The bottom three panels show the zoom-in view of the spectra within different wavelength ranges. All spectra in the bottom three panels have been scaled to have the same integrated luminosity between 3800 to 9000 \AA\, as the BINOSPEC spectrum taken on 2022 January 12. }
\label{fig:spec_fast_decline}
\end{figure}

\begin{figure}
\centering
\includegraphics[width=\textwidth]{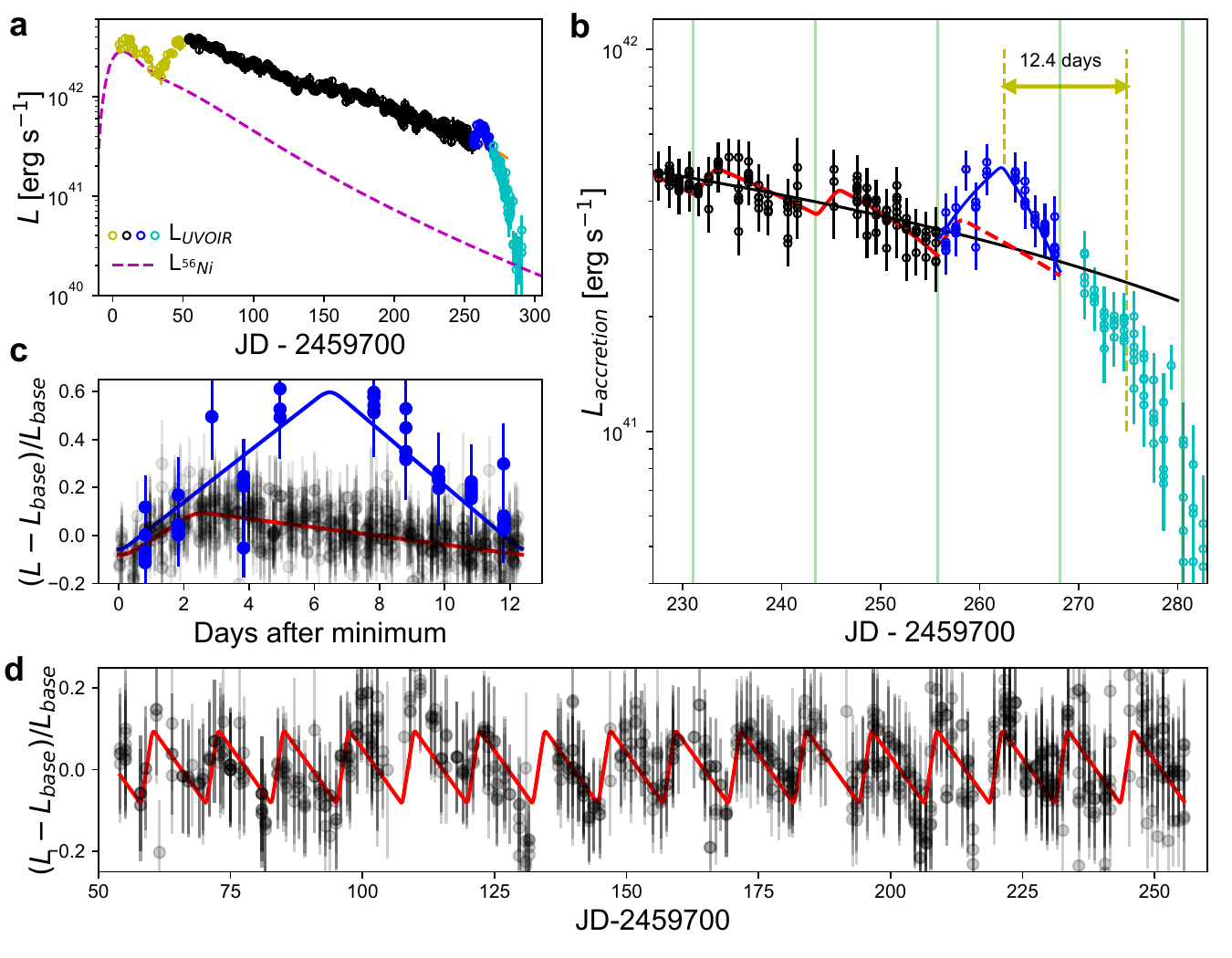}
\caption{Pseudo-bolometric light curve of SN\,2022jli. {\bf (a)} The points show the pseudo-bolometric light curve from 3750 \AA\, to 25000 \AA. These data points share the same colors with the other panels to indicate different phases. The black shows the gradual decline phase during which the constant relative undulation is detected. The last bump before the fast-declining phase is shown in blue, and the fast-declining phase is shown in cyan. The magenta line shows the radioactive decay model with 0.15 M$_\odot$ $^{56}$Ni.  {\bf (b)} Zoom-in view of the accretion-powered pseudo-bolometric light curve, $L_{accretion} = L_{UVOIR} - L_{^{56}Ni}$, before and during the fast-declining phase. The black solid line shows the linear fit to data between 200 and 260 days after JD$=$2,459,700. The red solid line shows the best-fit undulation model to the data, while the red dashed line shows the extrapolation of the undulation model if the SN follows the previous undulations. The vertical green lines mark the 12.4-day periods with the left three lines at the minima of the undulation profiles. {\bf (c)} The undulation profiles of the gradual-declining phase and the last bump. {\bf (d)} The relative undulation of the accretion-powered pseudo-bolometric light curve. The red line shows the empirical undulation model adopting the same empirical undulation profile as in panel (c). }
\label{fig:Lbol_undulation}
\end{figure}

\begin{figure}
\centering
\includegraphics[width=\textwidth]{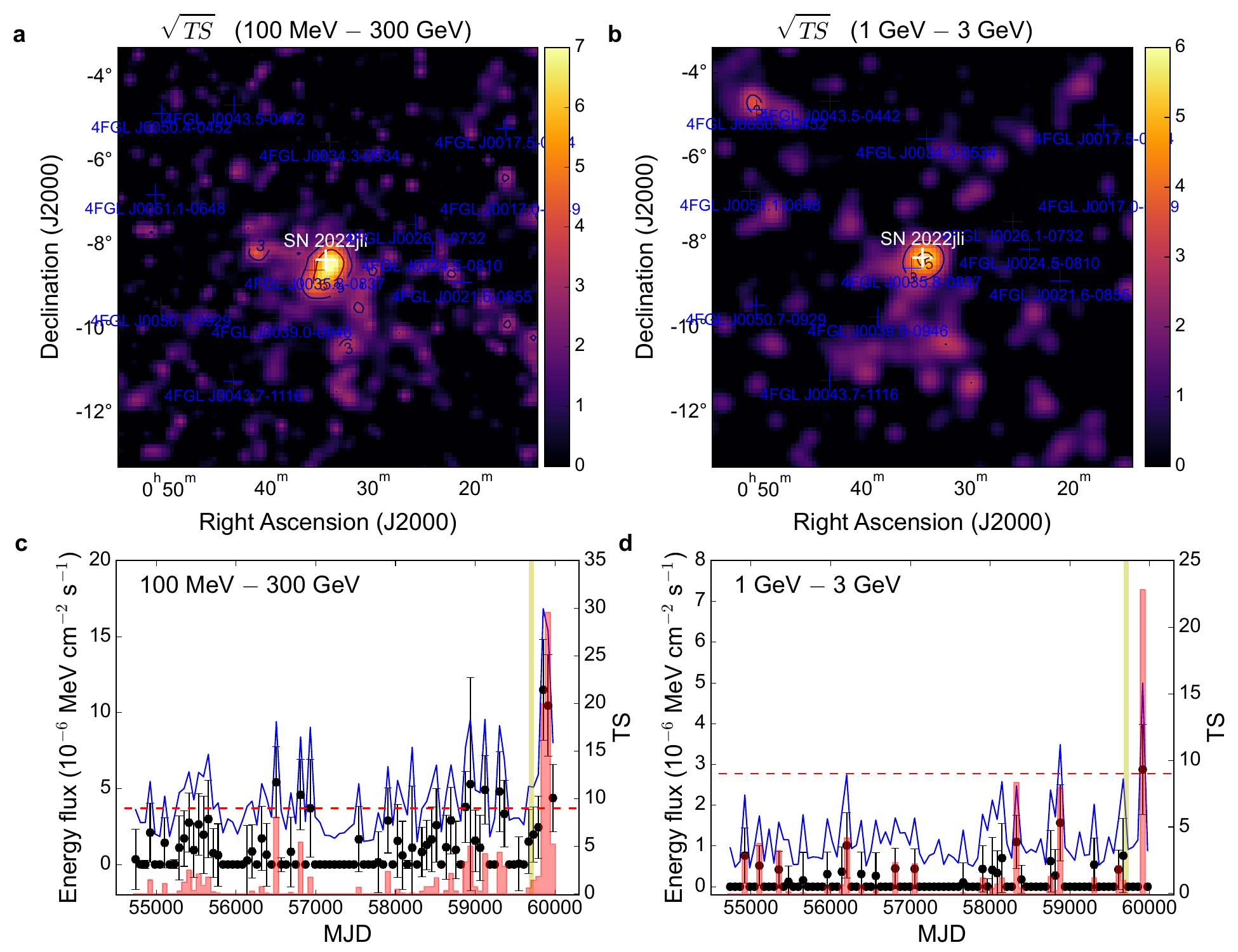}
\caption{Detection and light curve of the new $\gamma$-ray source. The top panels show the Test Statistic (TS) map of the region of interest in the direction of SN\,2022jli. {\bf (a)} The result of data observed from 2022 May 1 to 2023 March 1 in the broad energy band (100\,MeV\,--\,300\,GeV);  {\bf (b)} The result of data observed from 2022 November 1 to 2023 January 1 in the narrow energy band (1\,--\,3\,GeV). The sources from the 4FGL-DR3 catalog shown with blue plus symbols have been modeled and subtracted from the map. The bottom panels show the $\gamma$-ray light curves of the detected source at the position of SN\,2022jli with a bin size of 2 months. {\bf (c)} The energy range of 100 MeV to 300 GeV;  {\bf (d)} The energy range of 1 GeV to 3 GeV. The black points show the measured energy flux from the likelihood modeling with the Fermi-LAT analysis tool. The blue lines give the upper limit of energy flux within a 95\% confidence interval. The red histograms show the Test Statistics values on the right axis. The horizontal dashed red line marks TS=9. The vertical yellow lines mark the discovery time of SN\,2022jli. }
\label{fig:fermi_TS_lc}
\end{figure}

\begin{figure}
\centering
\includegraphics[width=0.7\columnwidth]{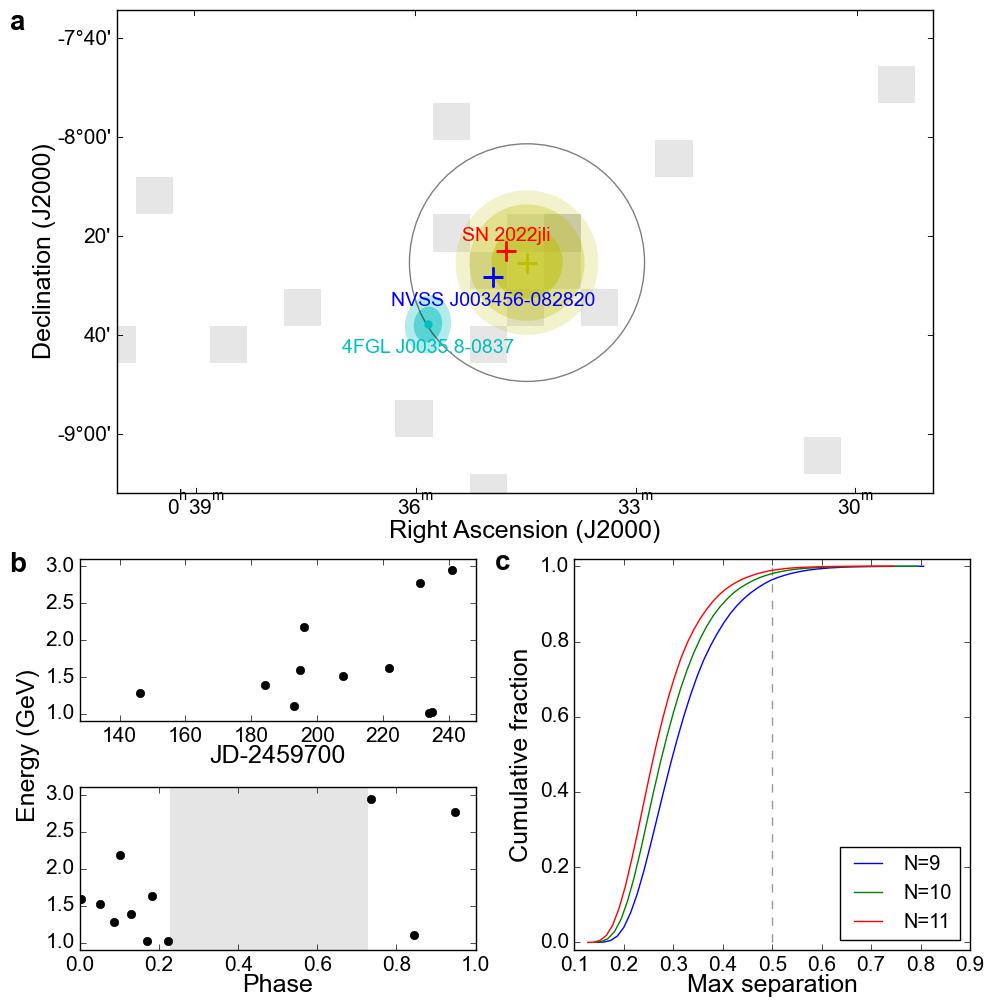}
\caption{ Localization and potential periodicity of the new $\gamma$-ray source. {\bf (a)} The background mosaic gray pixels show the count map of 1\,--\,3 GeV photons detected between 2022 November 1 and 2023 January 1. The pixel size is $0.125^\circ \times 0.125^\circ$. The yellow plus symbol shows the best-localized position of the new $\gamma$-ray source, and the surrounding yellow contours show the corresponding 68\%, 95\%, and 99\% confidence area. SN\,2022jli, the red plus symbol, is within the 68\% uncertainty region of the detected $\gamma$-ray source. The blazar candidate NVSS J003456-082820 is shown with the blue plus symbol and is within the 95\% uncertainty region of the new $\gamma$-ray source. In the central $1.625^\circ \times 1.625^\circ$ field, there is one detected $\gamma$-ray source from the LAT 12-year Source Catalog (4FGL-DR3), 4FGL J0035.8-0837.  The black circle has a radius of 0.4$^\circ$ corresponding roughly to the 50\% containment radius of the averaged PSF over the energy range between 1 GeV and 3 GeV.  {\bf (b)}: The distribution of 1\,--\,3\,GeV photons of the new $\gamma$-ray source. The top panel shows the photon energy and detection time. The bottom panel shows the distribution after folding the light curve with a period of 12.4 days. The reference time (phase=0) corresponds to the minimum of the optical undulation profile. Most of the $\gamma$-ray photons come from the rising phase of the optical bump. The 11 photons are within the half-containment radius shown in panel (a). {\bf (c)}: The cumulative distribution of the maximum separation between any two photons that would be achieved by drawing N photons randomly distributed in a time range of 120 days.}
\label{fig:fermi_localization_periodicity}
\end{figure}

\begin{figure}
\centering
\includegraphics[width=0.8\textwidth]{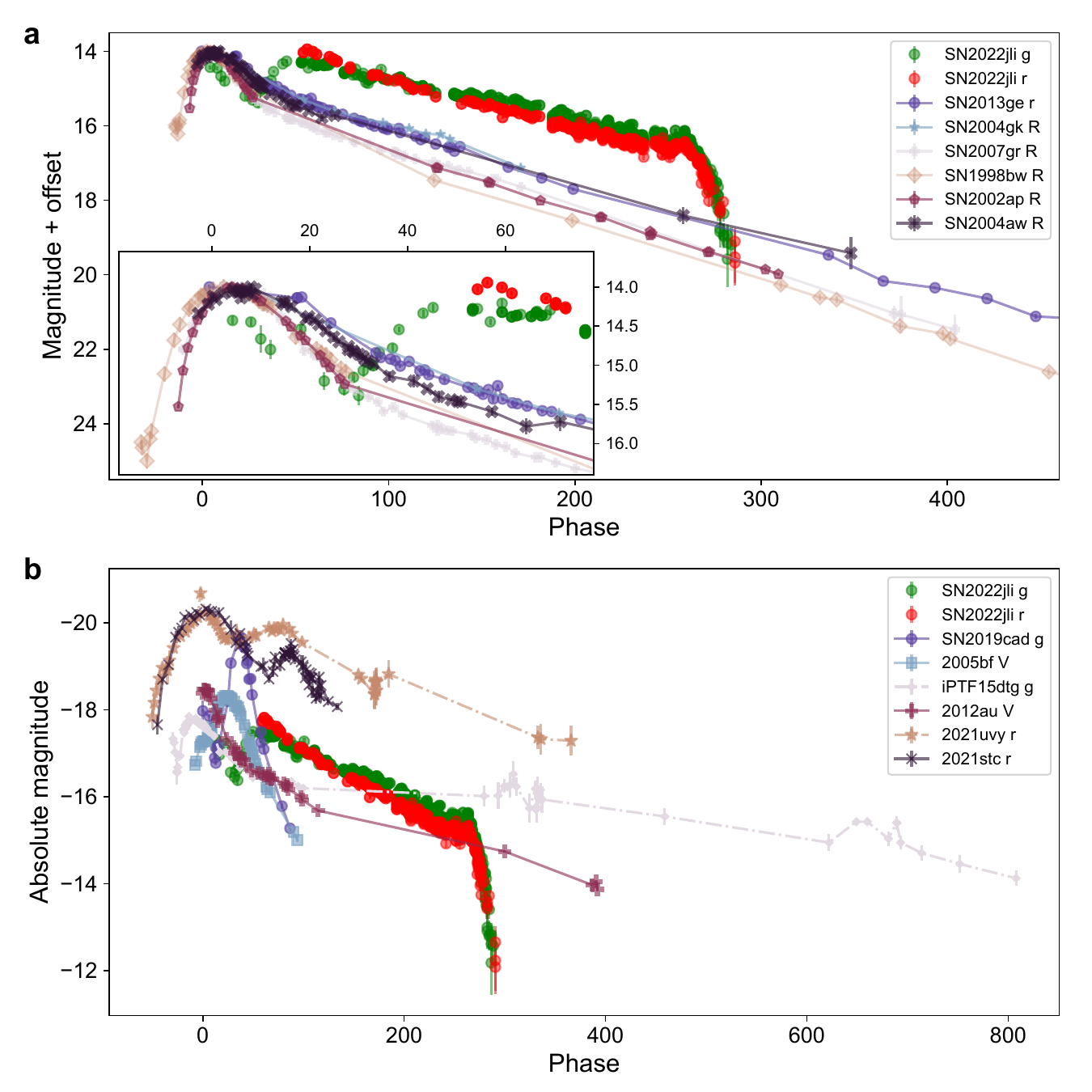}
\caption{Light curve of SN 2022jli compared with those of other supernovae. {\bf (a)} Comparison with Type Ic supernovae dominantly powered by radioactive decay showing clear exponential decay tails. All the comparison supernovae have been shifted to have a peak at 14th magnitude. The inset panel shows a zoom-in view around the peak light. {\bf (b)} Comparison with the long-lasting SN 2012au (SN Ib), and other supernovae with double-peaked light curves.}
\label{fig:lc_compare}
\end{figure}

\begin{figure}
\centering
\includegraphics[width=\textwidth]{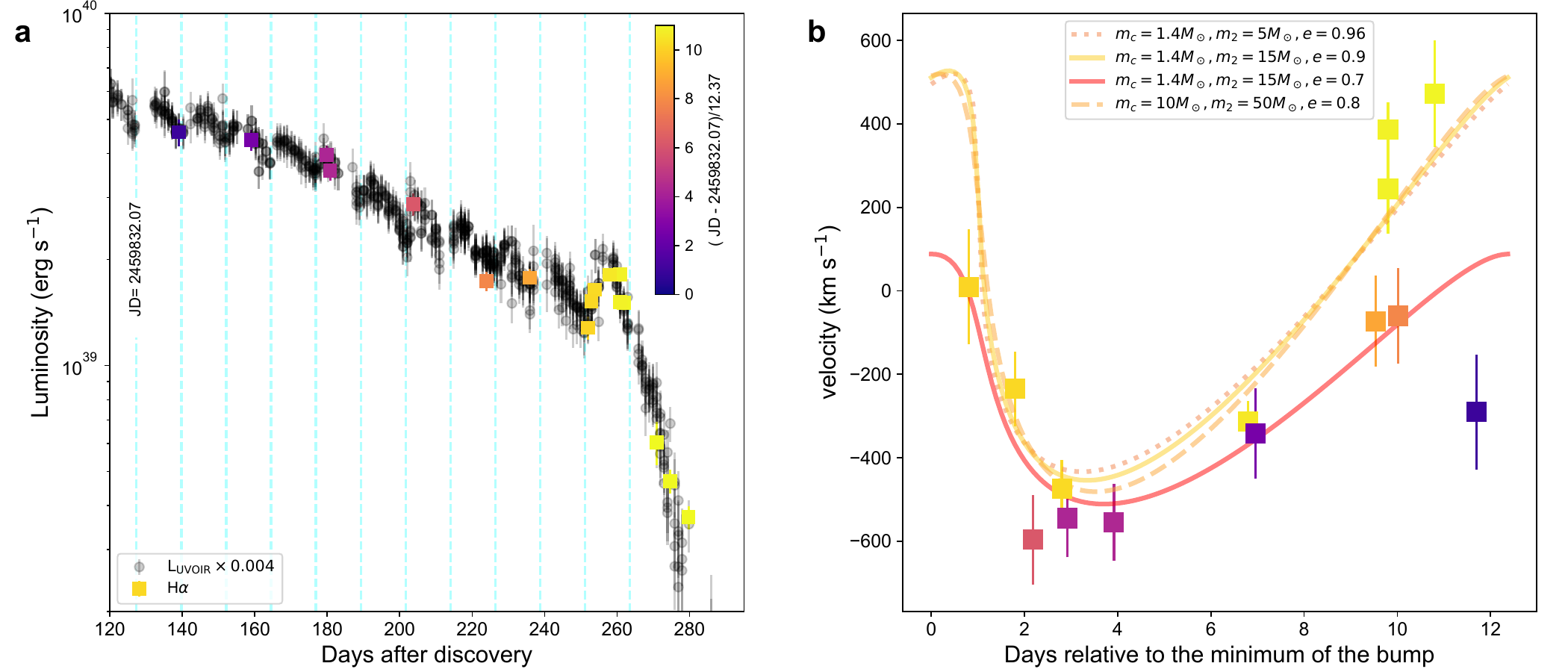}
\caption{Evolution of the accretion-powered H$\alpha$ emission. {\bf (a)} The line luminosity of the H$\alpha$ emission compared with the pseudo-bolometric luminosity. {\bf (b)} The velocity of the H$\alpha$ emission. The data points share the same color as in panel (a), indicating the phase of the corresponding spectrum. The lines show the orbital velocity model with specific orbital parameters in the legend. The orbital parameters include compact remnant mass $m_c$, companion star mass $m_2$, and orbital eccentricity $e$. The errorbars are $1\sigma$ confidence intervals.} 
\label{fig:Ha_luminosity_velocity}
\end{figure}

\end{extended_data}

\clearpage

\begin{supplement}

\setcounter{figure}{0}
\setcounter{table}{0}
\captionsetup[table]{name={\bf Supplementary Information Table}}
\captionsetup[figure]{name={\bf Supplementary Information Figure}}

\begin{figure}[h]
\centering
\includegraphics[width=\textwidth]{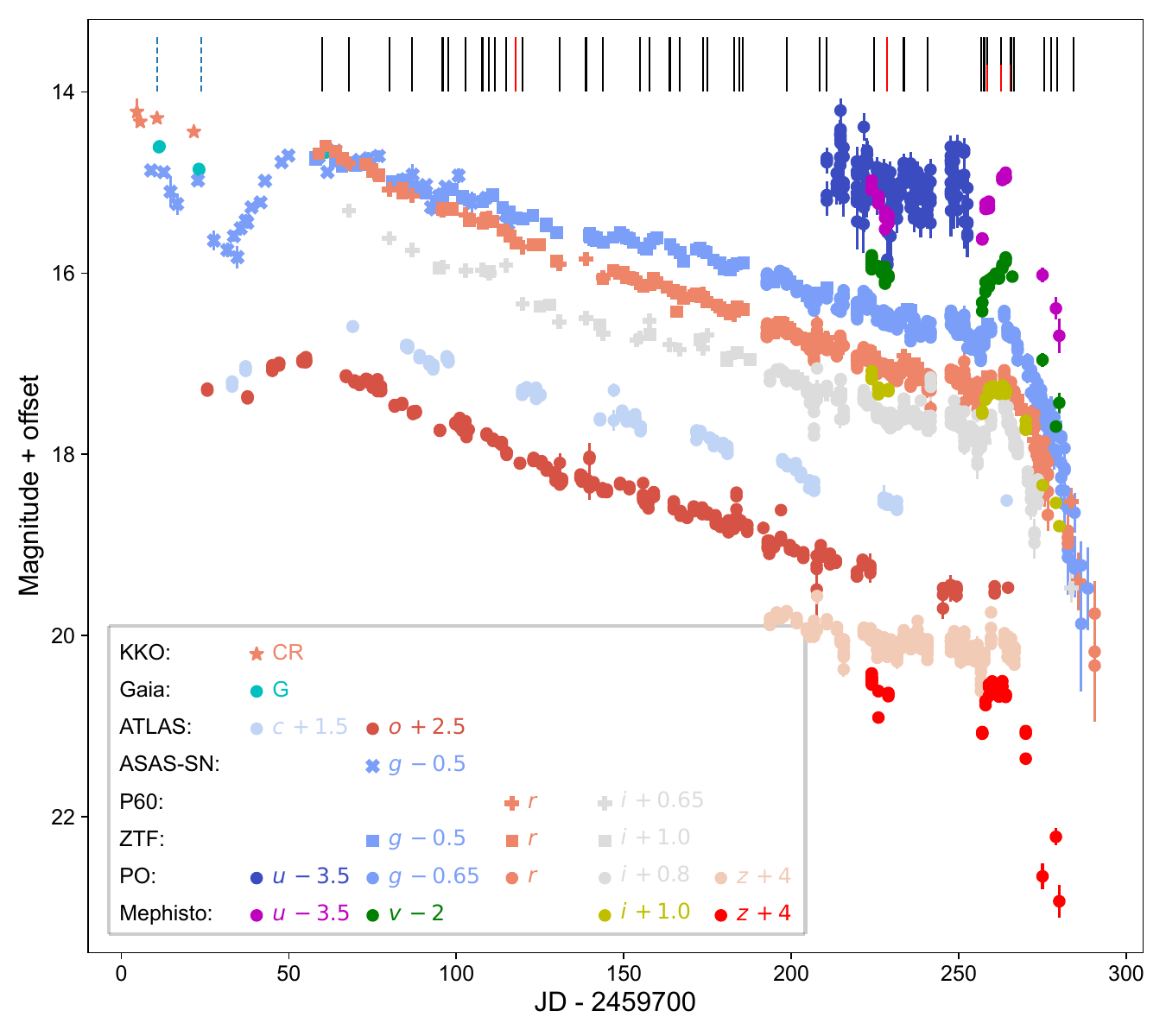}
\caption{Light curve of SN\,2022jli. The vertical lines on top show the epochs of the spectra. The epochs of the NIR spectra are shown in red color, and the optical spectra are shown in black. The first two dashed lines indicate the two classification spectra obtained from the TNS. The offsets applied to different bands are given in the legend.}
\label{fig:lc_sn2022jli}
\end{figure}

\begin{figure}[h]
\centering
\includegraphics[width=0.8\textwidth]{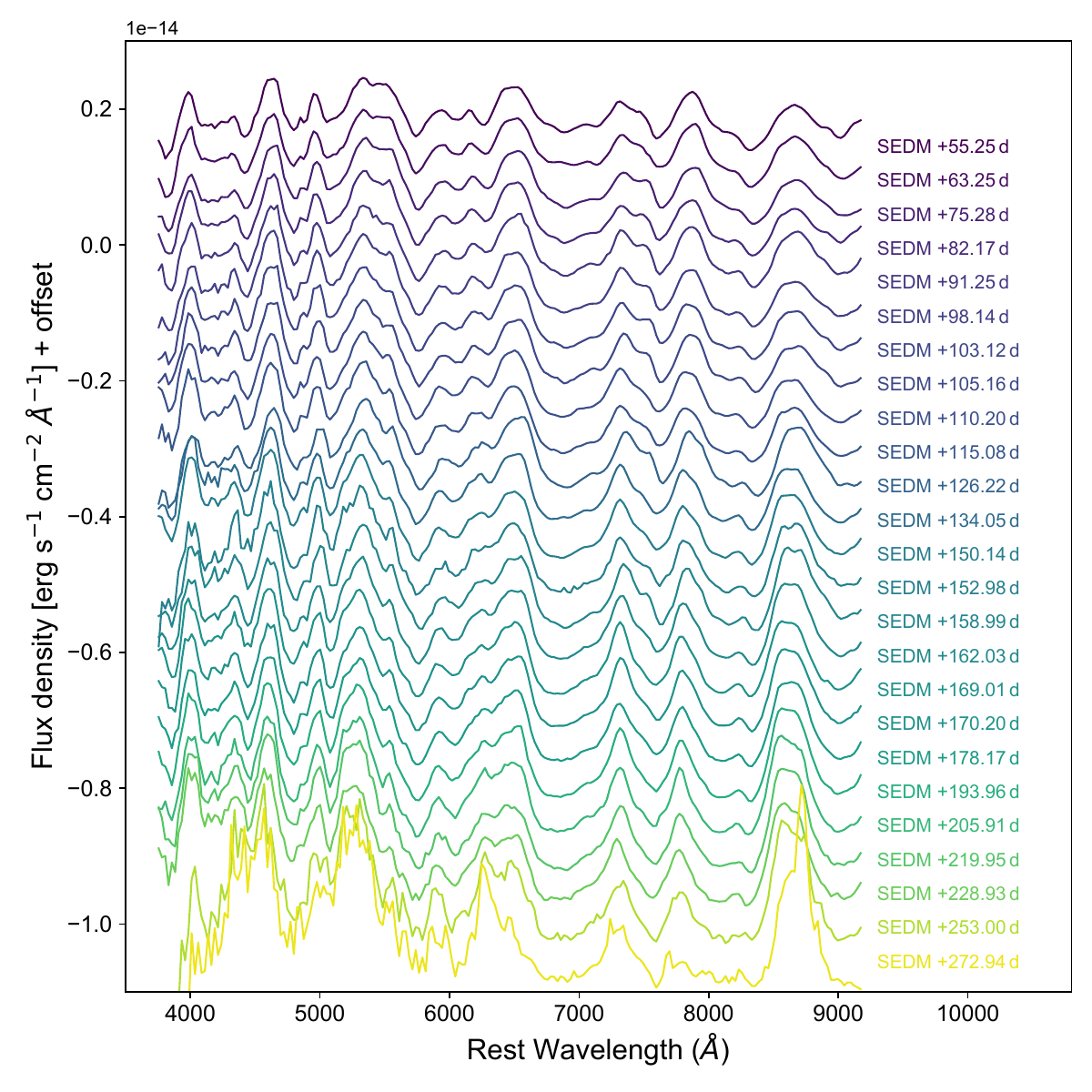}
\caption{Spectra of SN\,2022jli taken with P60/SEDM. The phases are reported relative to the discovery time of JD = 2459704.67.}
\label{fig:spec_opt_sedm}
\end{figure}

\begin{figure}[h]
\centering
\includegraphics[width=\textwidth]{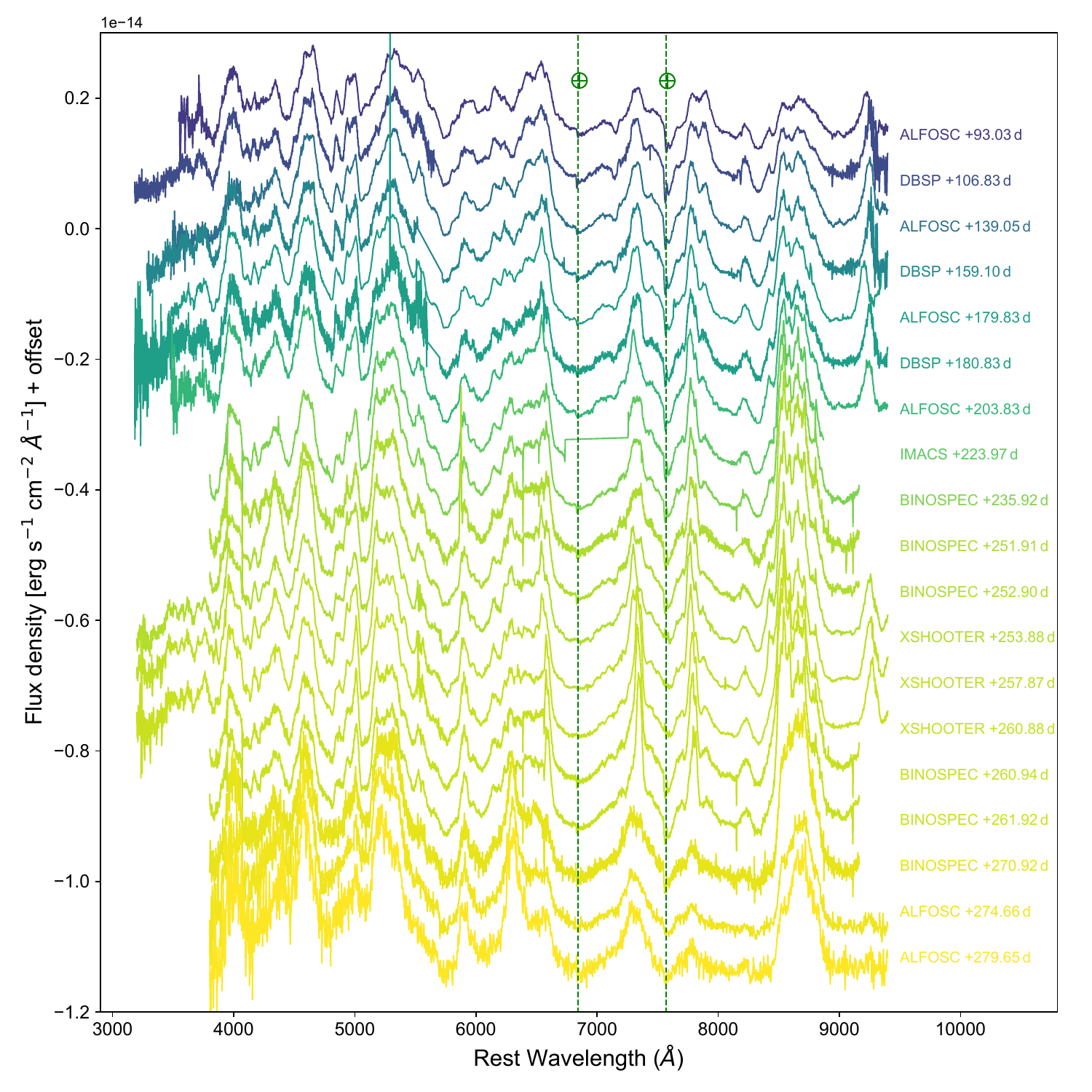}
\caption{Spectral evolution of SN\,2022jli. The phases are reported relative to the discovery time of JD = 2459704.67. The locations of telluric features are marked with a vertical green dashed line.}
\label{fig:spec_opt_nonsedm}
\end{figure}

\begin{figure}[h]
\centering
\includegraphics[width=\textwidth]{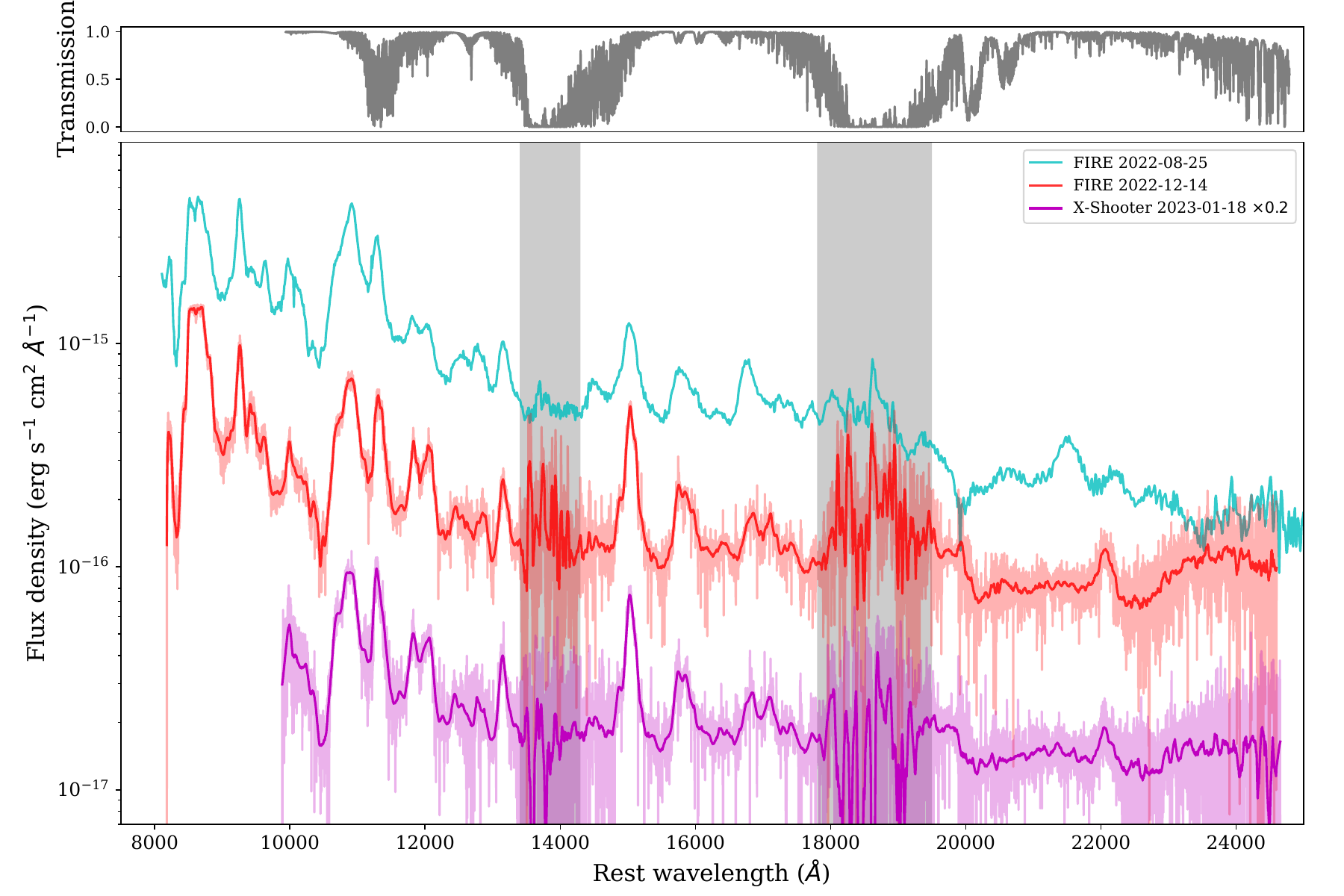}
\caption{NIR spectra of SN\,2022jli at phase +113, +224, and +258 days after discovery time. To separate the FIRE spectrum on 2022 December 14 and the X-shooter spectrum on 2023 January 18, the X-shooter spectrum has been multiplied by a scale factor of 0.2. The upper panel shows the transmission curve used to correct the X-Shooter 2023 January 18 spectrum, which gives an idea of how the telluric absorption affected the observed spectrum in different wavelengths, especially the total absorption around 1.4 $\mu$m (between $J$ and $H$ band) and 1.85 $\mu$m (between $H$ and $K$ band) result in large residuals marked by the shaded grey area. }
\label{fig:spec_nir}
\end{figure}

\begin{figure}[h]
\centering
\includegraphics[width=0.7\textwidth]{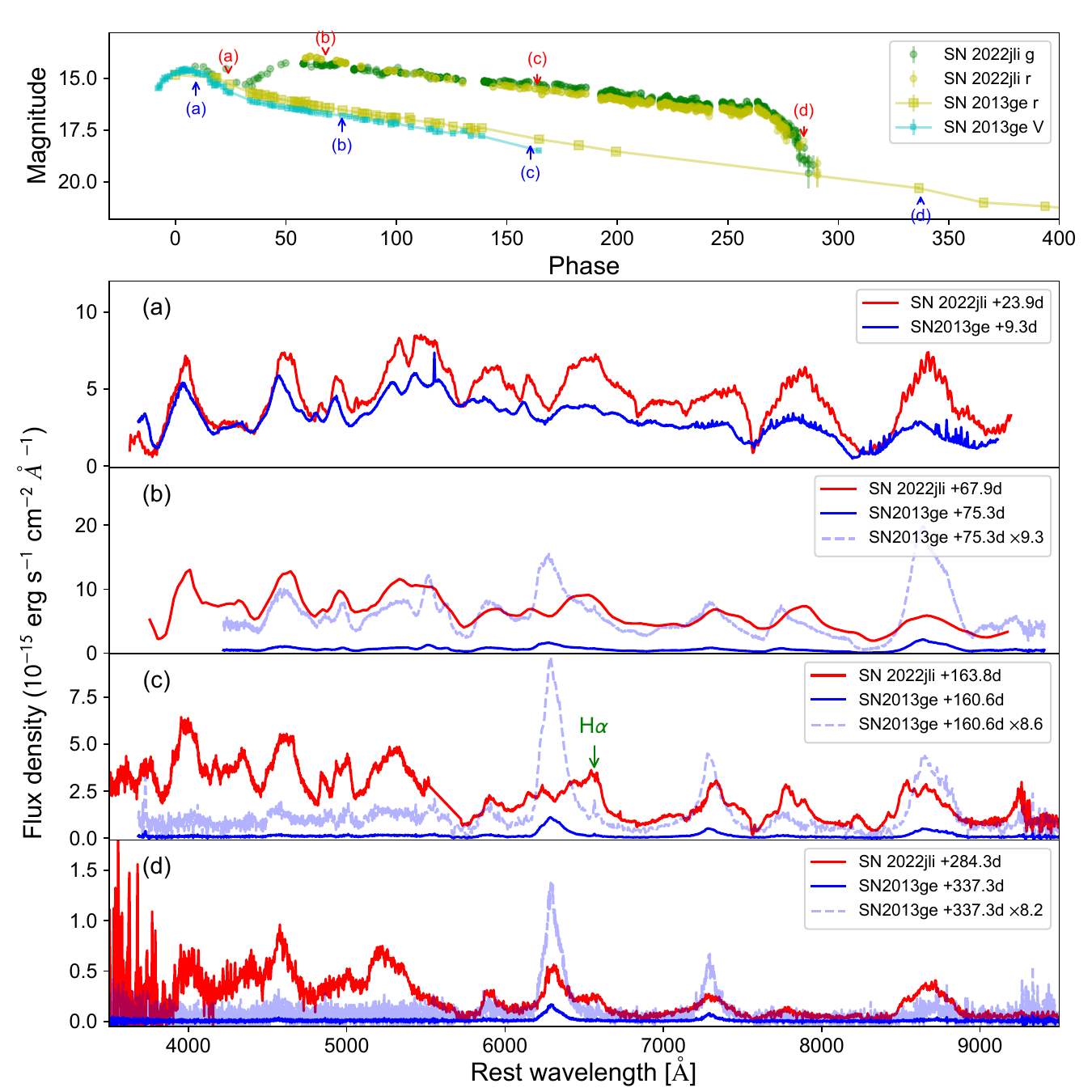}
\caption{Comparison between SN 2022jli and the normal Type Ib/c SN\,2013ge. Top: $g$- and $r$-band light curves of SN 20222jli and $V$- and $r$-band light curves of SN\,2013ge. All the light curves have been corrected for extinction from both the Milky Way and the host galaxy, with total extinction of E$(B-V)_{\rm{2022jli}} = 0.289$ mag and E$(B-V)_{\rm{2013ge}} = 0.067$ mag. Bottom: (a), (b), (c), and (d) subpanels show the spectra of SN\,2022jli, and SN\,2013 at different phases, as indicated with the arrows in the top panel. In (b), (c), and (d) subpanels, the spectra of SN\,2013ge shown as dashed lines have been scaled to match the $r$-band magnitude of SN\,2022jli at the comparison epoch.}
\label{fig:compare_2013ge}
\end{figure}

\begin{figure}[h]
\centering
\includegraphics[width=0.7\textwidth]{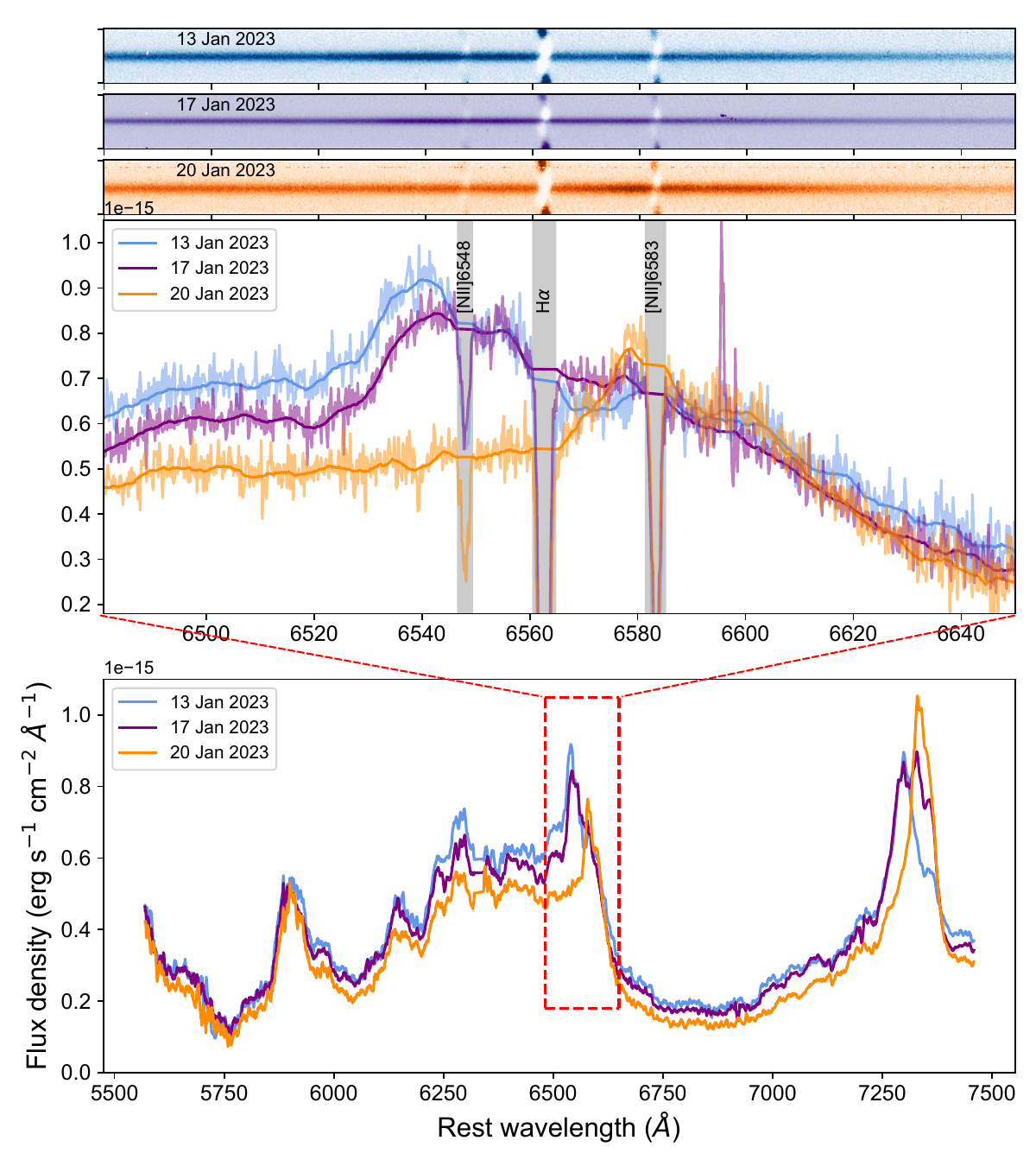}
\caption{X-Shooter spectra of SN\,2022jli in the 5500 -- 7500\,\AA\, region. The bottom panel shows the three X-Shooter spectra after removing artifacts due to over-subtraction of host galaxy emission lines and smoothing. The central panel shows the zoom-in view of the spectra near the H$\alpha$ region shown in the box. The top three panels show the wavelength and flux calibrated 2D spectra where the over-subtraction of host emission lines (\ionf{N}{ii} $\lambda\lambda 6548, 6583$, H$\alpha$) is clear.}
\label{fig:xsh_Ha}
\end{figure}

\begin{figure}[h]
\centering
\includegraphics[width=0.8\textwidth]{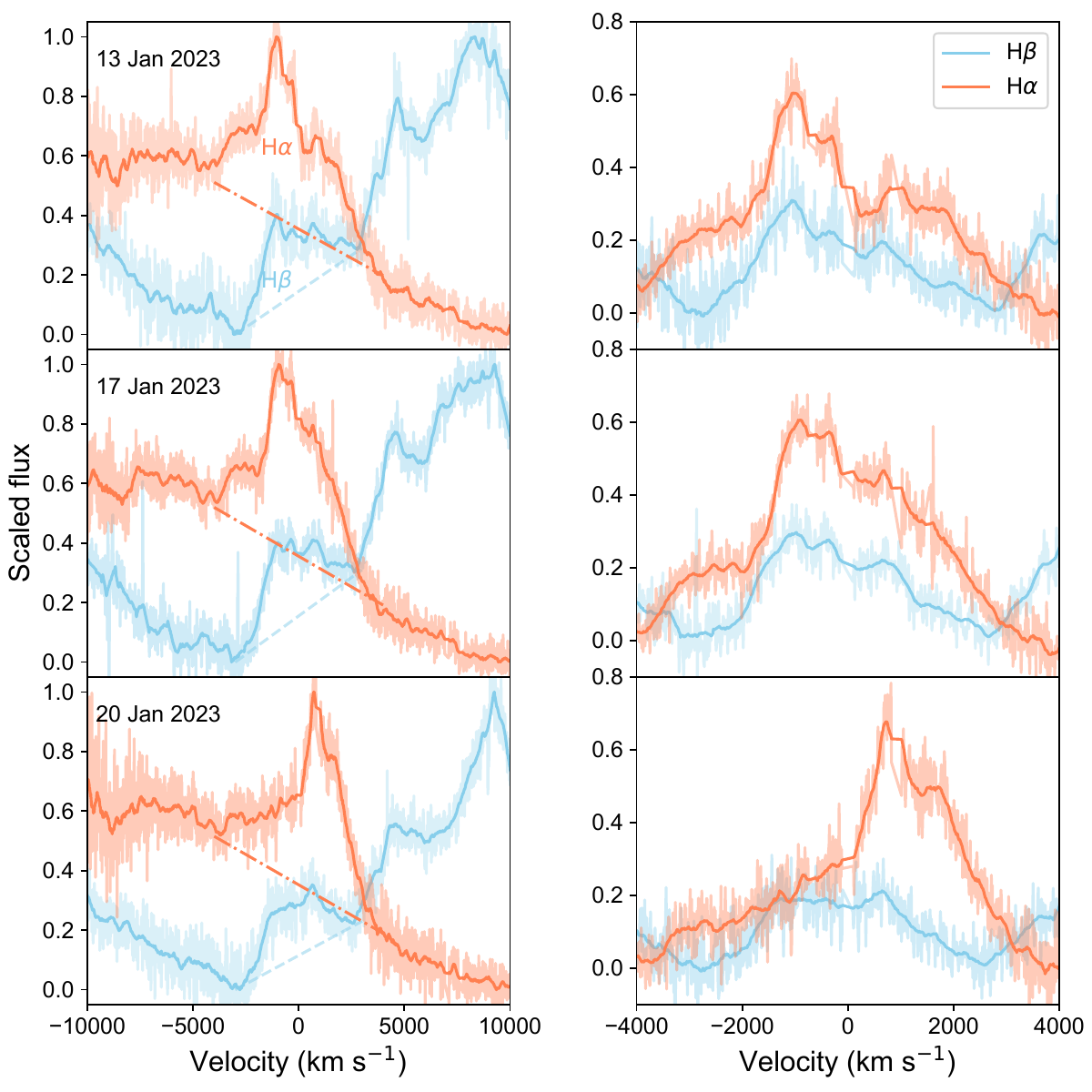}
\caption{X-Shooter spectra of SN\,2022jli around H$\alpha$ (in coral) and H$\beta$ region (in sky blue) showing similar profile and coevolution between these two emission features. Left: Normalized spectra in velocity space. The velocity was calculated to the rest-frame wavelength of H$\alpha$ (6562.79~\AA) and H$\beta$ (4861.35~\AA). The spectra were normalized such that the lowest flux is 0 and the highest flux is 1 within the velocity range from $-$10,000 km s$^{-1}$ to 10,000 km s$^{-1}$. Right: The pseudo-continuum (indicated with the dashed line in the left panels) subtracted spectra of the emission feature.}
\label{fig:HaHb_coevolution}
\end{figure}

\begin{figure}[h]
\centering
\includegraphics[width=\textwidth]{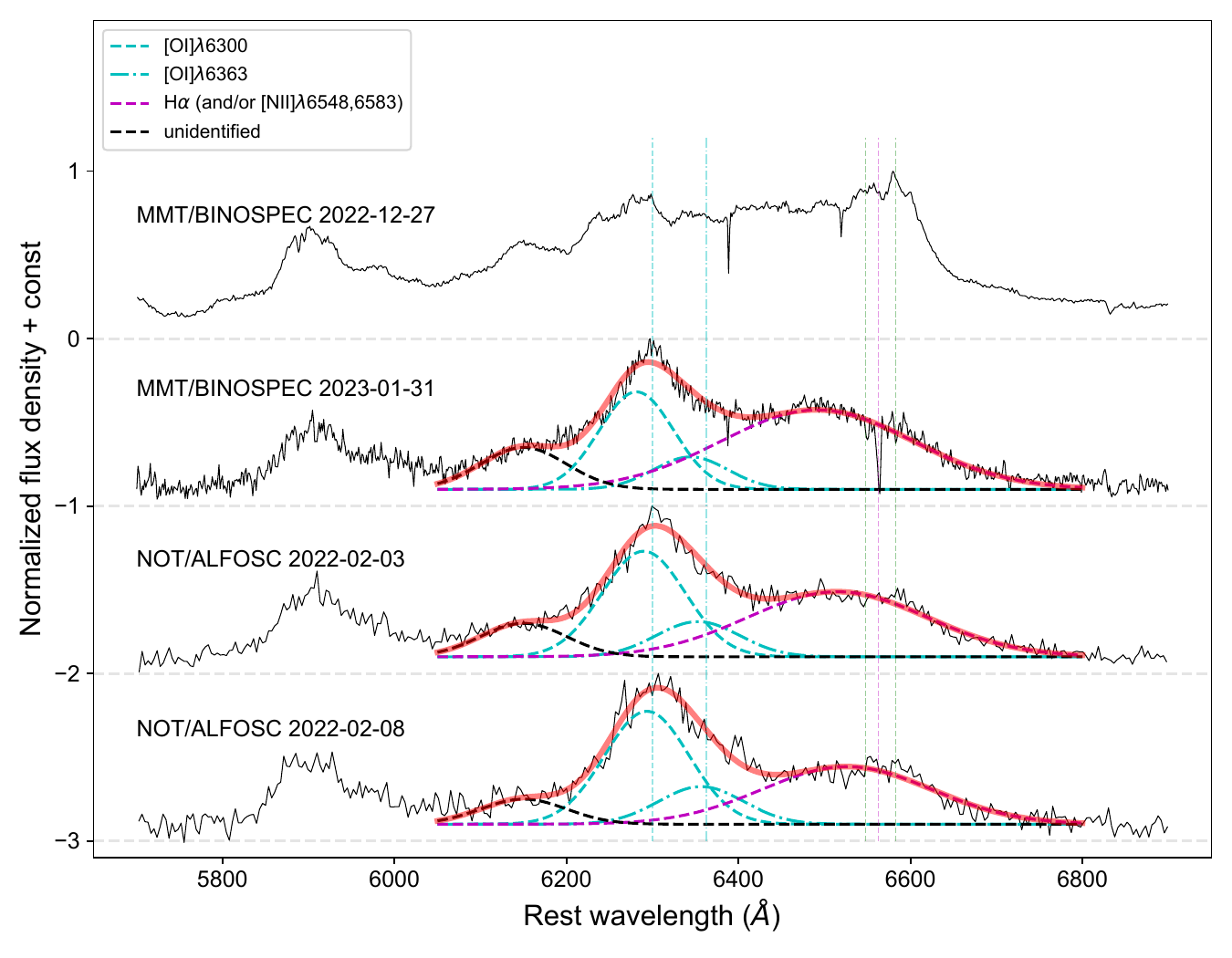}
\caption{Spectral decomposition of SN\,2022jli around 6400 \AA\ which are attributed to emission lines of \ionf{O}{i} $\lambda\lambda6300,6363$, H$\alpha$, and/or \ionf{N}{ii} $\lambda\lambda6548,6583$. The rest-frame wavelengths of the relevant lines are indicated by the vertical dashed line in cyan (\ionf{O}{i} $\lambda\lambda6300,6363$), magenta (H$\alpha$), and green (\ionf{N}{ii} $\lambda\lambda6548,6583$). The spectra have been normalized by the maximum flux within the region shown in the plot and have been shifted in flux for clarity. The decomposition fitting is heuristic, and the model adopted for the fitting is composed of four Gaussian components, whereof two are intended for \ionf{O}{i} $\lambda\lambda6300,6363$ (shown as cyan lines), and the other one is intended for the component on the red side of \ionf{O}{i} lines which might originate from H$\alpha$, and/or \ionf{N}{ii} $\lambda\lambda6548,6583$ (shown as magenta lines), and the last component on the blue side and red side of the \ionf{O}{i} lines (shown as the black lines) was added to improve the fitting. The two Gaussian components for \ionf{O}{i} $\lambda\lambda6300,6363$ share the same width and have flux ratio of $\frac{f(\ionf{{\rm O}}{i}\lambda 6300)}{f(\ionf{{\rm O}}{i}\lambda6363)}=3$. The fitting results shown as the red lines reproduce the observed spectra well. The fittings were performed for three spectra taken after the fast decline of the light curve, and one spectrum taken before the fast decline is shown for comparison. }
\label{fig:specfit}
\end{figure}

\begin{figure}[h]
\centering
\includegraphics[width=0.7\textwidth]{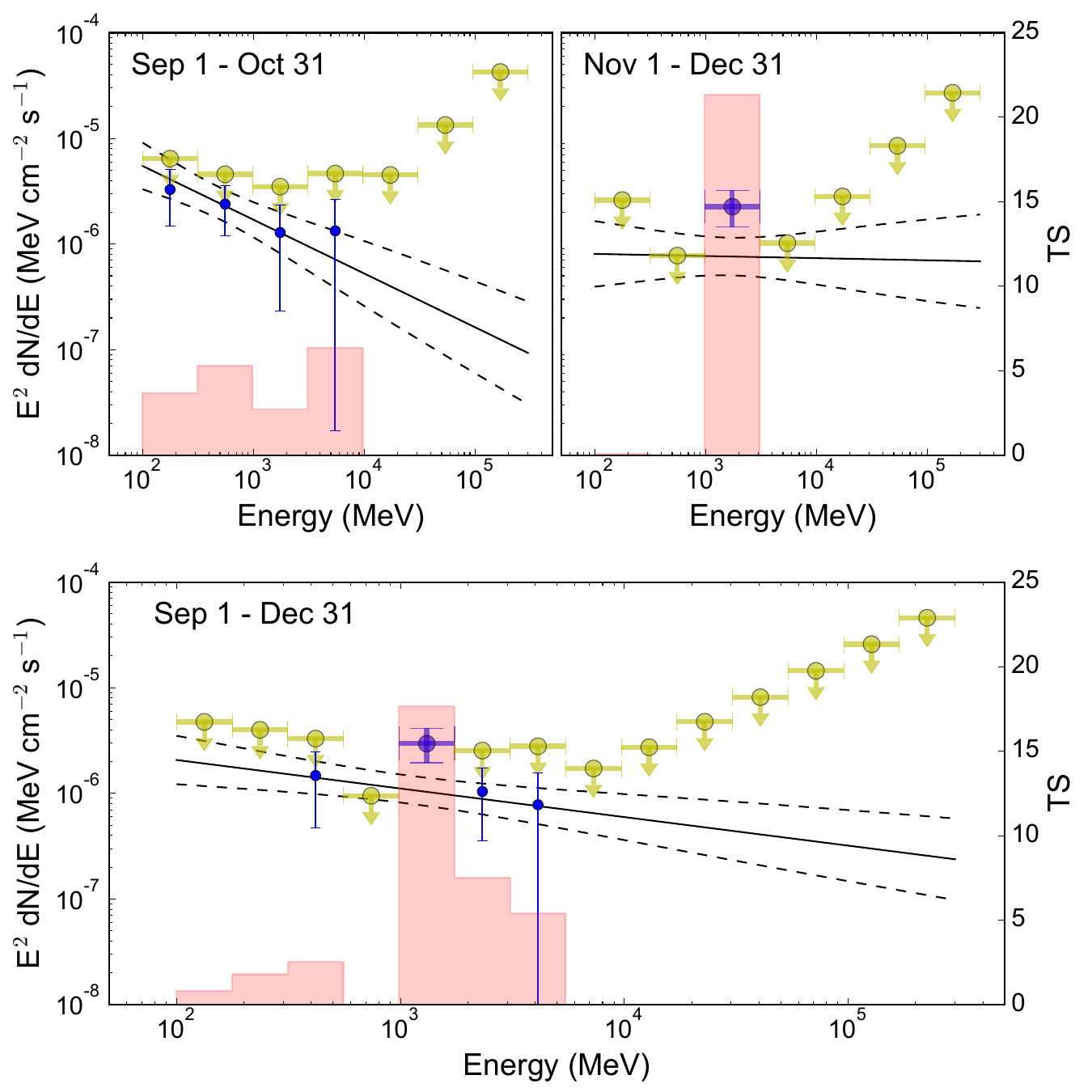}
\caption{Spectral energy distribution (SED) of the new $\gamma$-ray source. The energy range (100\,MeV\,--\,300\,GeV) is evenly divided in logarithmic space. The test statistic (TS) for each energy band is shown with the red bars, and the values are shown on the right axis. The blue points are the measured energy fluxes of sources with TS $> 2$ in individual energy bins, and the yellow points show the 95\% confidence upper limits. The black solid line is the power law model with the dotted line as uncertainty.  The top left panel is the SED for data in September and October 2022, where no source is detected in individual energy bands with significance above 95\% confidence. The top right is the SED for data in November and December 2022, which shows the source is significantly detected and only detected in the 1\,--\,3\,GeV energy band. The bottom panel is the SED for the data taken from September 1 to December 31.}
\label{fig:fermi_sed}
\end{figure}

\begin{figure}[h]
\centering
\includegraphics[width=\textwidth]{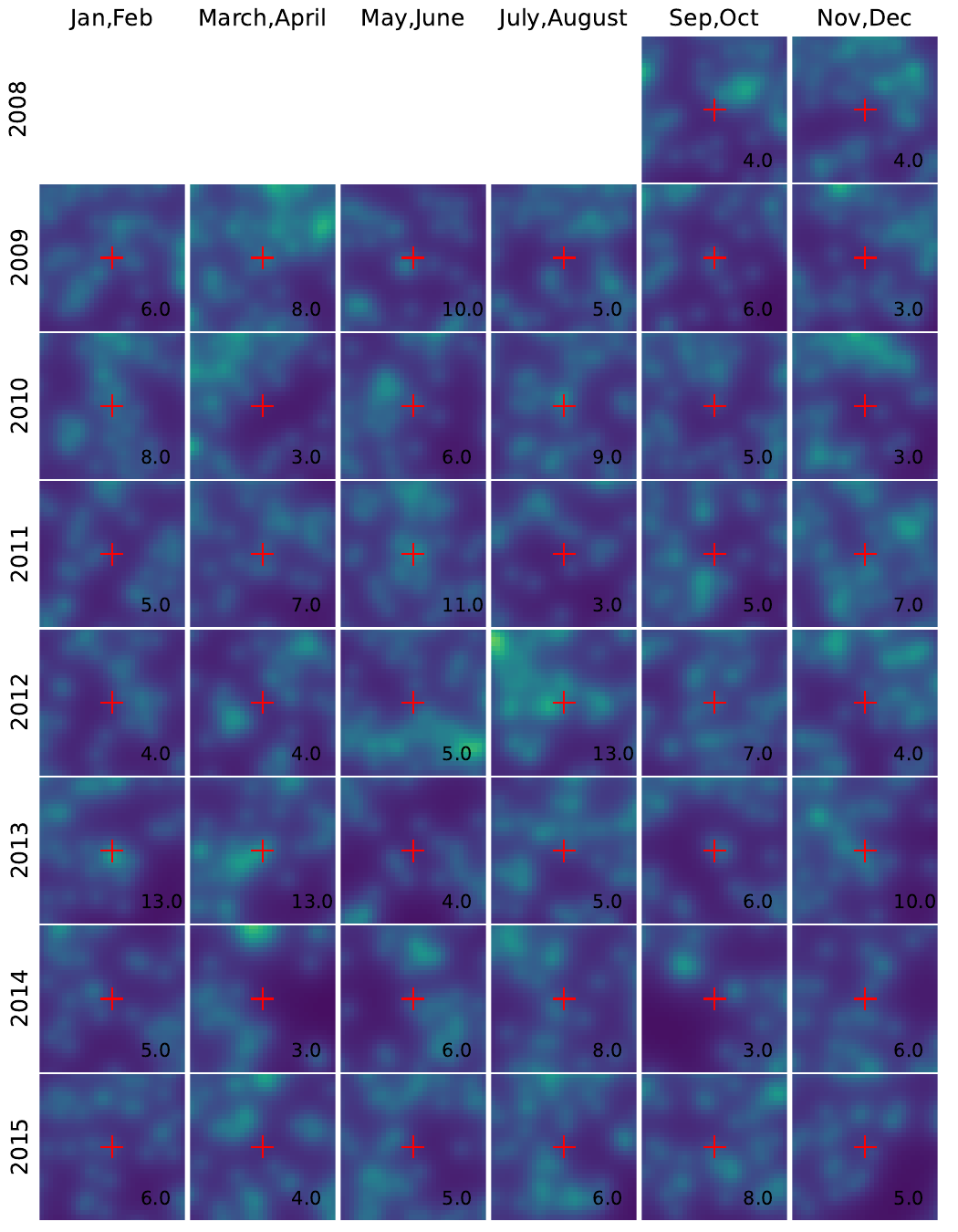}
\caption{S-map grid of the field of SN\,2022jli. Each stamp is built out of 2-month Fermi-LAT data of photons within energy band 1\,GeV to 3\,GeV, with the year and months of the data shown in the left and top of the figure, e.g., the top right corner stamp corresponds to data taken between 2008 November 1 to 2009 January 1. The field of view of each stamp is $4^{\circ}\times4^{\circ}$. The red plus symbol indicates the position of SN\,2022jli, and the value in the bottom right corner of each stamp shows the highest S value with $0.25^{\circ}$ radius of SN\,2022jli.}
\label{fig:Smap1_E1000to3000}
\end{figure}

\begin{figure}[h]
\centering
\includegraphics[width=\textwidth]{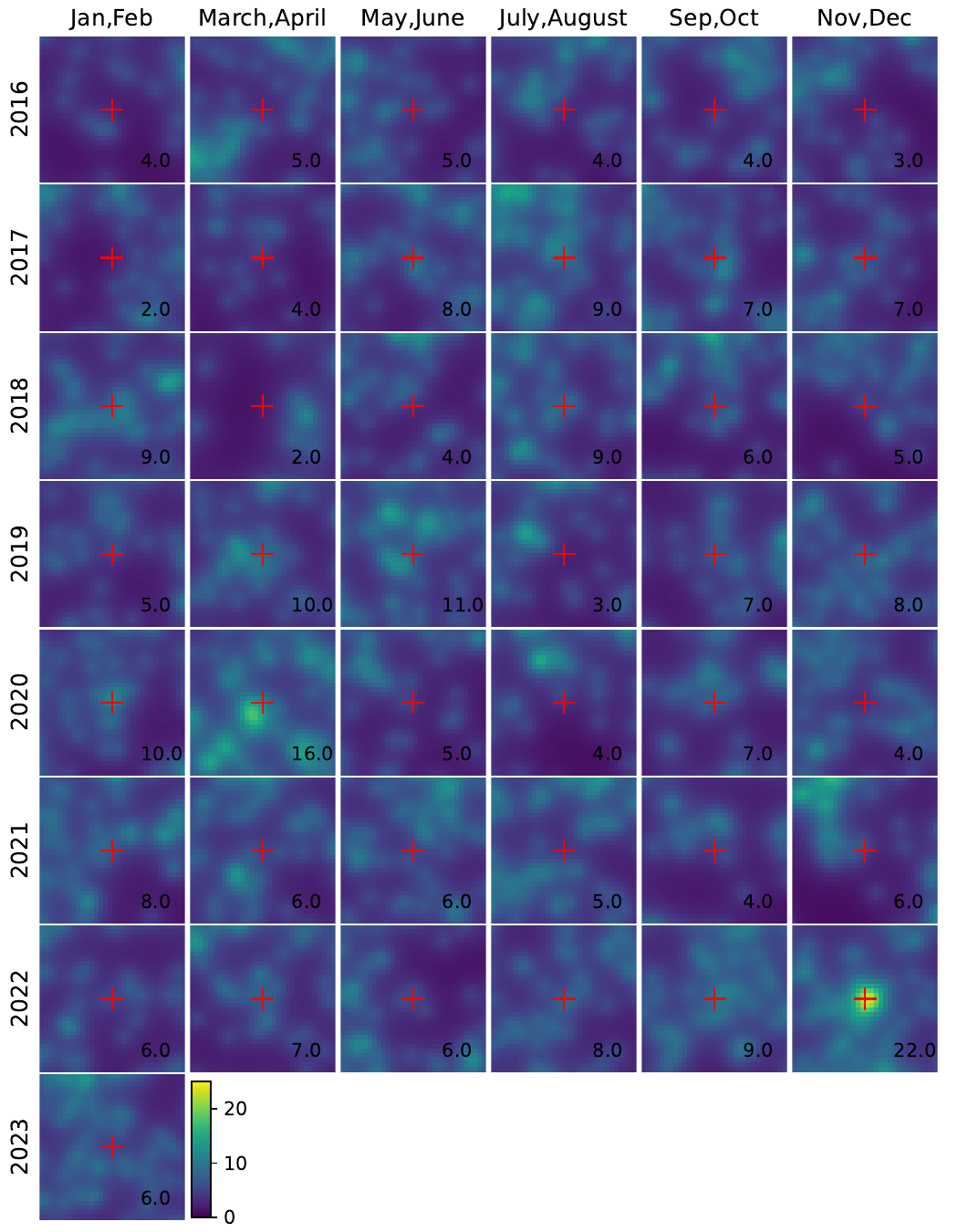}
\caption{S-map grid of SN\,2022jli field (continue of Supplementary Information Fig~\ref{fig:Smap1_E1000to3000}).}
\label{fig:Smap2_E1000to3000}
\end{figure}

\begin{figure}[h]
\centering
\includegraphics[width=\textwidth]{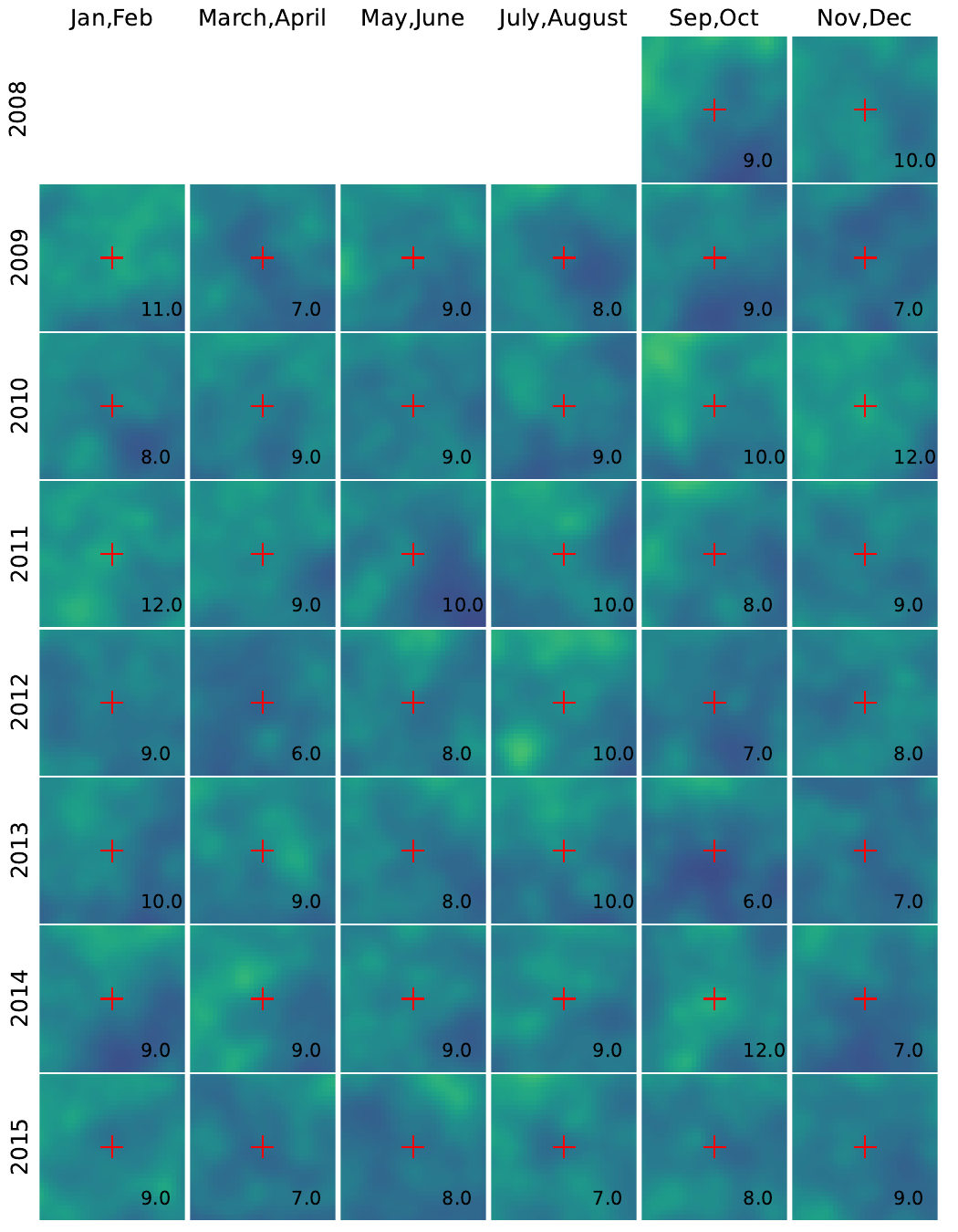}
\caption{S-map grid of SN\,2022jli field. The same as Supplementary Information Fig~\ref{fig:Smap1_E1000to3000} except the energy of photons is in the range of 300\,MeV to 1\,GeV.}
\label{fig:Smap1_E300to1000}
\end{figure}

\begin{figure}[h]
\centering
\includegraphics[width=\textwidth]{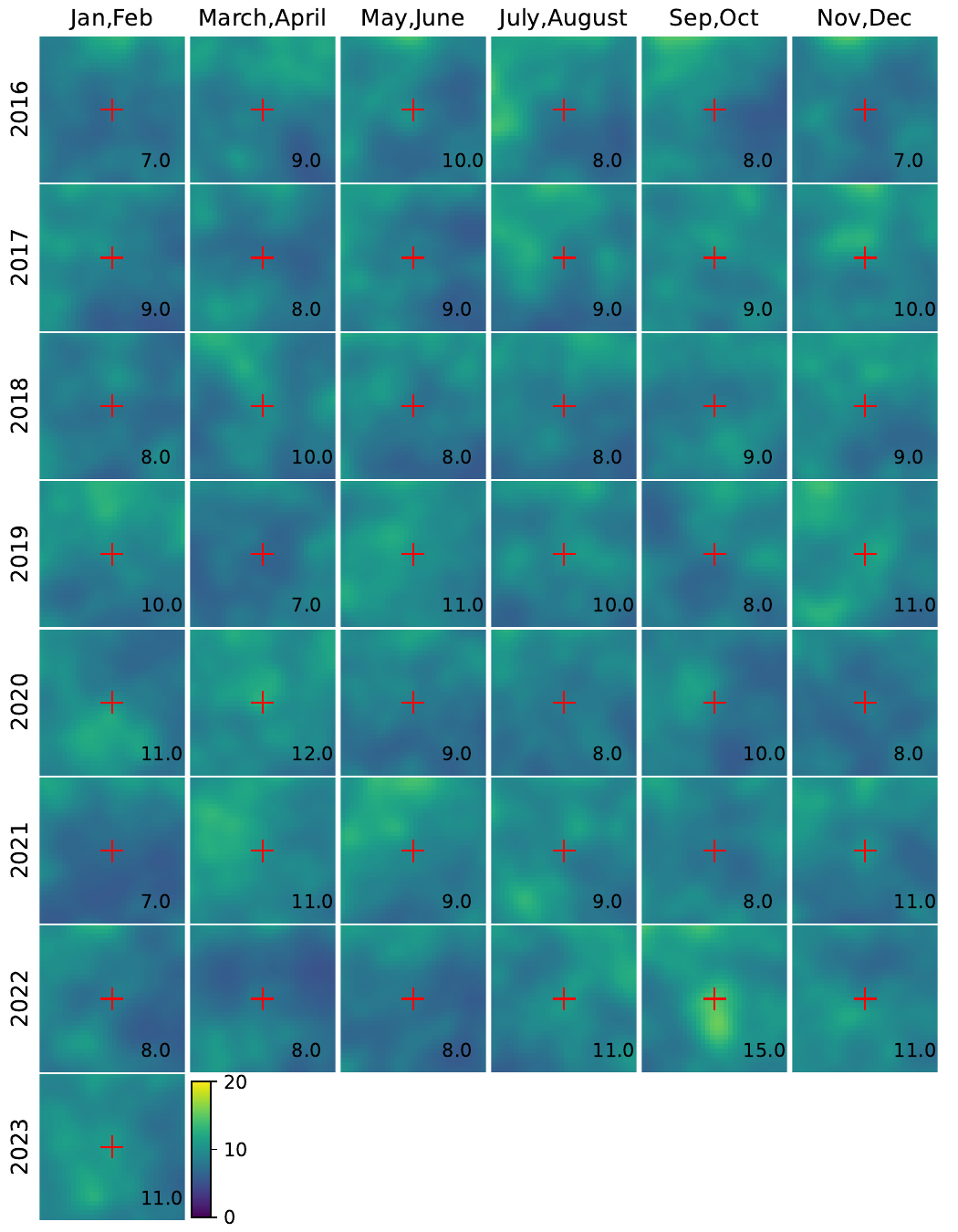}
\caption{S map of SN\,2022jli field (continue of Supplementary Information Fig~\ref{fig:Smap1_E300to1000})}
\label{fig:Smap2_E300to1000}
\end{figure}

\begin{figure}[h]
\centering
\includegraphics[width=\textwidth]{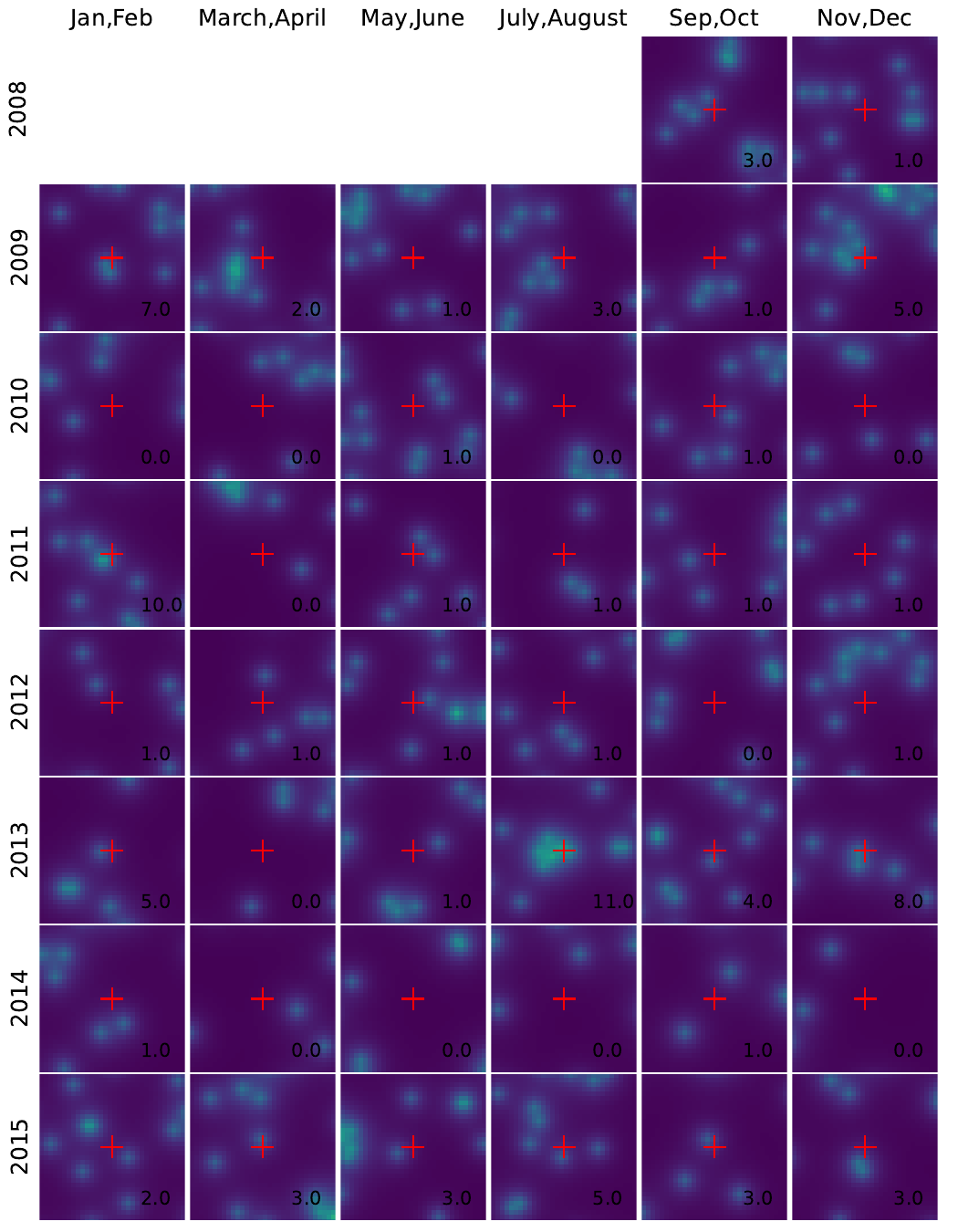}
\caption{S-map grid of SN\,2022jli field. The same as Supplementary Information Fig~\ref{fig:Smap1_E1000to3000} except the energy of photons is in the range of 3\,GeV to 10\,GeV.}
\label{fig:Smap1_E3000to10000}
\end{figure}

\begin{figure}[h]
\centering
\includegraphics[width=\textwidth]{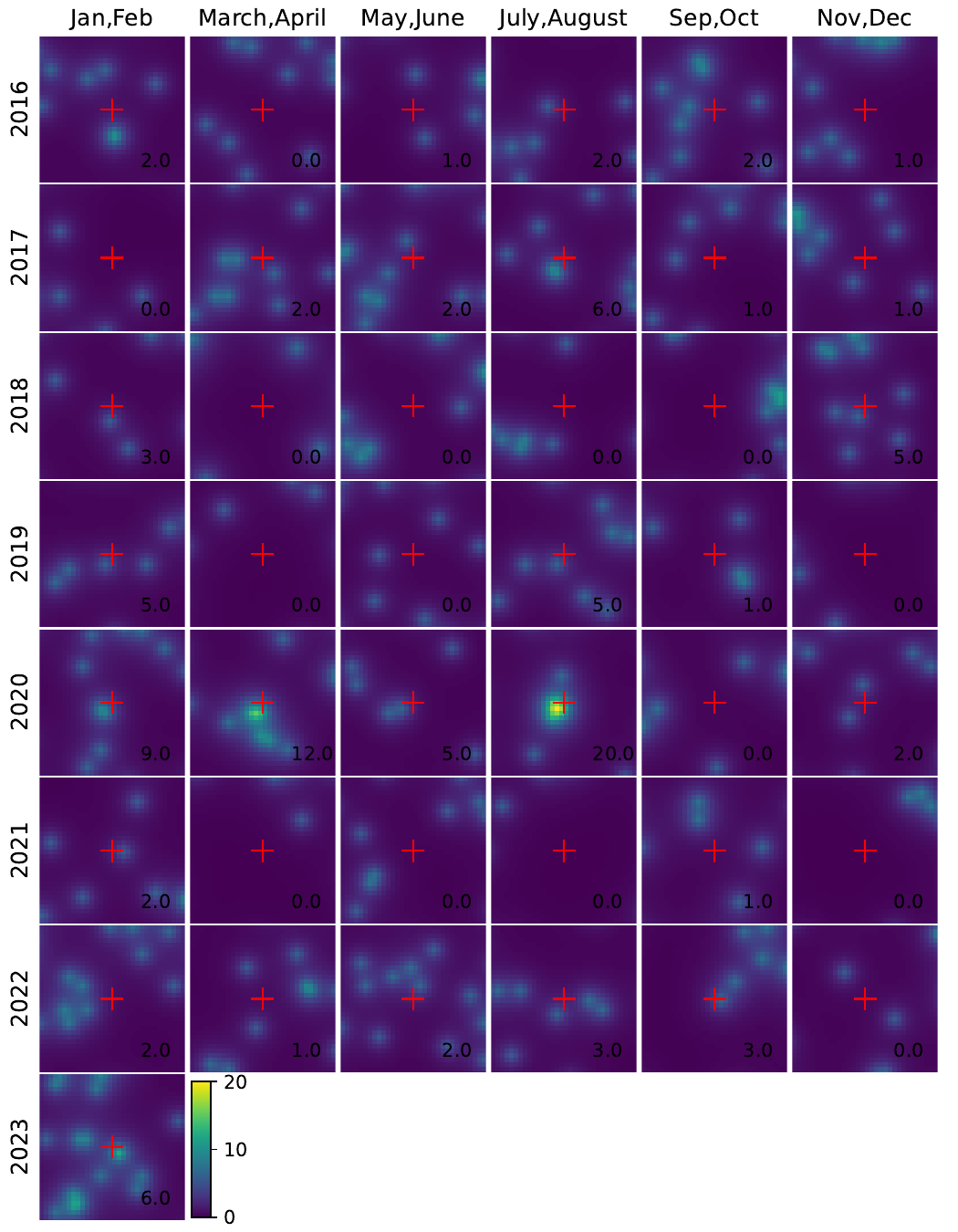}
\caption{S-map of SN\,2022jli field (continue of Supplementary Information Fig~\ref{fig:Smap1_E3000to10000})}
\label{fig:Smap2_E3000to10000}
\end{figure}

\begin{figure}[h]
\centering
\includegraphics[width=0.9\columnwidth]{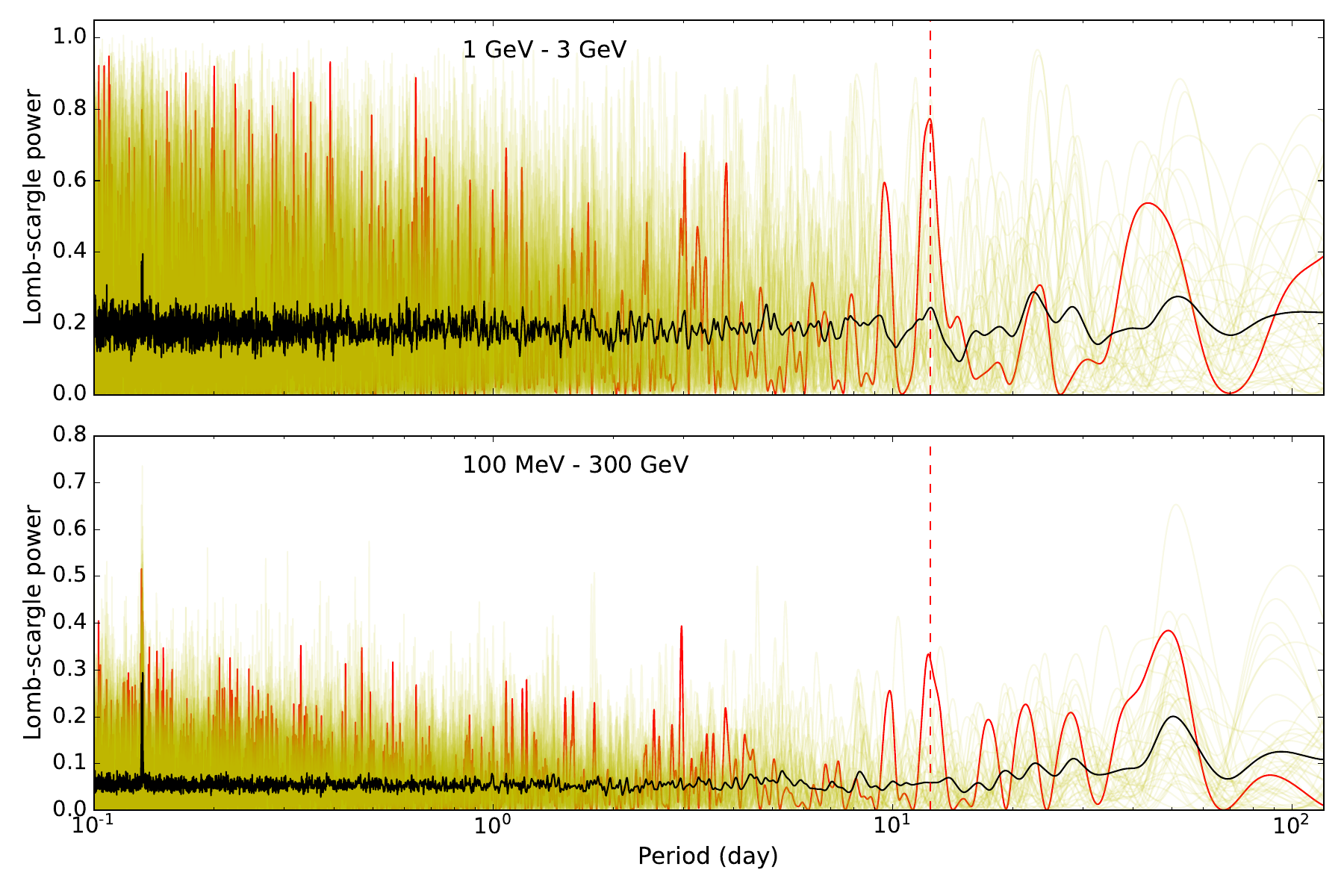}
\caption{The Lomb-Scargle periodogram of $\gamma$-ray photon arrival time. The Lomb-Scargle power spectra come from a series of unit values at the photon arrival time. The upper panel shows the result of photons in a narrow energy range of 1\,GeV to 3\,GeV, and the bottom panel shows the result from a broad energy range of 100\,MeV to 300\,GeV. The red line is for photons within $0.4^\circ$ radius of the new $\gamma$-ray source. The yellow lines are for photons within a specific radius of the 42 cataloged sources in 4FGL-DR3 within $10^{\circ}$ radius of SN\,2022jli. The radius for the 4FGL-DR3 source is chosen to have the same number of photons as the new $\gamma$-ray source, i.e., 11 photons for 1\,GeV to 3\,GeV, and 37 photons for 100\,MeV to 300\,GeV. The black lines show the averaged power spectra of the 42 4FGL-DR3 sources, which provide information on the window function of the \fermilat observation of the region. The period around 0.132 days ($\sim 190$ minutes) in the averaged spectrum is clear, corresponding to typical observatory sky-survey profiles which run for two orbits, with the observatory rocked to the north for the first orbit and to the south for the second. As a result, the SN\,2022jli region has been observed every two orbits of {\tt Fermi}-LAT observation. There is no significant periodic signal at 12.4 days, which means the $\gamma$-ray photon's association to the 12.4-day period of SN\,2022jli shown in Extended Data Fig.~\ref{fig:fermi_localization_periodicity} could not be due to the survey profile. The peaks around 12.4 days in the power spectra of the new $\gamma$-ray source indicated by the vertical dashed line in red color, although not significant alone, are consistent with the result found in Extended Data Fig.~\ref{fig:fermi_localization_periodicity}.}
\label{fig:gamma_period}
\end{figure}

\end{supplement}

\clearpage

\begin{deluxetable}{cccc}
\tablecaption{SN\,2022jli photometry}
\tablehead{\colhead{Epoch} & \colhead{AB Magnitude} & \colhead{Source } & \colhead{Filter}
\\ (JD-2,459,700) &  & & }
\tabletypesize{\footnotesize}  
\tablewidth{0pt}
\startdata
8.9154 & 15.366$\pm$0.053 & ASAS-SN & g \\
12.6611 & 15.386$\pm$0.053 & ASAS-SN & g \\
14.6775 & 15.602$\pm$0.171 & ASAS-SN & g \\
16.6692 & 15.737$\pm$0.120 & ASAS-SN & g \\
22.8739 & 15.477$\pm$0.060 & ASAS-SN & g \\
25.6585 & 14.786$\pm$0.007 & ATLAS & o \\
25.6657 & 14.775$\pm$0.007 & ATLAS & o \\
25.6826 & 14.796$\pm$0.012 & ATLAS & o \\
27.6503 & 16.140$\pm$0.113 & ASAS-SN & g \\
31.6665 & 16.245$\pm$0.078 & ASAS-SN & g \\
33.0868 & 15.751$\pm$0.013 & ATLAS & c \\
33.0953 & 15.696$\pm$0.011 & ATLAS & c \\
33.1092 & 15.707$\pm$0.013 & ATLAS & c \\
33.1134 & 15.718$\pm$0.017 & ATLAS & c \\
33.6136 & 16.092$\pm$0.063 & ASAS-SN & g \\
34.6683 & 16.324$\pm$0.125 & ASAS-SN & g \\
35.6096 & 16.005$\pm$0.082 & ASAS-SN & g \\
37.0850 & 15.568$\pm$0.011 & ATLAS & c \\
37.1016 & 15.540$\pm$0.011 & ATLAS & c \\
37.1042 & 15.920$\pm$0.072 & ASAS-SN & g \\
37.1044 & 15.529$\pm$0.010 & ATLAS & c \\
37.1150 & 15.579$\pm$0.018 & ATLAS & c \\
37.6023 & 15.944$\pm$0.066 & ASAS-SN & g \\
37.6553 & 14.867$\pm$0.007 & ATLAS & o \\
37.6739 & 14.861$\pm$0.008 & ATLAS & o \\
37.6854 & 14.882$\pm$0.015 & ATLAS & o \\
38.8515 & 15.774$\pm$0.051 & ASAS-SN & g \\
41.5805 & 15.720$\pm$0.053 & ASAS-SN & g \\
42.9147 & 15.482$\pm$0.039 & ASAS-SN & g \\
45.0839 & 14.578$\pm$0.009 & ATLAS & o \\
45.0959 & 14.517$\pm$0.009 & ATLAS & o \\
45.1065 & 14.563$\pm$0.008 & ATLAS & o \\
45.1101 & 14.538$\pm$0.008 & ATLAS & o \\
47.0924 & 14.489$\pm$0.009 & ATLAS & o \\
47.0965 & 14.520$\pm$0.009 & ATLAS & o \\
47.1034 & 14.494$\pm$0.008 & ATLAS & o \\
47.1159 & 14.496$\pm$0.019 & ATLAS & o \\
47.8300 & 15.275$\pm$0.047 & ASAS-SN & g \\
49.8793 & 15.202$\pm$0.049 & ASAS-SN & g \\
54.0836 & 14.482$\pm$0.005 & ATLAS & o \\
54.0868 & 14.464$\pm$0.006 & ATLAS & o \\
54.0966 & 14.463$\pm$0.005 & ATLAS & o \\
54.1023 & 14.454$\pm$0.005 & ATLAS & o \\
55.0878 & 14.434$\pm$0.006 & ATLAS & o \\
55.0925 & 14.483$\pm$0.006 & ATLAS & o \\
55.1004 & 14.475$\pm$0.006 & ATLAS & o \\
55.1018 & 14.464$\pm$0.006 & ATLAS & o \\
55.1105 & 14.455$\pm$0.006 & ATLAS & o \\
55.1119 & 14.427$\pm$0.006 & ATLAS & o \\
55.1183 & 14.481$\pm$0.009 & ATLAS & o \\
55.1197 & 14.467$\pm$0.011 & ATLAS & o \\
57.9530 & 15.244$\pm$0.024 & ZTF & g \\
57.9530 & 15.244$\pm$0.024 & ZTF & g \\
57.9540 & 15.220$\pm$0.029 & ZTF & g \\
57.9540 & 15.220$\pm$0.029 & ZTF & g \\
57.9540 & 15.220$\pm$0.029 & ZTF & g \\
58.9335 & 15.217$\pm$0.035 & ASAS-SN & g \\
\enddata
\label{tab:phot}
\tablecomments{Only part of the photometry is shown here for guidance regarding the form and content of the table.}
\end{deluxetable}

\begin{deluxetable}{lcccc}
\tablecaption{Summary of Spectroscopic Observations of SN\,2022jli}
\tabletypesize{\footnotesize}  
\tablewidth{0pt}
\tablehead{
\colhead{Date (UTC)} &
\colhead{Phase (d)} &
\colhead{Telescope} &
\colhead{Spectrograph} &
\colhead{Exp (s)}
}
\startdata 
2022-06-29.423 &  55.254 &       P60 &       SEDM &    1800 \\
2022-07-07.424 &  63.254 &       P60 &       SEDM &    1800 \\
2022-07-19.453 &  75.284 &       P60 &       SEDM &    1800 \\
2022-07-26.338 &  82.169 &       P60 &       SEDM &    1800 \\
2022-08-04.418 &  91.249 &       P60 &       SEDM &    1800 \\
2022-08-06.196 &  93.027 &       NOT &     ALFOSC &     600 \\
2022-08-11.310 &  98.141 &       P60 &       SEDM &    1800 \\
2022-08-16.288 & 103.119 &       P60 &       SEDM &    1800 \\
2022-08-18.325 & 105.155 &       P60 &       SEDM &    1800 \\
  2022-08-20.0 & 106.830 &      P200 &       DBSP &     300 \\
2022-08-23.371 & 110.202 &       P60 &       SEDM &    1800 \\
2022-08-26.824 & 113.154 &  Magellan &       FIRE &    507  \\
2022-08-28.254 & 115.084 &       P60 &       SEDM &    1800 \\
2022-09-08.387 & 126.217 &       P60 &       SEDM &    1800 \\
2022-09-16.216 & 134.046 &       P60 &       SEDM &    1800 \\
2022-09-21.221 & 139.052 &       NOT &     ALFOSC &     1200 \\
2022-10-02.305 & 150.136 &       P60 &       SEDM &    1800 \\
2022-10-05.149 & 152.980 &       P60 &       SEDM &    1800 \\
2022-10-11.159 & 158.989 &       P60 &       SEDM &    1800 \\
2022-10-11.274 & 159.104 &      P200 &       DBSP &    600 \\
2022-10-14.199 & 162.029 &       P60 &       SEDM &    1800 \\
2022-10-21.179 & 169.009 &       P60 &       SEDM &    1800 \\
2022-10-22.372 & 170.202 &       P60 &       SEDM &    1800 \\
2022-10-30.337 & 178.168 &       P60 &       SEDM &    1800 \\
2022-10-31.997 & 179.828 &       NOT &     ALFOSC &    1200 \\
  2022-11-02.0 & 180.830 &      P200 &       DBSP &     420 \\
2022-11-15.133 & 193.964 &       P60 &       SEDM &    1800 \\
  2022-11-25.1 & 203.831 &       NOT &     ALFOSC &    1200 \\
 2022-11-27.78 & 205.909 &       P60 &       SEDM &    1800 \\
 2022-12-11.119 & 219.949 &      P60 &       SEDM &    1800 \\
2022-12-15.137 & 223.967 & Magellan &      IMACS &     900 \\
2022-12-15.560 & 223.890 & Magellan &       FIRE &     2832\\ 
 2022-12-20.98 & 228.929 &      P60 &       SEDM &    1800 \\
 2022-12-27.91 & 235.921 &      MMT &   BINOSPEC &    2400 \\
 2023-01-12.77 & 251.907 &      MMT &   BINOSPEC &     450 \\
 2023-01-13.71 & 252.902 &      MMT &   BINOSPEC &    1350 \\
2023-01-13.172 & 253.003 &      P60 &       SEDM &    1800 \\
 2023-01-14.50 & 253.881 &      VLT &   XSHOOTER &     900 \\
 2023-01-18.44 & 257.874 &      VLT &   XSHOOTER &     900 \\
 2023-01-21.45 & 260.875 &      VLT &   XSHOOTER &     900 \\
2023-01-21.111 & 260.941 &      MMT &   BINOSPEC &    1800 \\
 2023-01-22.90 & 261.921 &      MMT &   BINOSPEC &    2250 \\
 2023-01-31.89 & 270.920 &      MMT &   BINOSPEC &    1350 \\
2023-02-02.108 & 272.938 &      P60 &       SEDM &    1800 \\
2023-02-03.829 & 274.660 &      NOT &     ALFOSC &    1200 \\
2023-02-08.824 & 279.654 &      NOT &     ALFOSC &    1200 \\
\enddata
\label{tab:spec}
\tablecomments{The phase is calculated with respect to the discovery date of JD= 2459704.67.}
\end{deluxetable}

\clearpage

\begin{thebibliography}{}
\expandafter\ifx\csname url\endcsname\relax
  \def\url#1{\texttt{#1}}\fi
\expandafter\ifx\csname urlprefix\endcsname\relax\def\urlprefix{URL }\fi
\providecommand{\bibinfo}[2]{#2}
\providecommand{\eprint}[2][]{\url{#2}}

\end{thebibliography}


\noindent {\bf References}

\vspace{1cm}


\begin{thebibliography}{100}
\expandafter\ifx\csname url\endcsname\relax
  \def\url#1{\texttt{#1}}\fi
\expandafter\ifx\csname urlprefix\endcsname\relax\def\urlprefix{URL }\fi
\providecommand{\bibinfo}[2]{#2}
\providecommand{\eprint}[2][]{\url{#2}}



\bibitem{Heger2003}
\bibinfo{author}{{Heger}, A.}, \bibinfo{author}{{Fryer}, C.~L.}, \bibinfo{author}{{Woosley}, S.~E.}, \bibinfo{author}{{Langer}, N.} \& \bibinfo{author}{{Hartmann}, D.~H.}
\newblock \bibinfo{title}{{How Massive Single Stars End Their Life}}.
\newblock \emph{\bibinfo{journal}{\apj}} \textbf{\bibinfo{volume}{591}}, \bibinfo{pages}{288--300} (\bibinfo{year}{2003}).

\bibitem{Sana2012}
\bibinfo{author}{{Sana}, H.} \emph{et~al.}
\newblock \bibinfo{title}{{Binary Interaction Dominates the Evolution of Massive Stars}}.
\newblock \emph{\bibinfo{journal}{Science}} \textbf{\bibinfo{volume}{337}}, \bibinfo{pages}{444} (\bibinfo{year}{2012}).

\bibitem{Hirai2022}
\bibinfo{author}{{Hirai}, Ryosuke} \& \bibinfo{author}{{Podsiadlowski}, Philipp}
\newblock \bibinfo{title}{{Neutron stars colliding with binary companions: formation of hypervelocity stars, pulsar planets, bumpy superluminous supernovae and Thorne-Żytkow objects}}.
\newblock \emph{\bibinfo{journal}{\mnras}} \textbf{\bibinfo{volume}{517}}, \bibinfo{pages}{4544} (\bibinfo{year}{2022}).

\bibitem{Gal-Yam2017}
\bibinfo{author}{{Gal-Yam}, A.}
\newblock \bibinfo{title}{{Observational and Physical Classification of Supernovae}}.
\newblock In \bibinfo{editor}{{Alsabti}, A.~W.} \& \bibinfo{editor}{{Murdin}, P.} (eds.) \emph{\bibinfo{booktitle}{Handbook of Supernovae}}, \bibinfo{pages}{195} (\bibinfo{year}{2017}).

\bibitem{Nicholl2016}
\bibinfo{author}{{Nicholl}, M.} \emph{et~al.}
\newblock \bibinfo{title}{{SN 2015BN: A Detailed Multi-wavelength View of a Nearby Superluminous Supernova}}.
\newblock \emph{\bibinfo{journal}{\apj}} \textbf{\bibinfo{volume}{826}},\bibinfo{pages}{39} (\bibinfo{year}{2016}).

\bibitem{Yan2017}
\bibinfo{author}{{Yan}, L.} \emph{et~al.}
\newblock \bibinfo{title}{{Hydrogen-poor Superluminous Supernovae with Late-time H{\ensuremath{\alpha}} Emission: Three Events From the Intermediate Palomar Transient Factory}}.
\newblock \emph{\bibinfo{journal}{\apj}} \textbf{\bibinfo{volume}{848}}, \bibinfo{pages}{6} (\bibinfo{year}{2017}).

\bibitem{Hosseinzadeh2022}
\bibinfo{author}{{Hosseinzadeh}, G.} \emph{et~al.}
\newblock \bibinfo{title}{{Bumpy Declining Light Curves Are Common in Hydrogen-poor Superluminous Supernovae}}.
\newblock \emph{\bibinfo{journal}{\apj}} \textbf{\bibinfo{volume}{933}}, \bibinfo{pages}{14} (\bibinfo{year}{2022}).

\bibitem{West2023}
\bibinfo{author}{{West}, S.~L.} \emph{et~al.}
\newblock \bibinfo{title}{{SN 2020qlb: A hydrogen-poor superluminous supernova with well-characterized light curve undulations}}. \newblock \emph{\bibinfo{journal}{\aap}} \textbf{\bibinfo{volume}{670}}, \bibinfo{pages}{A7} (\bibinfo{year}{2023}).

\bibitem{ChenZH2023}
\bibinfo{author}{{Chen}, Z.~H.} \emph{et~al.}
\newblock \bibinfo{title}{{The Hydrogen-poor Superluminous Supernovae from the Zwicky Transient Facility Phase I Survey. II. Light-curve Modeling and Characterization of Undulations}}. \newblock \emph{\bibinfo{journal}{\apj}} \textbf{\bibinfo{volume}{943}}, \bibinfo{pages}{42} (\bibinfo{year}{2023}).

\bibitem{Bonanos2016}
\bibinfo{author}{{Bonanos}, A.~.Z.} \& \bibinfo{author}{{Boumis}, P.}
\newblock \bibinfo{title}{{Evidence for rapid variability in the optical light curve of the Type Ia SN 2014J}}.
\newblock \emph{\bibinfo{journal}{\aap}} \textbf{\bibinfo{volume}{585}}, \bibinfo{pages}{A19} (\bibinfo{year}{2016}).

\bibitem{Gal-Yam2009}
\bibinfo{author}{{Gal-Yam}, A.} \emph{et~al.}
\newblock \bibinfo{title}{{Supernova 2007bi as a pair-instability explosion}}.
\newblock \emph{\bibinfo{journal}{\nat}} \textbf{\bibinfo{volume}{462}}, \bibinfo{pages}{624--627} (\bibinfo{year}{2009}).


\bibitem{Matheson2001}
\bibinfo{author}{{Matheson}, T.}, \bibinfo{author}{{Filippenko}, A.~V.}, \bibinfo{author}{{Li}, W.}, \bibinfo{author}{{Leonard}, D.~C.} \& \bibinfo{author}{{Shields}, J.~C.}
\newblock \bibinfo{title}{{Optical Spectroscopy of Type IB/C Supernovae}}.
\newblock \emph{\bibinfo{journal}{\aj}} \textbf{\bibinfo{volume}{121}}, \bibinfo{pages}{1648--1675} (\bibinfo{year}{2001}).

\bibitem{Mazzali2004}
\bibinfo{author}{{Mazzali}, P.~A.} \emph{et~al.}
\newblock \bibinfo{title}{{Properties of Two Hypernovae Entering the Nebular Phase: SN 1997ef and SN 1997dq}}.
\newblock \emph{\bibinfo{journal}{\apj}} \textbf{\bibinfo{volume}{614}}, \bibinfo{pages}{858--863} (\bibinfo{year}{2004}).

\bibitem{Milisavljevic2013}
\bibinfo{author}{{Milisavljevic}, D.} \emph{et~al.}
\newblock \bibinfo{title}{{SN 2012au: A Golden Link between Superluminous Supernovae and Their Lower-luminosity Counterparts}}.\newblock \emph{\bibinfo{journal}{\apjl}} \textbf{\bibinfo{volume}{770}}, \bibinfo{pages}{L38} (\bibinfo{year}{2013}).

\bibitem{Taddia2019}
\bibinfo{author}{{Taddia}, F.} \emph{et~al.}
\newblock \bibinfo{title}{{The luminous late-time emission of the type-Ic supernova iPTF15dtg - evidence for powering from a magnetar?}} \newblock \emph{\bibinfo{journal}{\aap}} \textbf{\bibinfo{volume}{621}}, \bibinfo{pages}{A64} (\bibinfo{year}{2019}).

\bibitem{Zdziarski1989}
\bibinfo{author}{{Zdziarski}, A.~A.} \& \bibinfo{author}{{Svensson}, R.}
\newblock \bibinfo{title}{{Absorption of X-Rays and Gamma Rays at Cosmological Distances}}.
\newblock \emph{\bibinfo{journal}{\apj}} \textbf{\bibinfo{volume}{344}}, \bibinfo{pages}{551} (\bibinfo{year}{1989}).

\bibitem{Acharyya2023}
\bibinfo{author}{{Acharyya}, A.} \emph{et~al.}
\newblock \bibinfo{title}{{VERITAS and Fermi-LAT Constraints on the Gamma-Ray Emission from Superluminous Supernovae SN2015bn and SN2017egm}}. \newblock \emph{\bibinfo{journal}{\apj}} \textbf{\bibinfo{volume}{945}}, \bibinfo{pages}{30} (\bibinfo{year}{2023}).

\bibitem{Chatzopoulos2012}
\bibinfo{author}{{Chatzopoulos}, E.} \& \bibinfo{author}{{Wheeler}, J.~C.}
\newblock \bibinfo{title}{{Hydrogen-poor Circumstellar Shells from Pulsational Pair-instability Supernovae with Rapidly Rotating Progenitors}}. \newblock \emph{\bibinfo{journal}{\apj}} \textbf{\bibinfo{volume}{760}}, \bibinfo{pages}{154} (\bibinfo{year}{2012}).

\bibitem{Chevalier2017}
\bibinfo{author}{{Chevalier}, R.~A.} \& \bibinfo{author}{{Fransson}, C.}
\newblock \bibinfo{title}{{Thermal and Non-thermal Emission from Circumstellar Interaction}}.
\newblock In \bibinfo{editor}{{Alsabti}, A.~W.} \& \bibinfo{editor}{{Murdin}, P.} (eds.) \emph{\bibinfo{booktitle}{Handbook of Supernovae}}, \bibinfo{pages}{875} (\bibinfo{year}{2017}).

\bibitem{Chen2017}
\bibinfo{author}{{Chen}, T.~W.} \emph{et~al.}
\newblock \bibinfo{title}{{The evolution of superluminous supernova LSQ14mo and its interacting host galaxy system}}.
\newblock \emph{\bibinfo{journal}{\aap}} \textbf{\bibinfo{volume}{602}}, \bibinfo{pages}{A9} (\bibinfo{year}{2017}).

\bibitem{Lau2022}
\bibinfo{author}{{Lau}, R.~M.} \emph{et~al.}
\newblock \bibinfo{title}{{Nested dust shells around the Wolf-Rayet binary WR 140 observed with JWST}}.
\newblock \emph{\bibinfo{journal}{Nature Astronomy}} \textbf{\bibinfo{volume}{6}}, \bibinfo{pages}{1308--1316}(\bibinfo{year}{2022}).

\bibitem{Ofek2014}
\bibinfo{author}{{Ofek}, E.~O.} \emph{et~al.}
\newblock \bibinfo{title}{{SN 2010jl: Optical to Hard X-Ray Observations Reveal an Explosion Embedded in a Ten Solar Mass Cocoon}}. \newblock \emph{\bibinfo{journal}{\apj}} \textbf{\bibinfo{volume}{781}}, \bibinfo{pages}{42} (\bibinfo{year}{2014}).

\bibitem{Zhu2023}
\bibinfo{author}{{Zhu}, J.} \emph{et~al.}
\newblock \bibinfo{title}{{SN2017egm: A Helium-rich Superluminous Supernova with Multiple Bumps in the Light Curves}}.
\newblock \emph{\bibinfo{journal}{\apj}} \textbf{\bibinfo{volume}{949}}, \bibinfo{pages}{23} (\bibinfo{year}{2023}).

\bibitem{Michel1988}
\bibinfo{author}{{Michel}, F.~C.}
\newblock \bibinfo{title}{{Neutron star disk formation from supernova fall-back and possible observational consequences}}.
\newblock \emph{\bibinfo{journal}{\nat}} \textbf{\bibinfo{volume}{333}}, \bibinfo{pages}{644--645} (\bibinfo{year}{1988}).

\bibitem{Chevalier1989}
\bibinfo{author}{{Chevalier}, R.~A.}
\newblock \bibinfo{title}{{Neutron Star Accretion in a Supernova}}. 
\newblock \emph{\bibinfo{journal}{\apj}} \textbf{\bibinfo{volume}{346}}, \bibinfo{pages}{847} (\bibinfo{year}{1989}).

\bibitem{Zhang2008}
\bibinfo{author}{{Zhang}, W.}, \bibinfo{author}{{Woosley}, S.~E.} \& \bibinfo{author}{{Heger}, A.}
\newblock \bibinfo{title}{{Fallback and Black Hole Production in Massive Stars}}. 
\newblock \emph{\bibinfo{journal}{\apj}} \textbf{\bibinfo{volume}{679}}, \bibinfo{pages}{639--654} (\bibinfo{year}{2008}).

\bibitem{Dexter2013}
\bibinfo{author}{{Dexter}, J.} \& \bibinfo{author}{{Kasen}, D.}
\newblock \bibinfo{title}{{Supernova Light Curves Powered by Fallback
  Accretion}}.
\newblock \emph{\bibinfo{journal}{\apj}} \textbf{\bibinfo{volume}{772}},
  \bibinfo{pages}{30} (\bibinfo{year}{2013}).

\bibitem{Moriya2018}
\bibinfo{author}{{Moriya}, T.~J.}, \bibinfo{author}{{Nicholl}, M.} \& \bibinfo{author}{{Guillochon}, J.}
\newblock \bibinfo{title}{{Systematic Investigation of the Fallback Accretion-powered Model for Hydrogen-poor Superluminous Supernovae}}.
\newblock \emph{\bibinfo{journal}{\apj}} \textbf{\bibinfo{volume}{867}}, \bibinfo{pages}{113} (\bibinfo{year}{2018}).

\bibitem{Moriya2019}
\bibinfo{author}{{Moriya}, T.~J.}, \bibinfo{author}{{M{\"u}ller}, B.}, \bibinfo{author}{{Chan}, C.}, \bibinfo{author}{{Heger}, A.} \& \bibinfo{author}{{Blinnikov}, S.~I.}
\newblock \bibinfo{title}{{Fallback Accretion-powered Supernova Light Curves Based on a Neutrino-driven Explosion Simulation of a 40 M $_{{\ensuremath{\odot}}}$ Star}}.
\newblock \emph{\bibinfo{journal}{\apj}} \textbf{\bibinfo{volume}{880}}, \bibinfo{pages}{21} (\bibinfo{year}{2019}).

\bibitem{Zanin2016}
\bibinfo{author}{{Zanin}, R.} \emph{et~al.}
\newblock \bibinfo{title}{{Gamma rays detected from Cygnus X-1 with likely jet origin}}.
\newblock \emph{\bibinfo{journal}{\aap}} \textbf{\bibinfo{volume}{596}}, \bibinfo{pages}{A55} (\bibinfo{year}{2016}).

\bibitem{Akashi2020}
\bibinfo{author}{{Akashi}, M.} \& \bibinfo{author}{{Soker}, N.}
\newblock \bibinfo{title}{{Simulating Jets from a Neutron Star Companion Hours after a Core-collapse Supernova}}.
\newblock \emph{\bibinfo{journal}{\apj}} \textbf{\bibinfo{volume}{901}}, \bibinfo{pages}{53} (\bibinfo{year}{2020}).

\bibitem{Hober2022}
\bibinfo{author}{{Hober}, O.}, \bibinfo{author}{{Bear}, E.} \& \bibinfo{author}{{Soker}, N.}
\newblock \bibinfo{title}{{Feeding post-core collapse supernova explosion jets with an inflated main sequence companion}}.
\newblock \emph{\bibinfo{journal}{\mnras}} \textbf{\bibinfo{volume}{516}}, \bibinfo{pages}{1846--1854} (\bibinfo{year}{2022}).

\bibitem{Renzo2019}
\bibinfo{author}{{Renzo}, M.} \emph{et~al.}
\newblock \bibinfo{title}{{Massive runaway and walkaway stars. A study of the kinematical imprints of the physical processes governing the evolution and explosion of their binary progenitors}}.
\newblock \emph{\bibinfo{journal}{\aap}} \textbf{\bibinfo{volume}{624}}, \bibinfo{pages}{A66} (\bibinfo{year}{2019}).

\bibitem{Chrimes2022}
\bibinfo{author}{{Chrimes}, A.~A.} \emph{et~al.}
\newblock \bibinfo{title}{{Where are the magnetar binary companions? Candidates from a comparison with binary population synthesis predictions}}.
\newblock \emph{\bibinfo{journal}{\mnras}} \textbf{\bibinfo{volume}{513}}, \bibinfo{pages}{3550--3563} (\bibinfo{year}{2022}).

\end{thebibliography}

\noindent{\bf References}

\vspace{1cm}
\begin{thebibliography}{10}
\expandafter\ifx\csname url\endcsname\relax
  \def\url#1{\texttt{#1}}\fi
\expandafter\ifx\csname urlprefix\endcsname\relax\def\urlprefix{URL }\fi
\providecommand{\bibinfo}[2]{#2}
\providecommand{\eprint}[2][]{\url{#2}}
\makeatletter
\addtocounter{\@listctr}{34}
\makeatother

\bibitem{Monard2022}
\bibinfo{author}{{Monard}, L.}
\newblock \bibinfo{title}{{Transient Discovery Report for 2022-05-05}}.
\newblock \emph{\bibinfo{journal}{Transient Name Server Discovery Report}} \textbf{\bibinfo{volume}{2022-1198}}, \bibinfo{pages}{1} (\bibinfo{year}{2022}).

\bibitem{Tonry2018}
\bibinfo{author}{{Tonry}, J.~L.} \emph{et~al.}
\newblock \bibinfo{title}{{ATLAS: A High-cadence All-sky Survey System}}.
\newblock \emph{\bibinfo{journal}{\pasp}} \textbf{\bibinfo{volume}{130}},
  \bibinfo{pages}{064505} (\bibinfo{year}{2018}).

\bibitem{Hodgkin2021}
\bibinfo{author}{{Hodgkin}, S.~T.} \emph{et~al.}
\newblock \bibinfo{title}{{Gaia Early Data Release 3. Gaia photometric science
  alerts}}.
\newblock \emph{\bibinfo{journal}{\aap}} \textbf{\bibinfo{volume}{652}},
  \bibinfo{pages}{A76} (\bibinfo{year}{2021}).

\bibitem{Chambers2016}
\bibinfo{author}{{Chambers}, K.~C.} \emph{et~al.}
\newblock \bibinfo{title}{{The Pan-STARRS1 Surveys}}.
\newblock \emph{\bibinfo{journal}{arXiv e-prints}}
  \bibinfo{pages}{arXiv:1612.05560} (\bibinfo{year}{2016}).

\bibitem{Bellm2019a}
\bibinfo{author}{{Bellm}, E.~C.} \emph{et~al.}
\newblock \bibinfo{title}{{The Zwicky Transient Facility: System Overview,
  Performance, and First Results}}.
\newblock \emph{\bibinfo{journal}{\pasp}} \textbf{\bibinfo{volume}{131}},
  \bibinfo{pages}{018002} (\bibinfo{year}{2019}).

\bibitem{Graham2019}
\bibinfo{author}{{Graham}, M.~J.} \emph{et~al.}
\newblock \bibinfo{title}{{The Zwicky Transient Facility: Science Objectives}}.
\newblock \emph{\bibinfo{journal}{\pasp}} \textbf{\bibinfo{volume}{131}},
  \bibinfo{pages}{078001} (\bibinfo{year}{2019}).

\bibitem{Mould2000}
\bibinfo{author}{{Mould}, J.~R.} \emph{et~al.}
\newblock \bibinfo{title}{{The Hubble Space Telescope Key Project on the
  Extragalactic Distance Scale. XXVIII. Combining the Constraints on the Hubble
  Constant}}.
\newblock \emph{\bibinfo{journal}{\apj}} \textbf{\bibinfo{volume}{529}},
  \bibinfo{pages}{786--794} (\bibinfo{year}{2000}).

\bibitem{Komatsu2011}
\bibinfo{author}{{Komatsu}, E.} \emph{et~al.}
\newblock \bibinfo{title}{{Seven-year Wilkinson Microwave Anisotropy Probe (WMAP) Observations: Cosmological Interpretation}}.
\newblock \emph{\bibinfo{journal}{\apjs}} \textbf{\bibinfo{volume}{192}}, \bibinfo{pages}{18} (\bibinfo{year}{2011}).

\bibitem{Riess2022}
\bibinfo{author}{{Riess}, A.~G.} \emph{et~al.}
\newblock \bibinfo{title}{{A Comprehensive Measurement of the Local Value of the Hubble Constant with 1 km s$^{-1}$ Mpc$^{-1}$ Uncertainty from the Hubble Space Telescope and the SH0ES Team}}.
\newblock \emph{\bibinfo{journal}{\apjl}} \textbf{\bibinfo{volume}{934}},
  \bibinfo{pages}{L7} (\bibinfo{year}{2022}).

\bibitem{Nasonova2011}
\bibinfo{author}{{Nasonova}, O.~G.}, \bibinfo{author}{{de Freitas Pacheco},
  J.~A.} \& \bibinfo{author}{{Karachentsev}, I.~D.}
\newblock \bibinfo{title}{{Hubble flow around Fornax cluster of galaxies}}.
\newblock \emph{\bibinfo{journal}{\aap}} \textbf{\bibinfo{volume}{532}},
  \bibinfo{pages}{A104} (\bibinfo{year}{2011}).

\bibitem{Tully2013}
\bibinfo{author}{{Tully}, R.~B.} \emph{et~al.}
\newblock \bibinfo{title}{{Cosmicflows-2: The Data}}.
\newblock \emph{\bibinfo{journal}{\aj}} \textbf{\bibinfo{volume}{146}},
  \bibinfo{pages}{86} (\bibinfo{year}{2013}).

\bibitem{Erwin2017}
\bibinfo{author}{{Erwin}, P.} \& \bibinfo{author}{{Debattista}, V.~P.}
\newblock \bibinfo{title}{{The frequency and stellar-mass dependence of boxy/peanut-shaped bulges in barred galaxies}}.
\newblock \emph{\bibinfo{journal}{\mnras}} \textbf{\bibinfo{volume}{468}}, \bibinfo{pages}{2058--2080} (\bibinfo{year}{2017}).

\bibitem{Poznanski2012}
\bibinfo{author}{{Poznanski}, D.}, \bibinfo{author}{{Prochaska}, J.~X.} \&
  \bibinfo{author}{{Bloom}, J.~S.}
\newblock \bibinfo{title}{{An empirical relation between sodium absorption and
  dust extinction}}.
\newblock \emph{\bibinfo{journal}{\mnras}} \textbf{\bibinfo{volume}{426}},
  \bibinfo{pages}{1465--1474} (\bibinfo{year}{2012}).

\bibitem{Lan2015}
\bibinfo{author}{{Lan}, T.-W.}, \bibinfo{author}{{M{\'e}nard}, B.} \&
  \bibinfo{author}{{Zhu}, G.}
\newblock \bibinfo{title}{{Exploring the diffuse interstellar bands with the
  Sloan Digital Sky Survey}}.
\newblock \emph{\bibinfo{journal}{\mnras}} \textbf{\bibinfo{volume}{452}},
  \bibinfo{pages}{3629--3649} (\bibinfo{year}{2015}).

\bibitem{Fan2019}
\bibinfo{author}{{Fan}, H.} \emph{et~al.}
\newblock \bibinfo{title}{{The Apache Point Observatory Catalog of Optical
  Diffuse Interstellar Bands}}.
\newblock \emph{\bibinfo{journal}{\apj}} \textbf{\bibinfo{volume}{878}},
  \bibinfo{pages}{151} (\bibinfo{year}{2019}).

\bibitem{Dekany2020}
\bibinfo{author}{{Dekany}, R.} \emph{et~al.}
\newblock \bibinfo{title}{{The Zwicky Transient Facility: Observing System}}.
\newblock \emph{\bibinfo{journal}{\pasp}} \textbf{\bibinfo{volume}{132}},
  \bibinfo{pages}{038001} (\bibinfo{year}{2020}).

\bibitem{Masci2019}
\bibinfo{author}{{Masci}, F.~J.} \emph{et~al.}
\newblock \bibinfo{title}{{The Zwicky Transient Facility: Data Processing,
  Products, and Archive}}.
\newblock \emph{\bibinfo{journal}{\pasp}} \textbf{\bibinfo{volume}{131}},
  \bibinfo{pages}{018003} (\bibinfo{year}{2019}).


\bibitem{Zackay2016}
\bibinfo{author}{{Zackay}, B.}, \bibinfo{author}{{Ofek}, E.~O.} \&
  \bibinfo{author}{{Gal-Yam}, A.}
\newblock \bibinfo{title}{{Proper Image Subtraction{\textemdash}Optimal
  Transient Detection, Photometry, and Hypothesis Testing}}.
\newblock \emph{\bibinfo{journal}{\apj}} \textbf{\bibinfo{volume}{830}},
  \bibinfo{pages}{27} (\bibinfo{year}{2016}).

\bibitem{Cenko2006}
\bibinfo{author}{{Cenko}, S.~B.} \emph{et~al.}
\newblock \bibinfo{title}{{The Automated Palomar 60 Inch Telescope}}.
\newblock \emph{\bibinfo{journal}{\pasp}} \textbf{\bibinfo{volume}{118}},
  \bibinfo{pages}{1396--1406} (\bibinfo{year}{2006}).

\bibitem{Blagorodnova2018}
\bibinfo{author}{{Blagorodnova}, N.} \emph{et~al.}
\newblock \bibinfo{title}{{The SED Machine: A Robotic Spectrograph for Fast
  Transient Classification}}.
\newblock \emph{\bibinfo{journal}{\pasp}} \textbf{\bibinfo{volume}{130}},
  \bibinfo{pages}{035003} (\bibinfo{year}{2018}).

\bibitem{Fremling2016}
\bibinfo{author}{{Fremling}, C.} \emph{et~al.}
\newblock \bibinfo{title}{{PTF12os and iPTF13bvn. Two stripped-envelope
  supernovae from low-mass progenitors in NGC 5806}}.
\newblock \emph{\bibinfo{journal}{\aap}} \textbf{\bibinfo{volume}{593}},
  \bibinfo{pages}{A68} (\bibinfo{year}{2016}).

\bibitem{vanderWalt2019}
\bibinfo{author}{{van der Walt}, S.}, \bibinfo{author}{{Crellin-Quick}, A.} \&
  \bibinfo{author}{{Bloom}, J.}
\newblock \bibinfo{title}{{SkyPortal: An Astronomical Data Platform}}.
\newblock \emph{\bibinfo{journal}{The Journal of Open Source Software}}
  \textbf{\bibinfo{volume}{4}}, \bibinfo{pages}{1247} (\bibinfo{year}{2019}).

\bibitem{Coughlin2023}
\bibinfo{author}{{Coughlin}, M.~W.} \emph{et~al.}
\newblock \bibinfo{title}{{A data science platform to enable time-domain astronomy}}.
\newblock \emph{\bibinfo{journal}{\apjs}} \textbf{\bibinfo{volume}{267}},
  \bibinfo{pages}{31} (\bibinfo{year}{2023}).
  

\bibitem{Shappee2014}
\bibinfo{author}{{Shappee}, B.~J.} \emph{et~al.}
\newblock \bibinfo{title}{{The Man behind the Curtain: X-Rays Drive the UV through NIR Variability in the 2013 Active Galactic Nucleus Outburst in NGC 2617}}.
\newblock \emph{\bibinfo{journal}{\apj}} \textbf{\bibinfo{volume}{788}},
  \bibinfo{pages}{48} (\bibinfo{year}{2014}).

\bibitem{Kochanek2017}
\bibinfo{author}{{Kochanek}, C.~S.} \emph{et~al.}
\newblock \bibinfo{title}{{The All-Sky Automated Survey for Supernovae (ASAS-SN) Light Curve Server v1.0}}.
\newblock \emph{\bibinfo{journal}{\pasp}} \textbf{\bibinfo{volume}{129}},
  \bibinfo{pages}{104502} (\bibinfo{year}{2017}).

\bibitem{atlasfp}
\bibinfo{title}{\url{https://fallingstar-data.com/forcedphot/}}

\bibitem{Smith2020}
\bibinfo{author}{{Smith}, K.~W.} \emph{et~al.}
\newblock \bibinfo{title}{{Design and Operation of the ATLAS Transient Science Server}}.
\newblock \emph{\bibinfo{journal}{\pasp}} \textbf{\bibinfo{volume}{132}},
  \bibinfo{pages}{085002} (\bibinfo{year}{2020}).

\bibitem{gaiaalert}
\bibinfo{title}{\url{http://gsaweb.ast.cam.ac.uk/alerts/alert/Gaia22cbu/}}

\bibitem{sdssdr12}
\bibinfo{title}{\url{https://dr12.sdss.org/mosaics}}

\bibitem{Chen2022}
\bibinfo{author}{{Chen}, P.} \emph{et~al.}
\newblock \bibinfo{title}{{The First Data Release of CNIa0.02-A Complete Nearby (Redshift <0.02) Sample of Type Ia Supernova Light Curves}}.
\newblock \emph{\bibinfo{journal}{\apjs}} \textbf{\bibinfo{volume}{259}},
  \bibinfo{pages}{53} (\bibinfo{year}{2022}).

\bibitem{LiuXW2019}
\bibinfo{author}{{Liu}, X.}
\newblock \bibinfo{title}{{Multi-channel Photometric Survey Telescope - Mephisto}}.
\newblock In \emph{\bibinfo{booktitle}{Galactic Archaeology in the Gaia Era}},
  \bibinfo{pages}{14} (\bibinfo{year}{2019}).

\bibitem{Yuan2020}
\bibinfo{author}{{Yuan}, X.} \emph{et~al.}
\newblock \bibinfo{title}{{Development of the Multi-channel Photometric Survey telescope}}.
\newblock In \emph{\bibinfo{booktitle}{Society of Photo-Optical Instrumentation Engineers (SPIE) Conference Series}}, vol. \bibinfo{volume}{11445} of
  \emph{\bibinfo{series}{Society of Photo-Optical Instrumentation Engineers (SPIE) Conference Series}}, \bibinfo{pages}{114457M} (\bibinfo{year}{2020}).

\bibitem{wiserep}
\bibinfo{title}{\url{http://wiserep.weizmann.ac.il/}}

\bibitem{Yaron2012}
\bibinfo{author}{{Yaron}, O.} \& \bibinfo{author}{{Gal-Yam}, A.}
\newblock \bibinfo{title}{{WISeREP{\textemdash}An Interactive Supernova Data Repository}}.
\newblock \emph{\bibinfo{journal}{\pasp}} \textbf{\bibinfo{volume}{124}}, \bibinfo{pages}{668} (\bibinfo{year}{2012}).

\bibitem{BenAmi2012}
\bibinfo{author}{{Ben-Ami}, S.} \emph{et~al.}
\newblock \bibinfo{title}{{The SED Machine: a dedicated transient IFU
  spectrograph}}.
\newblock In \bibinfo{editor}{{McLean}, I.~S.}, \bibinfo{editor}{{Ramsay},
  S.~K.} \& \bibinfo{editor}{{Takami}, H.} (eds.)
  \emph{\bibinfo{booktitle}{Ground-based and Airborne Instrumentation for Astronomy IV}}, vol. \bibinfo{volume}{8446} of \emph{\bibinfo{series}{Society of Photo-Optical Instrumentation Engineers (SPIE) Conference Series}},
  \bibinfo{pages}{844686} (\bibinfo{year}{2012}).

\bibitem{Rigault2019}
\bibinfo{author}{{Rigault}, M.} \emph{et~al.}
\newblock \bibinfo{title}{{Fully automated integral field spectrograph pipeline for the SEDMachine: pysedm}}.
\newblock \emph{\bibinfo{journal}{\aap}} \textbf{\bibinfo{volume}{627}},
  \bibinfo{pages}{A115} (\bibinfo{year}{2019}).

\bibitem{Kim2022}
\bibinfo{author}{{Kim}, Y.~L.} \emph{et~al.}
\newblock \bibinfo{title}{{New Modules for the SEDMachine to Remove Contaminations from Cosmic Rays and Non-target Light: BYECR and CONTSEP}}.
\newblock \emph{\bibinfo{journal}{\pasp}} \textbf{\bibinfo{volume}{134}},
  \bibinfo{pages}{024505} (\bibinfo{year}{2022}).

\bibitem{foscgui}
\bibinfo{title}{\url{http://sngroup.oapd.inaf.it/foscgui.html}}

\bibitem{Oke1982a}
\bibinfo{author}{{Oke}, J.~B.} \& \bibinfo{author}{{Gunn}, J.~E.}
\newblock \bibinfo{title}{{An Efficient Low Resolution and Moderate Resolution Spectrograph for the Hale Telescope}}.
\newblock \emph{\bibinfo{journal}{\pasp}} \textbf{\bibinfo{volume}{94}},
  \bibinfo{pages}{586} (\bibinfo{year}{1982}).

\bibitem{dbspdrp}
\bibinfo{title}{\url{https://github.com/finagle29/dbsp\_drp}}

\bibitem{Prochaska2020b}
\bibinfo{author}{{Prochaska}, J.} \emph{et~al.}
\newblock \bibinfo{title}{{PypeIt: The Python Spectroscopic Data Reduction Pipeline}}.
\newblock \emph{\bibinfo{journal}{The Journal of Open Source Software}}
  \textbf{\bibinfo{volume}{5}}, \bibinfo{pages}{2308} (\bibinfo{year}{2020}).

\bibitem{Prochaska2020a}
\bibinfo{author}{{Prochaska}, J.~X.} \emph{et~al.}
\newblock \bibinfo{title}{{pypeit/PypeIt: Release 1.0.0}}
  (\bibinfo{year}{2020}).

\bibitem{Fabricant2019}
\bibinfo{author}{{Fabricant}, D.} \emph{et~al.}
\newblock \bibinfo{title}{{Binospec: A Wide-field Imaging Spectrograph for the MMT}}.
\newblock \emph{\bibinfo{journal}{\pasp}} \textbf{\bibinfo{volume}{131}},
  \bibinfo{pages}{075004} (\bibinfo{year}{2019}).

\bibitem{Kansky2019}
\bibinfo{author}{{Kansky}, J.} \emph{et~al.}
\newblock \bibinfo{title}{{Binospec Software System}}.
\newblock \emph{\bibinfo{journal}{\pasp}} \textbf{\bibinfo{volume}{131}},
  \bibinfo{pages}{075005} (\bibinfo{year}{2019}).

\bibitem{Simcoe2013}
\bibinfo{author}{{Simcoe}, R.~A.} \emph{et~al.}
\newblock \bibinfo{title}{{FIRE: A Facility Class Near-Infrared Echelle Spectrometer for the Magellan Telescopes}}.
\newblock \emph{\bibinfo{journal}{\pasp}} \textbf{\bibinfo{volume}{125}},
  \bibinfo{pages}{270} (\bibinfo{year}{2013}).

\bibitem{Dressler2011}
\bibinfo{author}{{Dressler}, A.} \emph{et~al.}
\newblock \bibinfo{title}{{IMACS: The Inamori-Magellan Areal Camera and Spectrograph on Magellan-Baade}}.
\newblock \emph{\bibinfo{journal}{\pasp}} \textbf{\bibinfo{volume}{123}},
  \bibinfo{pages}{288} (\bibinfo{year}{2011}).

\bibitem{Vernet2011}
\bibinfo{author}{{Vernet}, J.} \emph{et~al.}
\newblock \bibinfo{title}{{X-shooter, the new wide band intermediate resolution spectrograph at the ESO Very Large Telescope}}.
\newblock \emph{\bibinfo{journal}{\aap}} \textbf{\bibinfo{volume}{536}},
  \bibinfo{pages}{A105} (\bibinfo{year}{2011}).


\bibitem{astroscrappy}
\bibinfo{title}{\url{https://github.com/astropy/astroscrappy}}

\bibitem{vanDokkum2001}
\bibinfo{author}{{van Dokkum}, P.~G.}
\newblock \bibinfo{title}{{Cosmic-Ray Rejection by Laplacian Edge Detection}}.
\newblock \emph{\bibinfo{journal}{\pasp}} \textbf{\bibinfo{volume}{113}},
  \bibinfo{pages}{1420--1427} (\bibinfo{year}{2001}).

\bibitem{Goldoni2006}
\bibinfo{author}{{Goldoni}, P.} \emph{et~al.}
\newblock \bibinfo{title}{{Data reduction software of the X-shooter
  spectrograph}}.
\newblock In \bibinfo{editor}{{McLean}, I.~S.} \& \bibinfo{editor}{{Iye}, M.}
  (eds.) \emph{\bibinfo{booktitle}{Society of Photo-Optical Instrumentation Engineers (SPIE) Conference Series}}, vol. \bibinfo{volume}{6269} of
  \emph{\bibinfo{series}{Society of Photo-Optical Instrumentation Engineers (SPIE) Conference Series}}, \bibinfo{pages}{62692K} (\bibinfo{year}{2006}).

\bibitem{Modigliani2010}
\bibinfo{author}{{Modigliani}, A.} \emph{et~al.}
\newblock \bibinfo{title}{{The X-shooter pipeline}}.
\newblock In \bibinfo{editor}{{Silva}, D.~R.}, \bibinfo{editor}{{Peck}, A.~B.} \& \bibinfo{editor}{{Soifer}, B.~T.} (eds.)
  \emph{\bibinfo{booktitle}{Observatory Operations: Strategies, Processes, and Systems III}}, vol. \bibinfo{volume}{7737} of \emph{\bibinfo{series}{Society of Photo-Optical Instrumentation Engineers (SPIE) Conference Series}},
  \bibinfo{pages}{773728} (\bibinfo{year}{2010}).

\bibitem{Freudling2013}
\bibinfo{author}{{Freudling}, W.} \emph{et~al.}
\newblock \bibinfo{title}{{Automated data reduction workflows for astronomy. The ESO Reflex environment}}.
\newblock \emph{\bibinfo{journal}{\aap}} \textbf{\bibinfo{volume}{559}},
  \bibinfo{pages}{A96} (\bibinfo{year}{2013}).

\bibitem{Smette2015}
\bibinfo{author}{{Smette}, A.} \emph{et~al.}
\newblock \bibinfo{title}{{Molecfit: A general tool for telluric absorption correction. I. Method and application to ESO instruments}}.
\newblock \emph{\bibinfo{journal}{\aap}} \textbf{\bibinfo{volume}{576}},
  \bibinfo{pages}{A77} (\bibinfo{year}{2015}).

\bibitem{VanderPlas2015}
\bibinfo{author}{{VanderPlas}, J.~T.} \& \bibinfo{author}{{Ivezi{\'c}},
  {\v{Z}}.}
\newblock \bibinfo{title}{{Periodograms for Multiband Astronomical Tim Series}}.
\newblock \emph{\bibinfo{journal}{\apj}} \textbf{\bibinfo{volume}{812}}, \bibinfo{pages}{18} (\bibinfo{year}{2015}).

\bibitem{Lomb1976}
\bibinfo{author}{{Lomb}, N.~R.}
\newblock \bibinfo{title}{{Least-Squares Frequency Analysis of Unequally Spaced Data}}.
\newblock \emph{\bibinfo{journal}{\apss}} \textbf{\bibinfo{volume}{39}},
  \bibinfo{pages}{447--462} (\bibinfo{year}{1976}).

\bibitem{Scargle1982}
\bibinfo{author}{{Scargle}, J.~D.}
\newblock \bibinfo{title}{{Studies in astronomical time series analysis. II. Statistical aspects of spectral analysis of unevenly spaced data.}}
\newblock \emph{\bibinfo{journal}{\apj}} \textbf{\bibinfo{volume}{263}},
  \bibinfo{pages}{835--853} (\bibinfo{year}{1982}).

\bibitem{VanderPlas2018}
\bibinfo{author}{{VanderPlas}, J.~T.}
\newblock \bibinfo{title}{{Understanding the Lomb-Scargle Periodogram}}.
\newblock \emph{\bibinfo{journal}{\apjs}} \textbf{\bibinfo{volume}{236}},
  \bibinfo{pages}{16} (\bibinfo{year}{2018}).

\bibitem{gatspy}
\bibinfo{title}{\url{http:/www.astroml.org/gatspy/}}

\bibitem{Lyman2014}
\bibinfo{author}{{Lyman}, J.~D.}, \bibinfo{author}{{Bersier}, D.} \&
  \bibinfo{author}{{James}, P.~A.}
\newblock \bibinfo{title}{{Bolometric corrections for optical light curves of core-collapse supernovae}}.
\newblock \emph{\bibinfo{journal}{\mnras}} \textbf{\bibinfo{volume}{437}},
  \bibinfo{pages}{3848--3862} (\bibinfo{year}{2014}).

\bibitem{Schlafly2011}
\bibinfo{author}{{Schlafly}, E.~F.} \& \bibinfo{author}{{Finkbeiner}, D.~P.}
\newblock \bibinfo{title}{{Measuring Reddening with Sloan Digital Sky Survey Stellar Spectra and Recalibrating SFD}}.
\newblock \emph{\bibinfo{journal}{\apj}} \textbf{\bibinfo{volume}{737}},
  \bibinfo{pages}{103} (\bibinfo{year}{2011}).

\bibitem{Cardelli1989}
\bibinfo{author}{{Cardelli}, J.~A.}, \bibinfo{author}{{Clayton}, G.~C.} \&
  \bibinfo{author}{{Mathis}, J.~S.}
\newblock \bibinfo{title}{{The Relationship between Infrared, Optical, and Ultraviolet Extinction}}.
\newblock \emph{\bibinfo{journal}{\apj}} \textbf{\bibinfo{volume}{345}},
  \bibinfo{pages}{245} (\bibinfo{year}{1989}).

\bibitem{Sharon2020}
\bibinfo{author}{{Sharon}, A.} \& \bibinfo{author}{{Kushnir}, D.}
\newblock \bibinfo{title}{{The {\ensuremath{\gamma}}-ray deposition histories of core-collapse supernovae}}.
\newblock \emph{\bibinfo{journal}{\mnras}} \textbf{\bibinfo{volume}{496}},
  \bibinfo{pages}{4517--4545} (\bibinfo{year}{2020}).

\bibitem{Folatelli2006}
\bibinfo{author}{{Folatelli}, G.} \emph{et~al.}
\newblock \bibinfo{title}{{SN 2005bf: A Possible Transition Event between Type Ib/c Supernovae and Gamma-Ray Bursts}}.
\newblock \emph{\bibinfo{journal}{\apj}} \textbf{\bibinfo{volume}{641}},
  \bibinfo{pages}{1039--1050} (\bibinfo{year}{2006}).

\bibitem{Taddia2018}
\bibinfo{author}{{Taddia}, F.} \emph{et~al.}
\newblock \bibinfo{title}{{PTF11mnb: First analog of supernova 2005bf. Long-rising, double-peaked supernova Ic from a massive progenitor}}.
\newblock \emph{\bibinfo{journal}{\aap}} \textbf{\bibinfo{volume}{609}},
  \bibinfo{pages}{A106} (\bibinfo{year}{2018}).

\bibitem{Gutierrez2021}
\bibinfo{author}{{Guti{\'e}rrez}, C.~P.} \emph{et~al.}
\newblock \bibinfo{title}{{The double-peaked Type Ic supernova 2019cad: another SN 2005bf-like object}}.
\newblock \emph{\bibinfo{journal}{\mnras}} \textbf{\bibinfo{volume}{504}},
  \bibinfo{pages}{4907--4922} (\bibinfo{year}{2021}).

\bibitem{Gomez2021}
\bibinfo{author}{{Gomez}, S.} \emph{et~al.}
\newblock \bibinfo{title}{{The Luminous and Double-peaked Type Ic Supernova 2019stc: Evidence for Multiple Energy Sources}}.
\newblock \emph{\bibinfo{journal}{\apj}} \textbf{\bibinfo{volume}{913}},
  \bibinfo{pages}{143} (\bibinfo{year}{2021}).

\bibitem{Chugai2022}
\bibinfo{author}{{Chugai}, N.~N.} \& \bibinfo{author}{{Utrobin}, V.~P.}
\newblock \bibinfo{title}{{Origin of post-maximum bump in luminous Type Ic supernova 2019stc}}.
\newblock \emph{\bibinfo{journal}{\mnras}} \textbf{\bibinfo{volume}{512}},
  \bibinfo{pages}{L71--L73} (\bibinfo{year}{2022}).

\bibitem{Gomez2022}
\bibinfo{author}{{Gomez}, S.}, \bibinfo{author}{{Berger}, E.}, \bibinfo{author}{{Nicholl}, M.}, \bibinfo{author}{{Blanchard}, P.~K.} \& \bibinfo{author}{{Hosseinzadeh}, G.}
\newblock \bibinfo{title}{{Luminous Supernovae: Unveiling a Population Between Superluminous and Normal Core-collapse Supernovae}}.
\newblock \emph{\bibinfo{journal}{\apj}} \textbf{\bibinfo{volume}{941}},
  \bibinfo{pages}{107} (\bibinfo{year}{2022}).


\bibitem{Kuncarayakti2023}
\bibinfo{author}{{Kuncarayakti}, H.} \emph{et~al.}
\newblock \bibinfo{title}{{The broad-lined Type-Ic supernova SN 2022xxf with extraordinary two-humped light curves}}.
\newblock \emph{\bibinfo{journal}{arXiv e-prints}}
  \bibinfo{pages}{arXiv:2303.16925} (\bibinfo{year}{2023}).

\bibitem{Das2023}
\bibinfo{author}{{Das}, K.~K.} \emph{et~al.}
\newblock \bibinfo{title}{{Probing pre-supernova mass loss in double-peaked Type Ibc supernovae from the Zwicky Transient Facility}}.
\newblock \emph{\bibinfo{journal}{arXiv e-prints}} \bibinfo{pages}{arXiv:2306.04698} (\bibinfo{year}{2023}).


\bibitem{Piro2015}
\bibinfo{author}{{Piro}, A.~L.}
\newblock \bibinfo{title}{{Using Double-peaked Supernova Light Curves to Study
  Extended Material}}.
\newblock \emph{\bibinfo{journal}{\apjl}} \textbf{\bibinfo{volume}{808}},
  \bibinfo{pages}{L51} (\bibinfo{year}{2015}).

\bibitem{Jin2021}
\bibinfo{author}{{Jin}, H.}, \bibinfo{author}{{Yoon}, S.-C.} \&
  \bibinfo{author}{{Blinnikov}, S.}
\newblock \bibinfo{title}{{The Effect of Circumstellar Matter on the
  Double-peaked Type Ic Supernovae and Implications for LSQ14efd, iPTF15dtg,
  and SN 2020bvc}}.
\newblock \emph{\bibinfo{journal}{\apj}} \textbf{\bibinfo{volume}{910}},
  \bibinfo{pages}{68} (\bibinfo{year}{2021}).

\bibitem{Orellana2022}
\bibinfo{author}{{Orellana}, M.} \& \bibinfo{author}{{Bersten}, M.~C.}
\newblock \bibinfo{title}{{Supernova double-peaked light curves from
  double-nickel distribution}}.
\newblock \emph{\bibinfo{journal}{\aap}} \textbf{\bibinfo{volume}{667}},
  \bibinfo{pages}{A92} (\bibinfo{year}{2022}).

\bibitem{Bersten2013}
\bibinfo{author}{{Bersten}, M.~C.}, \bibinfo{author}{{Tanaka}, M.},
  \bibinfo{author}{{Tominaga}, N.}, \bibinfo{author}{{Benvenuto}, O.~G.} \&
  \bibinfo{author}{{Nomoto}, K.}
\newblock \bibinfo{title}{{Early Ultraviolet/Optical Emission of The Type Ib SN
  2008D}}.
\newblock \emph{\bibinfo{journal}{\apj}} \textbf{\bibinfo{volume}{767}},
  \bibinfo{pages}{143} (\bibinfo{year}{2013}).

\bibitem{Walton2011}
\bibinfo{author}{{Walton}, D.~J.}, \bibinfo{author}{{Roberts}, T.~P.}, \bibinfo{author}{{Mateos}, S.} \& \bibinfo{author}{{Heard}, V.}
\newblock \bibinfo{title}{{2XMM ultraluminous X-ray source candidates in nearby galaxies}}.
\newblock \emph{\bibinfo{journal}{\mnras}} \textbf{\bibinfo{volume}{416}},
  \bibinfo{pages}{1844--1861} (\bibinfo{year}{2011}).

\bibitem{Israel2017}
\bibinfo{author}{{Israel}, G.~L.} \emph{et~al.}
\newblock \bibinfo{title}{{An accreting pulsar with extreme properties drives an ultraluminous x-ray source in NGC 5907}}.
\newblock \emph{\bibinfo{journal}{Science}} \textbf{\bibinfo{volume}{355}}, \bibinfo{pages}{817--819} (\bibinfo{year}{2017}).


\bibitem{Brightman2019}
\bibinfo{author}{{Brightman}, M.} \emph{et~al.}
\newblock \bibinfo{title}{{Breaking the limit: Super-Eddington accretion onto
  black holes and neutron stars}}.
\newblock \emph{\bibinfo{journal}{\baas}} \textbf{\bibinfo{volume}{51}},
  \bibinfo{pages}{352} (\bibinfo{year}{2019}).

\bibitem{Grzegorzek2022}
\bibinfo{author}{{Grzegorzek}, J.}
\newblock \bibinfo{title}{{PSH Transient Classification Report for
  2022-05-11}}.
\newblock \emph{\bibinfo{journal}{Transient Name Server Classification Report}}
  \textbf{\bibinfo{volume}{2022-1261}}, \bibinfo{pages}{1}
  (\bibinfo{year}{2022}).

\bibitem{Cosentino2022}
\bibinfo{author}{{Cosentino}, S.~P.} \emph{et~al.}
\newblock \bibinfo{title}{{ePESSTO+ Transient Classification Report for 2022-05-24}}.
\newblock \emph{\bibinfo{journal}{Transient Name Server Classification Report}}
  \textbf{\bibinfo{volume}{2022-1409}}, \bibinfo{pages}{1}
  (\bibinfo{year}{2022}).

\bibitem{Drout2016}
\bibinfo{author}{{Drout}, M.~R.} \emph{et~al.}
\newblock \bibinfo{title}{{The Double-peaked SN 2013ge: A Type Ib/c SN with an Asymmetric Mass Ejection or an Extended Progenitor Envelope}}.
\newblock \emph{\bibinfo{journal}{\apj}} \textbf{\bibinfo{volume}{821}}, \bibinfo{pages}{57} (\bibinfo{year}{2016}).

\bibitem{Hachinger2012}
\bibinfo{author}{{Hachinger}, S.} \emph{et~al.}
\newblock \bibinfo{title}{{How much H and He is 'hidden' in SNe Ib/c? - I.
  Low-mass objects}}.
\newblock \emph{\bibinfo{journal}{\mnras}} \textbf{\bibinfo{volume}{422}},
  \bibinfo{pages}{70--88} (\bibinfo{year}{2012}).

\bibitem{Dessart2020b}
\bibinfo{author}{{Dessart}, L.}, \bibinfo{author}{{Yoon}, S.-C.},
  \bibinfo{author}{{Aguilera-Dena}, D.~R.} \& \bibinfo{author}{{Langer}, N.}
\newblock \bibinfo{title}{{Supernovae Ib and Ic from the explosion of helium
  stars}}.
\newblock \emph{\bibinfo{journal}{\aap}} \textbf{\bibinfo{volume}{642}},
  \bibinfo{pages}{A106} (\bibinfo{year}{2020}).

\bibitem{Williamson2021}
\bibinfo{author}{{Williamson}, M.}, \bibinfo{author}{{Kerzendorf}, W.} \& \bibinfo{author}{{Modjaz}, M.}
\newblock \bibinfo{title}{{Modeling Type Ic Supernovae with TARDIS: Hidden Helium in SN 1994I?}}
\newblock \emph{\bibinfo{journal}{\apj}} \textbf{\bibinfo{volume}{908}}, \bibinfo{pages}{150} (\bibinfo{year}{2021}).

\bibitem{Tinyanont2023}
\bibinfo{author}{{Tinyanont}, S.} \emph{et~al.}
\newblock \bibinfo{title}{{Keck Infrared Transient Survey I: Survey Description and Data Release 1}}.
\newblock \emph{\bibinfo{journal}{arXiv e-prints}}
  \bibinfo{pages}{arXiv:2309.07102} (\bibinfo{year}{2023}).

\bibitem{Dessart2019}
\bibinfo{author}{{Dessart}, L.}
\newblock \bibinfo{title}{{Simulations of light curves and spectra for
  superluminous Type Ic supernovae powered by magnetars}}.
\newblock \emph{\bibinfo{journal}{\aap}} \textbf{\bibinfo{volume}{621}},
  \bibinfo{pages}{A141} (\bibinfo{year}{2019}).

\bibitem{Omand2023}
\bibinfo{author}{{Omand}, C. M.~B.} \& \bibinfo{author}{{Jerkstrand}, A.}
\newblock \bibinfo{title}{{Towards Nebular Spectral Modeling of Magnetar-Powered Supernovae}}.
\newblock \emph{\bibinfo{journal}{\aap}} \textbf{\bibinfo{volume}{673}},
  \bibinfo{pages}{A107} (\bibinfo{year}{2023}).
  

\bibitem{Budaj2005}
\bibinfo{author}{{Budaj}, J.}, \bibinfo{author}{{Richards}, M.~T.} \&
  \bibinfo{author}{{Miller}, B.}
\newblock \bibinfo{title}{{A Study of Synthetic and Observed
  H{\ensuremath{\alpha}} Spectra of TT Hydrae}}.
\newblock \emph{\bibinfo{journal}{\apj}} \textbf{\bibinfo{volume}{623}},
  \bibinfo{pages}{411--424} (\bibinfo{year}{2005}).

\bibitem{Miller2007}
\bibinfo{author}{{Miller}, B.}, \bibinfo{author}{{Budaj}, J.},
  \bibinfo{author}{{Richards}, M.}, \bibinfo{author}{{Koubsk{\'y}}, P.} \&
  \bibinfo{author}{{Peters}, G.~J.}
\newblock \bibinfo{title}{{Revealing the Nature of Algol Disks through Optical
  and UV Spectroscopy, Synthetic Spectra, and Tomography of TT Hydrae}}.
\newblock \emph{\bibinfo{journal}{\apj}} \textbf{\bibinfo{volume}{656}},
  \bibinfo{pages}{1075--1091} (\bibinfo{year}{2007}).

\bibitem{Atwood-Stone2012}
\bibinfo{author}{{Atwood-Stone}, C.}, \bibinfo{author}{{Miller}, B.~P.},
  \bibinfo{author}{{Richards}, M.~T.}, \bibinfo{author}{{Budaj}, J.} \&
  \bibinfo{author}{{Peters}, G.~J.}
\newblock \bibinfo{title}{{Modeling the Accretion Structure of AU Mon}}.
\newblock \emph{\bibinfo{journal}{\apj}} \textbf{\bibinfo{volume}{760}},
  \bibinfo{pages}{134} (\bibinfo{year}{2012}).

\bibitem{Patat1995}
\bibinfo{author}{{Patat}, F.}, \bibinfo{author}{{Chugai}, N.} \&
  \bibinfo{author}{{Mazzali}, P.~A.}
\newblock \bibinfo{title}{{Late-time H{\ensuremath{\alpha}} emission from the
  hydrogen shell of SN 1993J.}}
\newblock \emph{\bibinfo{journal}{\aap}} \textbf{\bibinfo{volume}{299}},
  \bibinfo{pages}{715} (\bibinfo{year}{1995}).

\bibitem{Maeda2007}
\bibinfo{author}{{Maeda}, K.} \emph{et~al.}
\newblock \bibinfo{title}{{The Unique Type Ib Supernova 2005bf at Nebular
  Phases: A Possible Birth Event of a Strongly Magnetized Neutron Star}}.
\newblock \emph{\bibinfo{journal}{\apj}} \textbf{\bibinfo{volume}{666}},
  \bibinfo{pages}{1069--1082} (\bibinfo{year}{2007}).

\bibitem{Taubenberger2011}
\bibinfo{author}{{Taubenberger}, S.} \emph{et~al.}
\newblock \bibinfo{title}{{The He-rich stripped-envelope core-collapse
  supernova 2008ax}}.
\newblock \emph{\bibinfo{journal}{\mnras}} \textbf{\bibinfo{volume}{413}},
  \bibinfo{pages}{2140--2156} (\bibinfo{year}{2011}).

\bibitem{Jerkstrand2015}
\bibinfo{author}{{Jerkstrand}, A.} \emph{et~al.}
\newblock \bibinfo{title}{{Late-time spectral line formation in Type IIb
  supernovae, with application to SN 1993J, SN 2008ax, and SN 2011dh}}.
\newblock \emph{\bibinfo{journal}{\aap}} \textbf{\bibinfo{volume}{573}},
  \bibinfo{pages}{A12} (\bibinfo{year}{2015}).

\bibitem{Fang2018}
\bibinfo{author}{{Fang}, Q.} \& \bibinfo{author}{{Maeda}, K.}
\newblock \bibinfo{title}{{The Origin of the Ha-like Structure in Nebular
  Spectra of Type IIb Supernovae}}.
\newblock \emph{\bibinfo{journal}{\apj}} \textbf{\bibinfo{volume}{864}},
  \bibinfo{pages}{47} (\bibinfo{year}{2018}).

\bibitem{Matheson2000a}
\bibinfo{author}{{Matheson}, T.} \emph{et~al.}
\newblock \bibinfo{title}{{Optical Spectroscopy of Supernova 1993J During Its
  First 2500 Days}}.
\newblock \emph{\bibinfo{journal}{\aj}} \textbf{\bibinfo{volume}{120}},
  \bibinfo{pages}{1487--1498} (\bibinfo{year}{2000}).

\bibitem{Matheson2000b}
\bibinfo{author}{{Matheson}, T.}, \bibinfo{author}{{Filippenko}, A.~V.},
  \bibinfo{author}{{Ho}, L.~C.}, \bibinfo{author}{{Barth}, A.~J.} \&
  \bibinfo{author}{{Leonard}, D.~C.}
\newblock \bibinfo{title}{{Detailed Analysis of Early to Late-Time Spectra of
  Supernova 1993J}}.
\newblock \emph{\bibinfo{journal}{\aj}} \textbf{\bibinfo{volume}{120}},
  \bibinfo{pages}{1499--1515} (\bibinfo{year}{2000}).

\bibitem{Dessart2021}
\bibinfo{author}{{Dessart}, L.}, \bibinfo{author}{{Hillier}, D.~J.},
  \bibinfo{author}{{Sukhbold}, T.}, \bibinfo{author}{{Woosley}, S.~E.} \&
  \bibinfo{author}{{Janka}, H.~T.}
\newblock \bibinfo{title}{{Nebular phase properties of supernova Ibc from
  He-star explosions}}.
\newblock \emph{\bibinfo{journal}{\aap}} \textbf{\bibinfo{volume}{656}},
  \bibinfo{pages}{A61} (\bibinfo{year}{2021}).

\bibitem{Marietta2000}
\bibinfo{author}{{Marietta}, E.}, \bibinfo{author}{{Burrows}, A.} \&
  \bibinfo{author}{{Fryxell}, B.}
\newblock \bibinfo{title}{{Type IA Supernova Explosions in Binary Systems: The
  Impact on the Secondary Star and Its Consequences}}.
\newblock \emph{\bibinfo{journal}{\apjs}} \textbf{\bibinfo{volume}{128}},
  \bibinfo{pages}{615--650} (\bibinfo{year}{2000}).

\bibitem{LiuZW2015}
\bibinfo{author}{{Liu}, Z.-W.} \emph{et~al.}
\newblock \bibinfo{title}{{The interaction of core-collapse supernova ejecta
  with a companion star}}.
\newblock \emph{\bibinfo{journal}{\aap}} \textbf{\bibinfo{volume}{584}},
  \bibinfo{pages}{A11} (\bibinfo{year}{2015}).

\bibitem{Dessart2020a}
\bibinfo{author}{{Dessart}, L.}, \bibinfo{author}{{Leonard}, D.~C.} \&
  \bibinfo{author}{{Prieto}, J.~L.}
\newblock \bibinfo{title}{{Spectral signatures of H-rich material stripped from
  a non-degenerate companion by a Type Ia supernova}}.
\newblock \emph{\bibinfo{journal}{\aap}} \textbf{\bibinfo{volume}{638}},
  \bibinfo{pages}{A80} (\bibinfo{year}{2020}).

\bibitem{Kollmeier2019}
\bibinfo{author}{{Kollmeier}, J.~A.} \emph{et~al.}
\newblock \bibinfo{title}{{H {\ensuremath{\alpha}} emission in the nebular
  spectrum of the Type Ia supernova ASASSN-18tb}}.
\newblock \emph{\bibinfo{journal}{\mnras}} \textbf{\bibinfo{volume}{486}},
  \bibinfo{pages}{3041--3046} (\bibinfo{year}{2019}).

\bibitem{Prieto2020}
\bibinfo{author}{{Prieto}, J.~L.} \emph{et~al.}
\newblock \bibinfo{title}{{Variable H{\ensuremath{\alpha}} Emission in the
  Nebular Spectra of the Low-luminosity Type Ia SN2018cqj/ATLAS18qtd}}.
\newblock \emph{\bibinfo{journal}{\apj}} \textbf{\bibinfo{volume}{889}},
  \bibinfo{pages}{100} (\bibinfo{year}{2020}).

\bibitem{Elias-Rosa2021}
\bibinfo{author}{{Elias-Rosa}, N.} \emph{et~al.}
\newblock \bibinfo{title}{{Nebular H{\ensuremath{\alpha}} emission in Type Ia
  supernova 2016jae}}.
\newblock \emph{\bibinfo{journal}{\aap}} \textbf{\bibinfo{volume}{652}},
  \bibinfo{pages}{A115} (\bibinfo{year}{2021}).

\bibitem{Yan2015}
\bibinfo{author}{{Yan}, L.} \emph{et~al.}
\newblock \bibinfo{title}{{Detection of Broad H{\ensuremath{\alpha}} Emission Lines in the Late-time Spectra of a Hydrogen-poor Superluminous Supernova}}.
\newblock \emph{\bibinfo{journal}{\apj}} \textbf{\bibinfo{volume}{814}},
  \bibinfo{pages}{108} (\bibinfo{year}{2015}).

\bibitem{Moriya2015}
\bibinfo{author}{{Moriya}, T.~J.}, \bibinfo{author}{{Liu}, Z.-W.}, \bibinfo{author}{{Mackey}, J.}, \bibinfo{author}{{Chen}, T.-W.} \& \bibinfo{author}{{Langer}, N.}
\newblock \bibinfo{title}{{Revealing the binary origin of Type Ic superluminous supernovae through nebular hydrogen emission}}.
\newblock \emph{\bibinfo{journal}{\aap}} \textbf{\bibinfo{volume}{584}},
  \bibinfo{pages}{L5} (\bibinfo{year}{2015}).

\bibitem{Murray2010}
\bibinfo{author}{{Murray}, C.~D.} \& \bibinfo{author}{{Correia}, A.~C.~M.}
\newblock \bibinfo{title}{{Keplerian Orbits and Dynamics of Exoplanets}}.
\newblock In \bibinfo{editor}{{Seager}, S.} (ed.)
  \emph{\bibinfo{booktitle}{Exoplanets}}, \bibinfo{pages}{15--23}
  (\bibinfo{year}{2010}).

\bibitem{Fruscione2006}
\bibinfo{author}{{Fruscione}, A.} \emph{et~al.}
\newblock \bibinfo{title}{{CIAO: Chandra's data analysis system}}.
\newblock In \bibinfo{editor}{{Silva}, D.~R.} \& \bibinfo{editor}{{Doxsey}, R.~E.} (eds.) \emph{\bibinfo{booktitle}{Society of Photo-Optical Instrumentation Engineers (SPIE) Conference Series}}, vol.
  \bibinfo{volume}{6270} of \emph{\bibinfo{series}{Society of Photo-Optical Instrumentation Engineers (SPIE) Conference Series}}, \bibinfo{pages}{62701V}
  (\bibinfo{year}{2006}).

\bibitem{Kalberla2005}
\bibinfo{author}{{Kalberla}, P.~M.~W.} \emph{et~al.}
\newblock \bibinfo{title}{{The Leiden/Argentine/Bonn (LAB) Survey of Galactic HI. Final data release of the combined LDS and IAR surveys with improved stray-radiation corrections}}.
\newblock \emph{\bibinfo{journal}{\aap}} \textbf{\bibinfo{volume}{440}},
  \bibinfo{pages}{775--782} (\bibinfo{year}{2005}).

\bibitem{Guver2009}
\bibinfo{author}{{G{\"u}ver}, T.} \& \bibinfo{author}{{{\"O}zel}, F.}
\newblock \bibinfo{title}{{The relation between optical extinction and hydrogen column density in the Galaxy}}.
\newblock \emph{\bibinfo{journal}{\mnras}} \textbf{\bibinfo{volume}{400}},
  \bibinfo{pages}{2050--2053} (\bibinfo{year}{2009}).

\bibitem{Alp2018}
\bibinfo{author}{{Alp}, D.} \emph{et~al.}
\newblock \bibinfo{title}{{X-Ray Absorption in Young Core-collapse Supernova Remnants}}.
\newblock \emph{\bibinfo{journal}{\apj}} \textbf{\bibinfo{volume}{864}},
  \bibinfo{pages}{175} (\bibinfo{year}{2018}).

\bibitem{Chandra2015}
\bibinfo{author}{{Chandra}, P.}, \bibinfo{author}{{Chevalier}, R.~A.},\bibinfo{author}{{Chugai}, N.}, \bibinfo{author}{{Fransson}, C.} \&\bibinfo{author}{{Soderberg}, A.~M.}
\newblock \bibinfo{title}{{X-Ray and Radio Emission from Type IIn Supernova SN 2010jl}}.
\newblock \emph{\bibinfo{journal}{\apj}} \textbf{\bibinfo{volume}{810}},
  \bibinfo{pages}{32} (\bibinfo{year}{2015}).

\bibitem{Immler2008}
\bibinfo{author}{{Immler}, S.} \emph{et~al.}
\newblock \bibinfo{title}{{Swift and Chandra Detections of Supernova 2006jc: Evidence for Interaction of the Supernova Shock with a Circumstellar Shell}}.
\newblock \emph{\bibinfo{journal}{\apjl}} \textbf{\bibinfo{volume}{674}},
  \bibinfo{pages}{L85} (\bibinfo{year}{2008}).

\bibitem{Chandra2012}
\bibinfo{author}{{Chandra}, P.} \emph{et~al.}
\newblock \bibinfo{title}{{Radio and X-Ray Observations of SN 2006jd: Another Strongly Interacting Type IIn Supernova}}.
\newblock \emph{\bibinfo{journal}{\apj}} \textbf{\bibinfo{volume}{755}},
  \bibinfo{pages}{110} (\bibinfo{year}{2012}).

\bibitem{Stritzinger2012}
\bibinfo{author}{{Stritzinger}, M.} \emph{et~al.}
\newblock \bibinfo{title}{{Multi-wavelength Observations of the Enduring Type IIn Supernovae 2005ip and 2006jd}}.
\newblock \emph{\bibinfo{journal}{\apj}} \textbf{\bibinfo{volume}{756}},
  \bibinfo{pages}{173} (\bibinfo{year}{2012}).

\bibitem{Harrison2013}
\bibinfo{author}{{Harrison}, F.~A.} \emph{et~al.}
\newblock \bibinfo{title}{{The Nuclear Spectroscopic Telescope Array (NuSTAR) High-energy X-Ray Mission}}.
\newblock \emph{\bibinfo{journal}{\apj}} \textbf{\bibinfo{volume}{770}},
  \bibinfo{pages}{103} (\bibinfo{year}{2013}).

\bibitem{Wilson2011}
\bibinfo{author}{{Wilson}, W.~E.} \emph{et~al.}
\newblock \bibinfo{title}{{The Australia Telescope Compact Array Broad-band Backend: description and first results}}.
\newblock \emph{\bibinfo{journal}{\mnras}} \textbf{\bibinfo{volume}{416}},
  \bibinfo{pages}{832--856} (\bibinfo{year}{2011}).

\bibitem{Sault1995}
\bibinfo{author}{{Sault}, R.~J.}, \bibinfo{author}{{Teuben}, P.~J.} \&
  \bibinfo{author}{{Wright}, M.~C.~H.}
\newblock \bibinfo{title}{{A Retrospective View of MIRIAD}}.
\newblock In \bibinfo{editor}{{Shaw}, R.~A.}, \bibinfo{editor}{{Payne}, H.~E.}
  \& \bibinfo{editor}{{Hayes}, J.~J.~E.} (eds.)
  \emph{\bibinfo{booktitle}{Astronomical Data Analysis Software and Systems
  IV}}, vol.~\bibinfo{volume}{77} of \emph{\bibinfo{series}{Astronomical Society of the Pacific Conference Series}}, \bibinfo{pages}{433}
  (\bibinfo{year}{1995}).
\newblock \eprint{astro-ph/0612759}.

\bibitem{Chevalier&Soker1989}
\bibinfo{author}{{Chevalier}, R.~A.} \& \bibinfo{author}{{Soker}, N.}
\newblock \bibinfo{title}{{Asymmetric Envelope Expansion of Supernova 1987A}}.
\newblock \emph{\bibinfo{journal}{\apj}} \textbf{\bibinfo{volume}{341}},
  \bibinfo{pages}{867} (\bibinfo{year}{1989}).

\bibitem{Langer2012}
\bibinfo{author}{{Langer}, N.}
\newblock \bibinfo{title}{{Presupernova Evolution of Massive Single and Binary Stars}}.
\newblock \emph{\bibinfo{journal}{\araa}} \textbf{\bibinfo{volume}{50}}, \bibinfo{pages}{107--164} (\bibinfo{year}{2012}).

\bibitem{Gal-Yam2014}
\bibinfo{author}{{Gal-Yam}, A.} \emph{et~al.}
\newblock \bibinfo{title}{{A Wolf-Rayet-like progenitor of SN 2013cu from spectral observations of a stellar wind}}.
\newblock \emph{\bibinfo{journal}{\nat}} \textbf{\bibinfo{volume}{509}},
  \bibinfo{pages}{471--474} (\bibinfo{year}{2014}).

\bibitem{Postnov2014}
\bibinfo{author}{{Postnov}, K.~A.} \& \bibinfo{author}{{Yungelson}, L.~R.}
\newblock \bibinfo{title}{{The Evolution of Compact Binary Star Systems}}.
\newblock \emph{\bibinfo{journal}{Living Reviews in Relativity}}
  \textbf{\bibinfo{volume}{17}}, \bibinfo{pages}{3} (\bibinfo{year}{2014}).

\bibitem{Hirai2018}
\bibinfo{author}{{Hirai}, R.}, \bibinfo{author}{{Podsiadlowski}, P.} \&
  \bibinfo{author}{{Yamada}, S.}
\newblock \bibinfo{title}{{Comprehensive Study of Ejecta-companion Interaction for Core-collapse Supernovae in Massive Binaries}}.
\newblock \emph{\bibinfo{journal}{\apj}} \textbf{\bibinfo{volume}{864}},
  \bibinfo{pages}{119} (\bibinfo{year}{2018}).

\bibitem{Dosopoulou2016}
\bibinfo{author}{{Dosopoulou}, Fani} \& \bibinfo{author}{{Kalogera}, Vicky}
\newblock \bibinfo{title}{{Orbital Evolution of Mass-transferring Eccentric Binary Systems. II. Secular Evolution}}.
\newblock \emph{\bibinfo{journal}{\apj}} \textbf{\bibinfo{volume}{825}}, \bibinfo{pages}{71} (\bibinfo{year}{2016}).

\bibitem{Ogata2021}
\bibinfo{author}{{Ogata}, M.}, \bibinfo{author}{{Hirai}, R.} \&
  \bibinfo{author}{{Hijikawa}, K.}
\newblock \bibinfo{title}{{Observability of inflated companion stars after supernovae in massive binaries}}.
\newblock \emph{\bibinfo{journal}{\mnras}} \textbf{\bibinfo{volume}{505}},
  \bibinfo{pages}{2485--2499} (\bibinfo{year}{2021}).

\bibitem{Maund2004}
\bibinfo{author}{{Maund}, J.~R.}, \bibinfo{author}{{Smartt}, S.~J.},
  \bibinfo{author}{{Kudritzki}, R.~P.}, \bibinfo{author}{{Podsiadlowski}, P.}
  \& \bibinfo{author}{{Gilmore}, G.~F.}
\newblock \bibinfo{title}{{The massive binary companion star to the progenitor of supernova 1993J}}.
\newblock \emph{\bibinfo{journal}{\nat}} \textbf{\bibinfo{volume}{427}},
  \bibinfo{pages}{129--131} (\bibinfo{year}{2004}).

\bibitem{Ryder2018}
\bibinfo{author}{{Ryder}, S.~D.} \emph{et~al.}
\newblock \bibinfo{title}{{Ultraviolet Detection of the Binary Companion to the Type IIb SN 2001ig}}.
\newblock \emph{\bibinfo{journal}{\apj}} \textbf{\bibinfo{volume}{856}},
  \bibinfo{pages}{83} (\bibinfo{year}{2018}).

\bibitem{Maund2016}
\bibinfo{author}{{Maund}, J.~R.}, \bibinfo{author}{{Pastorello}, A.},
  \bibinfo{author}{{Mattila}, S.}, \bibinfo{author}{{Itagaki}, K.} \&
  \bibinfo{author}{{Boles}, T.}
\newblock \bibinfo{title}{{The Possible Detection of a Binary Companion to a Type Ibn Supernova Progenitor}}.
\newblock \emph{\bibinfo{journal}{\apj}} \textbf{\bibinfo{volume}{833}}, \bibinfo{pages}{128} (\bibinfo{year}{2016}).

\bibitem{Sun2020}
\bibinfo{author}{{Sun}, N.-C.}, \bibinfo{author}{{Maund}, J.~R.},
  \bibinfo{author}{{Hirai}, R.}, \bibinfo{author}{{Crowther}, P.~A.} \&
  \bibinfo{author}{{Podsiadlowski}, P.}
\newblock \bibinfo{title}{{Origins of Type Ibn SNe 2006jc/2015G in interacting binaries and implications for pre-SN eruptions}}.
\newblock \emph{\bibinfo{journal}{\mnras}} \textbf{\bibinfo{volume}{491}},
  \bibinfo{pages}{6000--6019} (\bibinfo{year}{2020}).

\bibitem{Folatelli2014}
\bibinfo{author}{{Folatelli}, G.} \emph{et~al.}
\newblock \bibinfo{title}{{A Blue Point Source at the Location of Supernova 2011dh}}.
\newblock \emph{\bibinfo{journal}{\apjl}} \textbf{\bibinfo{volume}{793}},
  \bibinfo{pages}{L22} (\bibinfo{year}{2014}).

\bibitem{Maund2019}
\bibinfo{author}{{Maund}, J.~R.}
\newblock \bibinfo{title}{{The Origin of the Late-time Luminosity of Supernova 2011dh}}.
\newblock \emph{\bibinfo{journal}{\apj}} \textbf{\bibinfo{volume}{883}},
  \bibinfo{pages}{86} (\bibinfo{year}{2019}).

\bibitem{Tauris1998}
\bibinfo{author}{{Tauris}, T.~M.} \& \bibinfo{author}{{Takens}, R.~J.}
\newblock \bibinfo{title}{{Runaway velocities of stellar components originating
  from disrupted binaries via asymmetric supernova explosions}}.
\newblock \emph{\bibinfo{journal}{\aap}} \textbf{\bibinfo{volume}{330}},
  \bibinfo{pages}{1047--1059} (\bibinfo{year}{1998}).

\bibitem{Portegies2000}
\bibinfo{author}{{Portegies Zwart}, S.~F.}
\newblock \bibinfo{title}{{The Characteristics of High-Velocity O and B Stars
  Which Are Ejected from Supernovae in Binary Systems}}.
\newblock \emph{\bibinfo{journal}{\apj}} \textbf{\bibinfo{volume}{544}},
  \bibinfo{pages}{437--442} (\bibinfo{year}{2000}).

\bibitem{Kochanek2019}
\bibinfo{author}{{Kochanek}, C.~S.}, \bibinfo{author}{{Auchettl}, K.} \&
  \bibinfo{author}{{Belczynski}, K.}
\newblock \bibinfo{title}{{Stellar binaries that survive supernovae}}.
\newblock \emph{\bibinfo{journal}{\mnras}} \textbf{\bibinfo{volume}{485}},
  \bibinfo{pages}{5394--5410} (\bibinfo{year}{2019}).

\bibitem{Kochanek2021}
\bibinfo{author}{{Kochanek}, C.~S.}
\newblock \bibinfo{title}{{Supernovae producing unbound binaries and triples}}.
\newblock \emph{\bibinfo{journal}{\mnras}} \textbf{\bibinfo{volume}{507}},
  \bibinfo{pages}{5832--5846} (\bibinfo{year}{2021}).


\bibitem{Hameury2020}
\bibinfo{author}{{Hameury}, J.~M.}
\newblock \bibinfo{title}{{A review of the disc instability model for dwarf novae, soft X-ray transients and related objects}}.
\newblock \emph{\bibinfo{journal}{Advances in Space Research}} \textbf{\bibinfo{volume}{66}},
  \bibinfo{pages}{1004-1024} (\bibinfo{year}{2020}).


\bibitem{Atwood2009}
\bibinfo{author}{{Atwood}, W.~B.} \emph{et~al.}
\newblock \bibinfo{title}{{The Large Area Telescope on the Fermi Gamma-Ray Space Telescope Mission}}.
\newblock \emph{\bibinfo{journal}{\apj}} \textbf{\bibinfo{volume}{697}},
  \bibinfo{pages}{1071--1102} (\bibinfo{year}{2009}).

\bibitem{fssc}
\bibinfo{title}{\url{https://fermi.gsfc.nasa.gov/cgi-bin/ssc/LAT/LATDataQuery.cgi}}

\bibitem{Abdollahi2022}
\bibinfo{author}{{Abdollahi}, S.} \emph{et~al.}
\newblock \bibinfo{title}{{Incremental Fermi Large Area Telescope Fourth Source
  Catalog}}.
\newblock \emph{\bibinfo{journal}{\apjs}} \textbf{\bibinfo{volume}{260}},
  \bibinfo{pages}{53} (\bibinfo{year}{2022}).

\bibitem{gammabkg}
\bibinfo{title}{\url{https://fermi.gsfc.nasa.gov/ssc/data/access/lat/BackgroundModels.html}}


\bibitem{Ofek2018}
\bibinfo{author}{{Ofek}, E.~O.} \& \bibinfo{author}{{Zackay}, B.}
\newblock \bibinfo{title}{{Optimal Matched Filter in the Low-number Count
  Poisson Noise Regime and Implications for X-Ray Source Detection}}.
\newblock \emph{\bibinfo{journal}{\aj}} \textbf{\bibinfo{volume}{155}},
  \bibinfo{pages}{169} (\bibinfo{year}{2018}).

\bibitem{Dullo2020}
\bibinfo{author}{{Dullo}, B.~T.}, \bibinfo{author}{{Bouquin}, A. Y.~K.},
  \bibinfo{author}{{Gil de Paz}, A.}, \bibinfo{author}{{Knapen}, J.~H.} \&
  \bibinfo{author}{{Gorgas}, J.}
\newblock \bibinfo{title}{{The Black Hole Mass-Color Relations for Early- and
  Late-type Galaxies: Red and Blue Sequences}}.
\newblock \emph{\bibinfo{journal}{\apj}} \textbf{\bibinfo{volume}{898}},
  \bibinfo{pages}{83} (\bibinfo{year}{2020}).

\bibitem{Itoh2020}
\bibinfo{author}{{Itoh}, R.} \emph{et~al.}
\newblock \bibinfo{title}{{Blazar Radio and Optical Survey (BROS): A Catalog of
  Blazar Candidates Showing Flat Radio Spectrum and Their Optical
  Identification in Pan-STARRS1 Surveys}}.
\newblock \emph{\bibinfo{journal}{\apj}} \textbf{\bibinfo{volume}{901}},
  \bibinfo{pages}{3} (\bibinfo{year}{2020}).

\bibitem{Liodakis2019}
\bibinfo{author}{{Liodakis}, I.}, \bibinfo{author}{{Romani}, R.~W.},
  \bibinfo{author}{{Filippenko}, A.~V.}, \bibinfo{author}{{Kocevski}, D.} \&
  \bibinfo{author}{{Zheng}, W.}
\newblock \bibinfo{title}{{Probing Blazar Emission Processes with
  Optical/Gamma-Ray Flare Correlations}}.
\newblock \emph{\bibinfo{journal}{\apj}} \textbf{\bibinfo{volume}{880}},
  \bibinfo{pages}{32} (\bibinfo{year}{2019}).

\bibitem{Matz1988}
\bibinfo{author}{{Matz}, S.~M.} \emph{et~al.}
\newblock \bibinfo{title}{{Gamma-ray line emission from SN1987A}}.
\newblock \emph{\bibinfo{journal}{\nat}} \textbf{\bibinfo{volume}{331}},
  \bibinfo{pages}{416--418} (\bibinfo{year}{1988}).

\bibitem{Teegarden1989}
\bibinfo{author}{{Teegarden}, B.~J.}, \bibinfo{author}{{Barthelmy}, S.~D.},
  \bibinfo{author}{{Gehrels}, N.}, \bibinfo{author}{{Tueller}, J.} \&
  \bibinfo{author}{{Leventhal}, M.}
\newblock \bibinfo{title}{{Resolution of the 1,238-keV
  {\ensuremath{\gamma}}-ray line from supernova 1987A}}.
\newblock \emph{\bibinfo{journal}{\nat}} \textbf{\bibinfo{volume}{339}},
  \bibinfo{pages}{122--123} (\bibinfo{year}{1989}).

\bibitem{Churazov2015}
\bibinfo{author}{{Churazov}, E.} \emph{et~al.}
\newblock \bibinfo{title}{{Gamma-rays from Type Ia Supernova SN2014J}}.
\newblock \emph{\bibinfo{journal}{\apj}} \textbf{\bibinfo{volume}{812}},
  \bibinfo{pages}{62} (\bibinfo{year}{2015}).

\bibitem{Ackermann2015}
\bibinfo{author}{{Ackermann}, M.} \emph{et~al.}
\newblock \bibinfo{title}{{Search for Early Gamma-ray Production in Supernovae
  Located in a Dense Circumstellar Medium with the Fermi LAT}}.
\newblock \emph{\bibinfo{journal}{\apj}} \textbf{\bibinfo{volume}{807}},
  \bibinfo{pages}{169} (\bibinfo{year}{2015}).

\bibitem{Renault-Tinacci2018}
\bibinfo{author}{{Renault-Tinacci}, N.}, \bibinfo{author}{{Kotera}, K.},
  \bibinfo{author}{{Neronov}, A.} \& \bibinfo{author}{{Ando}, S.}
\newblock \bibinfo{title}{{Search for {\ensuremath{\gamma}}-ray emission from
  superluminous supernovae with the Fermi-LAT}}.
\newblock \emph{\bibinfo{journal}{\aap}} \textbf{\bibinfo{volume}{611}},
  \bibinfo{pages}{A45} (\bibinfo{year}{2018}).

\bibitem{Yuan2018}
\bibinfo{author}{{Yuan}, Q.} \emph{et~al.}
\newblock \bibinfo{title}{{Fermi Large Area Telescope Detection of Gamma-Ray
  Emission from the Direction of Supernova iPTF14hls}}.
\newblock \emph{\bibinfo{journal}{\apjl}} \textbf{\bibinfo{volume}{854}},
  \bibinfo{pages}{L18} (\bibinfo{year}{2018}).

\bibitem{Prokhorov2021}
\bibinfo{author}{{Prokhorov}, D.~A.}, \bibinfo{author}{{Moraghan}, A.} \&
  \bibinfo{author}{{Vink}, J.}
\newblock \bibinfo{title}{{Search for gamma rays from SNe with a variable-size
  sliding-time-window analysis of the Fermi-LAT data}}.
\newblock \emph{\bibinfo{journal}{\mnras}} \textbf{\bibinfo{volume}{505}},
  \bibinfo{pages}{1413--1421} (\bibinfo{year}{2021}).

\bibitem{Xi2020}
\bibinfo{author}{{Xi}, S.-Q.} \emph{et~al.}
\newblock \bibinfo{title}{{A Serendipitous Discovery of GeV Gamma-Ray Emission
  from Supernova 2004dj in a Survey of Nearby Star-forming Galaxies with
  Fermi-LAT}}.
\newblock \emph{\bibinfo{journal}{\apjl}} \textbf{\bibinfo{volume}{896}},
  \bibinfo{pages}{L33} (\bibinfo{year}{2020}).

\bibitem{Abdo2009}
\bibinfo{author}{{Fermi LAT Collaboration}} \emph{et~al.}
\newblock \bibinfo{title}{{Modulated High-Energy Gamma-Ray Emission from the
  Microquasar Cygnus X-3}}.
\newblock \emph{\bibinfo{journal}{Science}} \textbf{\bibinfo{volume}{326}},
  \bibinfo{pages}{1512} (\bibinfo{year}{2009}).

\bibitem{Piano2012}
\bibinfo{author}{{Piano}, G.} \emph{et~al.}
\newblock \bibinfo{title}{{The AGILE monitoring of Cygnus X-3: transient
  gamma-ray emission and spectral constraints}}.
\newblock \emph{\bibinfo{journal}{\aap}} \textbf{\bibinfo{volume}{545}},
  \bibinfo{pages}{A110} (\bibinfo{year}{2012}).

\bibitem{Modjaz2014}
\bibinfo{author}{{Modjaz}, M.} \emph{et~al.}
\newblock \bibinfo{title}{{Optical Spectra of 73 Stripped-envelope
  Core-collapse Supernovae}}.
\newblock \emph{\bibinfo{journal}{\aj}} \textbf{\bibinfo{volume}{147}},
  \bibinfo{pages}{99} (\bibinfo{year}{2014}).

\bibitem{Hunter2009}
\bibinfo{author}{{Hunter}, D.~J.} \emph{et~al.}
\newblock \bibinfo{title}{{Extensive optical and near-infrared observations of
  the nearby, narrow-lined type Ic SN 2007gr: days 5 to
  415}}.
\newblock \emph{\bibinfo{journal}{\aap}} \textbf{\bibinfo{volume}{508}},
  \bibinfo{pages}{371--389} (\bibinfo{year}{2009}).

\bibitem{Taubenberger2009}
\bibinfo{author}{{Taubenberger}, S.} \emph{et~al.}
\newblock \bibinfo{title}{{Nebular emission-line profiles of Type Ib/c
  supernovae - probing the ejecta asphericity}}.
\newblock \emph{\bibinfo{journal}{\mnras}} \textbf{\bibinfo{volume}{397}},
  \bibinfo{pages}{677--694} (\bibinfo{year}{2009}).

\bibitem{Jerkstrand2017}
\bibinfo{author}{{Jerkstrand}, A.} \emph{et~al.}
\newblock \bibinfo{title}{{Long-duration Superluminous Supernovae at Late
  Times}}.
\newblock \emph{\bibinfo{journal}{\apj}} \textbf{\bibinfo{volume}{835}},
  \bibinfo{pages}{13} (\bibinfo{year}{2017}).


\end{thebibliography}

\end{document}